\documentclass[10pt]{iopart}
\usepackage{url} 
\usepackage{pdflscape} 
\usepackage{amsmath}
\usepackage{amssymb} 
\usepackage{epsfig} 
\usepackage{amsfonts}
\usepackage{graphicx}
\usepackage{subfigure}
\usepackage{epsfig}
\usepackage{graphicx}
\usepackage{epsfig}
\usepackage{times,fancyhdr}
\usepackage{rotating}

\usepackage{enumitem}
\setlist[enumerate,2]{label=\roman*)}

\def\case#1/#2{\textstyle\frac{#1}{#2}}

\newcommand{\be}{\begin{equation}}
\newcommand{\ee}{\end{equation}}
\newcommand{\ben}{\begin{eqnarray}}
\newcommand{\een}{\end{eqnarray}}
\usepackage{mdframed}
\usepackage{grffile}
\usepackage{amssymb}
\usepackage{ulem}
\usepackage[T1]{fontenc}

\usepackage{subfigure}
\newtheorem{thm}{Theorem}[section]
\newtheorem{cor}{Corollary}[thm]

\newtheorem{defn}{Definition}

\providecommand{\U}[1]{\protect\rule{.1in}{.1in}}
\newcommand{\mincir}{\raise
-3.truept\hbox{\rlap{\hbox{$\sim$}}\raise4.truept\hbox{$<$}\ }}
\newcommand{\magcir}{\raise
-3.truept\hbox{\rlap{\hbox{$\sim$}}\raise4.truept\hbox{$>$}\ }}

\def\S{{}^3\!S_+}

\begin{document}

\title{Generalized scalar field cosmologies}

\author{Genly Leon}
\address{Departamento  de  Matem\'aticas,  Universidad  Cat\'olica  del  Norte, Avda. Angamos  0610,  Casilla  1280  Antofagasta,  Chile.}
\ead{genly.leon@ucn.cl}

\author{Felipe Orlando Franz Silva}
\address{Departamento  de  Matem\'aticas,  Universidad  Cat\'olica  del  Norte, Avda. Angamos  0610,  Casilla  1280  Antofagasta,  Chile.}
\ead{felipe.franz@alumnos.ucn.cl}

\begin{abstract}
In this paper, we use both local and global phase-space descriptions and averaging methods to find qualitative features of solutions for the FLRW and Bianchi I metrics in the context of scalar field cosmologies with arbitrary potentials and arbitrary couplings to matter. We prove new theorems and also retrieve previous results that can be seen as corollaries of the present results. We study the stability of the equilibrium points in a phase-space, as well as the dynamics in the regime where the scalar field diverges. We obtain equilibrium points that represent some solutions of cosmological interest, such as several types of scaling solutions, a kinetic dominated solution representing a stiff fluid, a solution dominated by an effective energy density of geometric origin, a quintessence scalar field dominated solution, the vacuum de Sitter solution associated to the minimum of the potential, and a non-interacting matter dominated solution.  All reveal a very rich cosmological phenomenology. Finally, we present some examples that violate one or more hypotheses of the Theorems proved, obtaining some counterexamples.  In particular, we incorporate small cosine-like corrections motivated by inflationary loop-quantum cosmology, and we study the oscillatory behavior. 
\end{abstract}


\maketitle

\section{Introduction}


Scalar fields are used to describe the gravitational field in scalar theories of gravitation. 
Some scalar field theories with special interest are the Scalar-tensor theories like Jordan theory \cite{Jordan:1958zz} as a generalization of the Kaluza-Klein theory, the Brans-Dicke theory \cite{Brans:1961sx}, Horndeski theories \cite{Horndeski:1974wa}, inflationary models \cite{Guth:1980zm}, extended quintessence, modified gravity, Ho\v{r}ava-Lifschitz and the Galileons, etc.,  \cite{Copeland:1993jj,Ibanez:1995zs,Chimento:1995da,Lidsey:1995np,Coley:1997nk,Copeland:1998fz,Coley:1999mj,Coley:1999uh,Coley:2000zw,Coley:2000yc,Coley:2003tf,Elizalde:2004mq,Capozziello:2005tf,Curbelo:2005dh,Gonzalez:2005ie,Gonzalez:2006cj,Lazkoz:2007mx,Elizalde:2008yf,Leon:2009dt,Leon:2009rc,Leon:2009ce,Leon:2010pu,Basilakos:2011rx,Xu:2012jf,Leon:2012mt,Leon:2013qh,Fadragas:2013ina,Kofinas:2014aka,Leon:2014yua,Paliathanasis:2014yfa,DeArcia:2015ztd,Paliathanasis:2015gga,Leon:2015via,Barrow:2016qkh,Barrow:2016wiy,Cruz:2017ecg,Paliathanasis:2017ocj,Alhulaimi:2017ocb,Dimakis:2017kwx,Giacomini:2017yuk,Karpathopoulos:2017arc,DeArcia:2018pjp,Tsamparlis:2018nyo,Paliathanasis:2018vru,Basilakos:2019dof,VanDenHoogen:2018anx,Leon:2018lnd,Leon:2018skk,Leon:2019mbo,Paliathanasis:2019qch,Leon:2019jnu,Paliathanasis:2019pcl,Barrow:2018zav,Quiros:2019ktw}. 

There are several studies in literature that provide both global and local dynamical systems analysis for scalar field cosmologies with arbitrary potentials and arbitrary couplings. 
In reference \cite{Foster:1998sk} they studied a very large and natural class of scalar field models having an arbitrary non-negative potential function $V(\phi)$ with a flat Friedmann-Lema\^{i}tre-Robertson-Walker (FLRW) metric; yielding to a simple and regular past asymptotic structure which corresponds to the exactly integrable massless scalar field cosmologies with the exception of a set that has zero measure.  This model was generalized in \cite{Miritzis:2003ym} for flat and negatively curved FLRW models by adding a perfect fluid matter source. In particular, for potentials having a local zero minimum, flat and negatively curved FLRW models are ever expanding and the energy density asymptotically approaches zero, and the scalar field asymptotically reaches the minimum of the potential. Additionally, it was commented that a closed FLRW model with ordinary matter can avoid re-collapse due to the presence of a scalar field with a non-negative potential.
The model by \cite{Miritzis:2003ym} was extended by \cite{Dania&Yunelsy,Leon:2008de} to a scalar field non-minimally coupled to matter (this scenario incidentally contains a particular realization the model of \cite{Giambo:2009byn}, which arises in the conformal frame of $f(R)$ theories non-minimally coupled to matter). It was proved that
under generic hypothesis the future attractor corresponds to the vacuum de Sitter solution by considering a generic potential  $V(\phi)$ and a generic coupling function $\chi(\phi)$. Also, it was proved that the scalar field diverges into the past extending results by \cite{Foster:1998sk,Miritzis:2003ym}. So, in order to study the dynamics close to the initial singularity, the limit $\phi\rightarrow \infty$ was studied by imposing some regularity
conditions on the potential and on the coupling function. Interestingly, for a general class of models which admit scaling solutions, as in \cite{Miritzis:2003ym}, the asymptotic structure of solutions towards the past is simple and regular, and it is independent of the features of the potential, the coupling function, and the background matter. The dynamics of a non-minimally coupled scalar field model in the case of a $F(\phi)R$ coupling with $F(\phi)=1-\xi \phi^2$ and the potentials $V(\phi)=V_0(1+\phi^p)^2$ and $V(\phi)=V_0e^{\lambda \phi^2}$ were studied in \cite{Shahalam:2019jgs}. Other non-minimally coupled scalar field models were studied in e.g.:  \cite{Nojiri:2019riz,Humieja:2019ywy,Matsumoto:2017gnx,Matsumoto:2015hua,Solomon:2015hja,Harko:2015pma,Minazzoli:2014xua,Skugoreva:2013ooa,Jamil:2012vb,Miritzis:2011zz,Hrycyna:2007gd}.

In reference \cite{Giambo:2008ck}, homogeneous FLRW cosmological models with a self-interacting scalar field source were studied, not only for the flat geometry but also for negatively and positively curved FLRW models. The analysis incorporates a wide class of interaction potentials, and only requires a scalar field potential to be bounded from below and divergent when the field diverges. Thus, incorporating positive potentials, that exhibit asymptotically polynomial or exponential behaviors. Potentials with a negative inferior bound lead asymptotically to anti de Sitter (AdS) solutions for such cosmologies.

In reference \cite{Giambo:2009byn}, the evolution of a cosmological model with a perfect fluid matter source {with energy density $\rho_m$, and pressure $p_m$, with an equation of state parameter $p_m=(\gamma-1) \rho_m$}, and with a scalar field non-minimally coupled to matter with an exponential coupling $\chi$ (in the sense of \cite{Dania&Yunelsy,Leon:2008de}) was studied for flat and negatively curved FLRW models. It was proved in \cite{Giambo:2009byn} that there exists a very generic class of potentials having an equilibrium point which corresponds to the non-negative local minimum for $V(\phi)$, which is asymptotically stable. The same happens for horizontal asymptotes approached from above by $V(\phi)$. Furthermore, in this reference there were classified all flat models for which one the matter constituents will eventually dominate. Particularly, if the barotropic matter index $\gamma$ is larger than 1. Generically, there is an energy transfer from the fluid to the scalar field which eventually dominates over the background matter.

In references \cite{Leon:2010ai,Leon:2014bta,Leon:2014rra,Fadragas:2014mra}, the original models by \cite{Foster:1998sk,Miritzis:2003ym} were gradually extended to more general scenarios. In \cite{Fadragas:2014mra}, they studied the flat FLRW models in the conformal (Einstein) frame of scalar-tensor gravity theories for arbitrary positive potentials and arbitrary coupling functions, and incorporating radiation in the matter content to obtain a more realistic scenario. In \cite{Leon:2010ai,Fadragas:2014mra}, a procedure for equations analysis in the limit $\phi \rightarrow \infty$ was implemented by using a suitable change of variables. The method has been exemplified for: (a) a double exponential potential $V(\phi)= V_1 e^{\alpha \phi}+ V_2 e^{\beta \phi}$, $\alpha$ and $\beta$ are constants that satisfy $0<\alpha<\beta$, and
a coupling function $\chi=\chi_0 e^{\frac{\lambda \phi}{4-3\gamma}}$, where $\lambda$ is a constant, discussed in \cite{Tzanni:2014eja}, and
(b) a general class of potentials containing the cases investigated in \cite{vandenHoogen:1999qq,Albrecht:1999rm,Copeland:1997et}, being the so-called Albrecht-Skordis potential $V(\phi)= e^{-\mu \phi}\left(A-(\phi-B)^2\right)$, and a power-law coupling $\chi(\phi)=\left(\frac{3\alpha}{8}\right)^{\frac{1}{\alpha}}\chi_0 \left(\phi-\phi_0\right)^{\frac{2}{\alpha}}$, with $\alpha>0$, constant, and $\phi_0\geq 0$, originally investigated in \cite{Leon:2008de} for a less general model. In \cite{Tzanni:2014eja}, a flat FLRW model with a perfect fluid source and a scalar field with double exponential potential which is non-minimally coupled to matter was studied. The coupling is derived from the formulation of the $f(R)$- gravity as an equivalent scalar-tensor theory. There were provided conditions for which $\rho_m\rightarrow 0, \dot \phi\rightarrow 0$ and $\phi \rightarrow +\infty$ as $t\rightarrow \infty$ (see  Proposition 1 of \cite{Tzanni:2014eja}), and conditions under which $H$ and $\phi$ blows-up in a finite time  (see  Proposition 2 of \cite{Tzanni:2014eja}).  In the reference \cite{Giambo:2019ymx}, the late time of a negatively curved FLRW model  with a perfect fluid matter source, {with energy density $\rho_m$, and pressure $p_m$, with an equation of state parameter $p_m=(\gamma-1) \rho_m$,} and a scalar field non-minimally coupled to matter was studied. Under mild assumptions on the potential, it was found that equilibria correspond to non-negative local minima of $V$ are asymptotically stable. For non-degenerated minima with zero critical value, it was proved that for $\gamma> 2/3$, there is a transfer of energy from the fluid and the scalar field to the energy density of the scalar curvature; in contrast to the previous bound $\gamma> 1$ for a flat FLRW model. Thus, if there is a scalar curvature,
it has a dominant effect on the late evolution of the universe and eventually dominates both the perfect fluid and the scalar field. The analysis was complemented with a case where $V$ is exponential and therefore the scalar field diverges to infinity.

In \cite{Cid:2017wtf}, was presented a generalized Brans-Dicke Lagrangian including a non-minimally coupled Gauss-Bonnet term without imposing the vanishing torsion condition. The cosmological consequences of this model were studied. In \cite{Paliathanasis:2018vru}, the existence of exact solutions and integrable dynamical systems in multi-scalar field cosmology, more specifically, in the so-called Chiral cosmology where non-linear terms exist in the kinetic term of the scalar fields was studied. Some exact analytic solutions for a system of N-scalar fields were presented. Some studies of cosmological effects of scalar fields and their effects in multiple-field
inflation are: \cite{Khlopov:1985jw,Sakharov:1993qh,Sakharov:1994pr,Rubin:2001yw,Khlopov:2002yi,Khlopov:2004sc,Khlopov:2008qy}.

This paper is organized as follows: in Section \ref{Model} we investigate a scalar field which has an arbitrary self-interacting potential, and it is non-minimally coupled to matter through an arbitrary coupling function in which we analyze the corresponding cosmology.  In particular, we study several specific cases and extend the results in literature. In Section \ref{SECT:Non_min}, we discuss our main results including the new Theorems \ref{Proposition I}, \ref{Theorem5.2}, \ref{Theorem5.3} and \ref{thmIIIFINAL}, and Corollaries \ref{Proposition II} and \ref{Proposition III} which are collectively referred as Leon \& Franz-Silva 2019. Some well-known results like Corollaries \ref{PropositionIb}, \ref{tm}, \ref{PropositionII}, \ref{thm2.1}, \ref{Prop4Miritzis} and \ref{thm2.2} are recovered. In Section \ref{SECT:6}, we provide a dynamical systems analysis for arbitrary potentials $V(\phi)$ and arbitrary couplings $\chi(\phi)$. We start by a local dynamical systems analysis using Hubble normalized equations in Section
\ref{SECT:6.1}, and then we proceed to a dynamical systems formulation in Section \ref{SECT:3.3} using global dynamical systems variables based on Alho \& Uggla's approach \cite{Alho:2014fha}.  Section \ref{SECT:3.4} is devoted to the asymptotic analysis as $\phi\rightarrow \infty$ for arbitrary $V(\phi)$ and $\chi(\phi)$. In particular, Section \ref{Section2.3} is devoted to equations analysis of a scalar field model with potential $V(\phi)= V_0 e^{-\lambda \phi}$ in a vacuum. In Sections \ref{Sect:2.4.1} and \ref{Sect:2.5.1} we present the asymptotic analysis as $\phi\rightarrow \infty$ for generalized harmonic potentials $V_1(\phi)=\mu^3 \left[\frac{\phi^2}{\mu} + b f \cos\left(\delta + \frac{\phi}{f}\right)\right]$, $b\neq 0$ and $V_2(\phi)= \mu ^3 \left[b f \left(\cos (\delta )-\cos \left(\delta +\frac{\phi }{f}\right)\right)+\frac{\phi ^2}{\mu}\right]$, $b\neq 0$, respectively, in a vacuum. These potentials incorporate cosine-like corrections with small phase motivated by inflationary loop-quantum cosmology \cite{Sharma:2018vnv}. Section \ref{SECT4} is devoted to an alternative dynamical systems formulation for a scalar-field cosmology with generalized harmonic potentials in a vacuum. More specific, in Sections \ref{Sect2.4}, and \ref{Sect.2.5}, we provide a qualitative analysis for a scalar-field cosmology with potentials $V_1(\phi)$ and $V_2(\phi)$ using a new formulation.  In Sections \ref{Sect:2.4.2} and \ref{Sect:2.5.2}, we investigate the oscillatory regime for the scalar field under the potentials $V_1(\phi)$ and $V_2(\phi)$, respectively. Finally, Section \ref{Sect:7} is devoted to our conclusions.

\section{Theorems on Asymptotic Behavior}
\label{Model}

The action for a general class of Scalar Tensor Theory of Gravity is written in the so-called Einstein frame (EF), and given by \cite{Kaloper:1997sh,Gonzalez:2007ht}:
\begin{align}&
\mathcal{L}_{\phi}=\int  d{ }^4 x \sqrt{|g|}\left\{\frac{1}{2} R-\frac{1}{2} g^{\mu
\nu}\nabla_\mu\phi\nabla_\nu\phi-V(\phi)+\chi(\phi)^{-2}
\mathcal{L}(\mu,\nabla\mu,\chi(\phi)^{-1}g_{\alpha\beta})\right\}.\label{eq1}
\end{align} We use a system of units in which $8\pi G=c=\hbar=1.$
In this equation $R$ is the curvature scalar, $\phi$ is the 
scalar field,  $\nabla_\alpha$
is the covariant derivative, $V(\phi)$ is the quintessence self-interaction potential,
$\chi(\phi)^{-2}$ is the coupling function, $\mathcal{L}$
is the matter Lagrangian, and $\mu$ is the collective name for the
matter degrees of freedom. 

The matter energy-momentum tensor is defined by:
\begin{equation}T_{\alpha
\beta}=-\frac{2}{\sqrt{|g|}}\frac{\delta}{\delta g^{\alpha
\beta}}\left\{\sqrt{|g|}
 \chi^{-2}\mathcal{L}(\mu,\nabla\mu,\chi^{-1}g_{\alpha
 \beta})\right\}.\label{Tab}\end{equation}
We define the ``energy exchange'' vector as: 
$$Q_\beta\equiv\nabla^\alpha T_{\alpha \beta}=-\frac{1}{2}T\frac{1}{\chi(\phi)}\frac{\mathrm{d}\chi(\phi)}{\mathrm{d}\phi}\nabla_{\beta}\phi,\;
 T=T^\alpha_\alpha,$$
 where $T$ is the trace of the energy-momentum tensor. 
 Additionally, we incorporate the geometric properties of the metric in the form of an effective function 
\begin{align}
&G_0(a)=\left\{ \begin{array}{cc}
-3\frac{k}{a^2}, k=0, \pm 1, & \text{FLRW}\\
\frac{\sigma_0^2}{a^6}, & \text{Bianchi I}
\end{array} 
\right.,
\end{align} and we obtain the equations of motion for a  scalar field cosmology with the scalar field non-minimally coupled to matter, given by:
\begin{subequations}
	\label{Non_min2}
	\begin{align}
	& \dot{H}=-\frac{1}{2}\left(\gamma \rho_m+y^2\right)+\frac{1}{6}a G_0'(a), \label{Rachd}\\
	&\dot{\rho_m}=-3\gamma H\rho_m-\frac{1}{2}(4-3\gamma)\rho_m y\frac{d\ln\chi(\phi)}{d \phi}, \label{consb}\\
	& \dot a= a H,\\
	&\dot y=-3 H y -\frac{d V(\phi)}{d\phi}+\frac{1}{2}(4-3\gamma)\rho_m \frac{d\ln\chi(\phi)}{d \phi} \label{13EQ},\\
	& \dot\phi=y, \\
	& 3H^2=\rho_m+\frac{1}{2}\dot\phi^2+V(\phi)+\Lambda+G_0(a),\label{Fried2b}
	\end{align}
\end{subequations}
where we assume  $\Lambda\geq 0$, and  $\gamma \in [1,2]$. Equation \eqref{Fried2b} allows us to define the phase space
\begin{equation}
\label{Fried2bB}
   \left\{(H, \rho_m, a, y, \phi)\in \mathbb{R}^5: 3H^2=\rho_m+\frac{1}{2}y^2+V(\phi)+\Lambda+G_0(a)\right\}.
\end{equation}
\\
In the present research, we mainly consider an inverse power-law non-negative ``geometric'' term $G_0(a)$; but due its generality, it can effectively behaves  like radiation fluid (see, e.g., \cite{Fadragas:2014mra}) where the energy density decays as $\propto a^{-4}$; or as  stiff fluid, where the energy density decays as $\propto a^{-6}$.  Any effective non-negative energy density that depends on the scale factor $a$ it can be considered as sub-cases of the present model as well.

\subsection{Main theorems}
\label{SECT:Non_min}

Firstly, we  study the cases  $G_0(a)=-3\frac{k}{a^2}, k=0, - 1$, and $G_0(a)=\frac{\sigma_0^2}{a^6}$, which are special cases of $G_0(a)= \frac{K^2}{a^p} \geq 0$, with $K=0$ for the flat FRW metric, $K^2=1, p=2$ for the negatively curved  FLRW metric, or $K^2=\sigma_0^2, p=6$ for the Bianchi I metric. The FLRW model  with positive curvature, $k=+1$, will be discussed in Section \ref{positive-k}.
We define $\Omega=\left\{(H, \rho_m, a, y, \phi)\in \mathbb{R}^5: 3H^2=\rho_m+\frac{1}{2}y^2+V(\phi)+\Lambda+\frac{K^2}{a^p}\right\}$, and assume $V(\phi)\geq 0$ is of class $C^2(\mathbb{R})$. Assuming that $V(\phi)$ has a local minimum at $\phi=0$, $V(0)=0$. In this case $(0,0,a_*,0,0)$, $a_*\rightarrow +\infty$, is an equilibrium configuration for the flow of \eqref{Non_min2}. Due to the set  $\left\{(H, \rho_m, a, y, \phi)\in \Omega: H=0\right\}$, is invariant for the flow of  \eqref{Non_min2}, $H$ does not change the sign; on the contrary, if there is an orbit with  $H(0)>0$ and $H(t_1)<0$ for some $t_1>0$, this solution will pass through the origin violating the existence and uniqueness of the solutions of a $C^1$ flow. 

\begin{thm}[Leon \& Franz-Silva, 2019]
\label{Proposition I}
Assuming 
\begin{enumerate}
\item  $V(\phi)\in C^2(\mathbb{R}), V(\phi)\geq 0$, and $V(\phi)=0$ if and only if $\phi=0$. 
\item  $V^{\prime}(\phi)$ is bounded on $A\subset\mathbb{R}$ if $V(\phi)$ is bounded on  $A$.
\item $\Lambda\geq 0$, and $G_0(a)\geq 0$ have a negative power-law functional form $G_0(a)=\frac{K^2}{a^p}, p>0$. 
\item $\chi(\phi)\in C^2(\mathbb{R})$, and for all $A\subset \mathbb{R}$ there is a non-negative constant $K_1$, possibly depending on $A$, such as $\left| \frac{\chi '(\phi)}{\chi (\phi)}\right|\leq K_1$ for all $\phi \in A$. 
\end{enumerate}
Then, $\lim_{t\rightarrow \infty} \left( \rho_m, y, \frac{K^2}{a^p}\right)=(0,0,0)$.
\end{thm}
\textbf{Proof}. Let the positive orbit $O^+(x_0)$, passing at the time $t_0$ through the regular point $x_0\in
\left\{(H, \rho_m, a, y, \phi)\in \Omega: H>0\right\}$. Since $H$ is positive and decreasing along $O^+(x_0)$, there exists  $\lim_{t\rightarrow \infty} H(t)$ and it is a non-negative number $\eta$. Furthermore, $H(t)\leq H(t_0)$ for all $t>t_0$. Then, 
 $\rho_m(t)+\frac{1}{2}y(t)^2+V(\phi(t))+\Lambda+\frac{K^2}{a(t)^p}= 3H^2\leq 3 H(t_0)^2$, for all $t>t_0$. All above terms are non-negative,  so it follows that $\rho_m, \frac{1}{2}y(t)^2, \Lambda, \frac{K^2}{a(t)^p}$ are bounded by  $3 H(t_0)^2$  for all $t>t_0$.
We define the set $A=\{\phi\in \mathbb{R}: V(\phi)\leq 3 H(t_0)^2\}$. Then, the orbit $O^+(x_0)$ is such that $\phi$  remains at the interior of $A$ for all $t>t_0$. 
 Given $G_0(a)=\frac{K^2}{a^p}, p>0$, the equation  \eqref{Rachd} can be written as  
\begin{equation}
 \dot{H}=-\frac{1}{2}\left(\gamma \rho_m+y^2\right)-\frac{K^2 p}{6 a^p},\label{Rachd2} 
\end{equation}
i.e., by integration, we have : 
\begin{equation*}
H(t_0)-H(t)=\int_{t_0}^t \left(\frac{1}{6} K^2 p a(s)^{-p}+\frac{1}{2}
   \gamma  \rho (s)+\frac{1}{2} y(s)^2\right) \, ds.
\end{equation*}
Taking the limit as $t\rightarrow +\infty$ we obtain 
\begin{equation*}
H(t_0)-\eta= H(t_0) -\lim_{t\rightarrow \infty} H(t)=\int_{t_0}^\infty \left(\frac{1}{6} K^2 p a(s)^{-p}+\frac{1}{2}
   \gamma  \rho_m (s)+\frac{1}{2} y(s)^2\right) \, ds.
\end{equation*}
From these equations we find the convergent improper integral
\begin{equation*}
\int_{t_0}^\infty \left(\frac{1}{6} K^2 p a(s)^{-p}+\frac{1}{2}
   \gamma  \rho_m (s)+\frac{1}{2} y(s)^2\right) \, ds<\infty.
   \end{equation*}
Defining $f(t)=\left(\frac{1}{6} K^2 p a(t)^{-p}+\frac{1}{2}
   \gamma  \rho_m (t)+\frac{1}{2} y(t)^2\right)$, we have \\
$\frac{d}{dt} f(t)  = -y V'(\phi)+H  \left(-\frac{1}{6} K^2 p^2 a^{-p}-\frac{3}{2} \left(\gamma ^2 \rho_m+2 y^2\right)\right)+\frac{(\gamma -2) (3 \gamma -4) \rho_m y 
   \chi '(\phi)}{4 \chi (\phi)}$. \\
Hence, $\left|\frac{d}{dt} f(t)\right|\leq  y |V'(\phi)|+\frac{1}{6}p^2 H \left| K^2 a^{-p}\right|+\frac{3}{2}\gamma^2 \rho_m H +3 y^2 H+ \left|\frac{(\gamma -2) (3 \gamma -4)}{4}\right| \rho_m y \left|\frac{
   \chi '(\phi)}{ \chi (\phi)}\right|\leq \sqrt{6}  H(t_{0})
   \left|V'(\phi (t))\right| +\frac{1}{2}  H(t_{0})^3 (9\gamma^2 + p^2 +36)  +  \frac{3 \sqrt{6}}{4} H(t_{0})^3 K_1 |(\gamma -2) (3 \gamma -4)|$, 
for all $t>t_0$ along the positive orbit $O^+(x_0)$. For deducing the above, we have used the results of $\rho_m, \frac{1}{2}y(t)^2$, and $\frac{K^2}{a(t)^p}$ which are bounded by $3 H(t_0)^2$  for all $t>t_0$, and the hypothesis for $\chi$. Finally, due to $V(\phi)$ which is bounded on $A$, $V^{\prime}(\phi)$, it also will be bounded on $A$ as well. Results: $\left|\frac{d}{dt} f(t)\right|<\infty$ along the positive orbit $O^+(x_0)$.  This $f(t)$ is non-negative; it has a bounded derivative along the orbit $O^+(x_0)$ and $\int_{t_0}^\infty f(s) \, ds$ is convergent.  Hence, we have
  $\lim_{t\rightarrow \infty} f(t)=0$, from which, along with the non-negativeness of each term of $f(t)$,  we have $\lim_{t\rightarrow \infty} \left( \rho_m, y, \frac{K^2}{a^p}\right)=(0,0,0)$.  $\blacksquare$
	\\
We will now show how our result generalizes previous theorems. 
\begin{enumerate}
    \item[(A)] 
 Setting  $\chi (\phi) \equiv 1$ (minimal coupling) in Theorem \ref{Proposition I} it follows:
\begin{cor}[Leon \& Franz-Silva, 2019]\label{PropositionI}
Let us assume that hypotheses i); ii) and iii) of Theorem \ref{Proposition I} are satisfied. Let be $\chi (\phi) \equiv 1$, then  $\lim_{t\rightarrow \infty} \left( \rho_m, y, \frac{K^2}{a^p}\right)=(0,0,0)$.
\end{cor}
\item[(B)] Setting $\chi (\phi) \equiv 1$, $\Lambda=0$, and $G_0(a)\equiv 0$ in Theorem \ref{Proposition I} it follows:  
\begin{cor}[Miritzis 2003. Proposition 2 of \cite{Miritzis:2003ym}] 
\label{PropositionIb}
Let us assume that hypotheses i) and ii) of Theorem \ref{Proposition I} are satisfied. 
Assuming $\chi (\phi) \equiv 1$, $\Lambda=0$, and  $G_0(a)\equiv 0$, then $\displaystyle{\lim_{t\rightarrow  \infty} (\rho_m, y) =(0, 0)}$.
\end{cor}
Corollary \ref{PropositionIb} is a particular case of Corollary \ref{PropositionI} for flat FLRW universe. 
\item[(C)] Setting $\chi (\phi) \equiv 1$ (minimal coupling), $\Lambda=0$, and $G_0(a)\equiv 0$ (flat FLRW universe), $\rho_m=0$ (vacuum)  in Theorem \ref{Proposition I} it follows: 
\begin{cor}[Corollary of Proposition 2 of \cite{Miritzis:2003ym}]\label{tm} Let us assume that hypotheses i) and ii) of Theorem \ref{Proposition I} are satisfied. 
Assuming $\chi (\phi) \equiv 1$, $\Lambda=0$, $G_0(a)\equiv 0$, and $\rho_m=0$, then $\displaystyle{\lim_{t\rightarrow  \infty} y =0}$. 
\end{cor}
\end{enumerate}
\begin{thm}[Leon \& Franz-Silva, 2019]
\label{Theorem5.2} Under the hypotheses 
\begin{enumerate}
\item $V(\phi)\in C^2(\mathbb{R}), V(\phi)\geq 0$, and $V(\phi)=0$ if and only if $\phi=0$. 
\item $V^{\prime}(\phi)$ is bounded on $A\subset\mathbb{R}$ if $V(\phi)$ is bounded on  $A$.
\item $\Lambda\geq 0$, and $G_0(a)\geq 0$ have a negative power-law functional form $G_0(a)=\frac{K^2}{a^p}, p>0$. 
\item $\chi(\phi)\in C^2(\mathbb{R})$, and for all $A\subset \mathbb{R}$ there is a non-negative constant $K_1$, possibly depending on $A$, such as $\left| \frac{\chi '(\phi)}{\chi (\phi)}\right|\leq K_1$ for all $\phi \in A$. And assuming 
\item $V^{\prime}(\phi)<0$ for $\phi<0$ and $V^{\prime}(\phi)>0$ for $\phi>0$,
\end{enumerate}
then,  $\lim_{t\rightarrow\infty}\phi \in \{-\infty, 0, +\infty\}$. 
\end{thm}
\textbf{Proof}. As before, we consider the positive orbit $O^+(x_0)$ passing at the time $t_0$ through the regular point $x_0\in
   \left\{(H, \rho_m, a, y, \phi)\in \Omega: H>0\right\}$.
Using the same argument in the proof of Theorem 
\ref{Proposition I}, $\exists \lim_{t\rightarrow\infty}
H(t)=\eta$ along the orbit $O^+(x_0)$. Under the  hypothesis (i), (ii), (iii) and (iv) (see Theorem \ref{Proposition I}), it follows $\lim_{t\rightarrow \infty} \left( \rho_m, y, \frac{K^2}{a^p}\right)=(0,0,0)$. 

If $3\eta^2=\Lambda$, then by the restriction  \eqref{Fried2bB} and using Theorem \ref{Proposition I} we have $\lim_{t\rightarrow\infty} V(\phi(t))=0$. As $V$ is continuous $V(\phi)=0\Leftrightarrow \phi=0$ this implies that 
$\lim_{t\rightarrow\infty} \phi(t)=0$.

Suppose that $3\eta^2>\Lambda$, then by the restriction  \eqref{Fried2bB} and using Theorem \ref{Proposition I}
$\lim_{t\rightarrow\infty} V(\phi(t))=3\eta^2-\Lambda> 0$. Hence, there is 
$t'$ such as $V(\phi)>(3\eta^2 -\Lambda)/2$ for all $t>t'$. From this fact, it follows that  $\phi$ cannot be zero for $t>t'$ due to  $\phi=0
\Leftrightarrow V(\phi)=0$. Then, the sign of $\phi$ is invariant for all $t>t'$.

Suppose $\phi$ is positive for all $t>t'$. Due to the fact that $V$ is an increasing function of $\phi$ in $(0,+\infty)$, we have
$\lim_{t\rightarrow\infty} V(\phi(t))=(3\eta^2 -\Lambda)\leq
\lim_{\phi\rightarrow\infty} V(\phi)$.  By continuity and monotonicity of $V$ it follows that this equality holds if and only if $\lim_{t\rightarrow\infty} \phi(t)=+\infty$.

If $\lim_{t\rightarrow\infty} V(\phi(t))<
\lim_{\phi\rightarrow\infty} V(\phi)$, then there exists
$\bar{\phi}> 0$ such as $$ \lim_{t\rightarrow\infty}
V(\phi(t))=V(\bar{\phi}).$$   Due to $V$ being strictly increasing and continuous, we have
$$ \lim_{t\rightarrow\infty} \phi=\bar{\phi}.$$  

 Taking the limit $t\rightarrow\infty$ on \eqref{13EQ} we have 
$$ \lim_{t\rightarrow\infty}\frac{d}{d
t}y=-V'(\bar{\phi})<0.$$   Hence, there exists $t''> t'$ such that
$\frac{d}{d t}y<-V'(\bar{\phi})/2$  for all $t\geq t'' $. This implies
$$ y(t)- y(t'')=\int_{t''}^{t}\left(\frac{d}{d t}y\right)
dt<-\frac{ V'(\bar{\phi})}{2}(t- t''),$$  that is, $y(t)$ takes negative values large enough as $t\rightarrow \infty$, which is impossible because $\lim_{t\rightarrow\infty}y(t)=0$. Henceforth, if $\phi>0$ for all $t>t'$, we have  
$\lim_{t\rightarrow\infty}\phi=+\infty$. In the same way, for $\phi<0$
for all $t>t'$, we have $\lim_{t\rightarrow\infty}\phi=-\infty$. $\blacksquare$

If initially
$3H(t_0)^2<\min\left\{\lim_{\phi\rightarrow\infty}V(\phi),
\lim_{\phi\rightarrow -\infty}V(\phi)\right\}$,  then, 
$\lim_{t\rightarrow\infty}H(t)=\sqrt{\frac{\Lambda}{3}}$. Indeed, from the above Theorem
$\lim_{t\rightarrow\infty}\phi$ equals $+\infty$, $0$ or
$-\infty$. If $\lim_{t\rightarrow\infty}\phi=+\infty$, from the restriction \eqref{Rachd2}, it follows
$$ 3 \eta^2-\Lambda=\lim_{t\rightarrow\infty}V(\phi(t))=
\lim_{\phi\rightarrow\infty} V(\phi)>3 H(t_0)^2.$$   This is impossible because $H(t)$ is decreasing and $H(t_0)\geq
\eta, \Lambda\geq 0$. In the same way, the assumption $\lim_{t\rightarrow\infty}\phi=-\infty$
leads to a contradiction. Then,  $\lim_{t\rightarrow\infty}\phi=0$
and this implies $\lim_{t\rightarrow\infty}V(\phi(t))=0$, and from \eqref{Rachd2} it follows $\lim_{t\rightarrow\infty}H(t)=\sqrt{\frac{\Lambda}{3}}$. 
	Now we explicitly emphasize how our result generalizes previous theorems: 
\begin{enumerate}
\item[(A)] Setting $\chi (\phi) \equiv 1$ (minimal coupling) in Theorem \ref{Theorem5.2} we obtain the following:
\begin{cor}[Leon \& Franz-Silva, 2019]\label{Proposition II}
Under the hypotheses (i), (ii), and (iii) and (v) of Theorem \ref{Theorem5.2}, and setting $\chi(\phi)\equiv 1$, then  $\lim_{t\rightarrow\infty}\phi \in \{-\infty, 0, +\infty\}$. 
\end{cor}

\item[(B)] Setting $\chi (\phi) \equiv 1$, $\Lambda=0$, and $G_0(a)\equiv 0$ in Theorem \ref{Theorem5.2} we obtain the following:
\begin{cor}[Miritzis 2003. Proposition 3, \cite{Miritzis:2003ym}] 
\label{PropositionII}
Under the hypotheses (i), (ii), and (v) of Theorem \ref{Theorem5.2}, and setting $\chi(\phi)\equiv 1$, $\Lambda=0$, and $G_0(a)\equiv 0$, 
then, $\displaystyle{\lim_{t\rightarrow  \infty} \phi(t) \in \lbrace -\infty , 0 , + \infty \rbrace }$. 
 \end{cor}
\item[(C)]
Setting $\chi (\phi) \equiv 1$ (minimal coupling),  $\Lambda=0$, $G_0(a)\equiv 0$ (flat FLRW universe), and $\rho_m=0$ (vacuum)  in the Theorem \ref{Theorem5.2} it follows: 
\begin{cor}[Corollary of Proposition 3 of \cite{Miritzis:2003ym}]\label{thm2.1} Under the hypotheses (i), (ii), and (v) of Theorem \ref{Theorem5.2}, and setting $\chi (\phi) \equiv 1$, $G_0(a)\equiv 0$, $\Lambda=0$, and $\rho_m=0$, then, $\displaystyle{\lim_{t\rightarrow  \infty} \phi(t) \in \lbrace -\infty , 0 , + \infty \rbrace }$. 
 \end{cor}
\end{enumerate}
\begin{thm}[Leon \& Franz-Silva, 2019]
\label{Theorem5.3}
Assuming
\begin{enumerate}
    \item $V(\phi)\in C^2(\mathbb{R})$, $V(\phi)\geq 0$, and $\lim_{\phi \rightarrow -\infty}V(\phi)=+\infty$.
    \item $V^{\prime}(\phi)$ is continuous and $V^{\prime}(\phi)<0$.
    \item $V^{\prime}(\phi)$ is bounded on $A\subset\mathbb{R}$ if $V(\phi)$ is bounded on  $A$.
    \item $\Lambda \geq 0$ and $G_0(a)=\frac{K^2}{a^p}, p>0$.
    \item $\chi(\phi)\in C^2(\mathbb{R})$ such as for all $A\subset \mathbb{R}$ there is a constant $K_1$, possibly depending on $A$, such as $\left| \frac{\chi '(\phi)}{\chi (\phi)}\right|<K_1$ for all $\phi \in A$.
\end{enumerate}
Then,  $\lim_{t\rightarrow \infty} \left( \rho_m, y, \frac{K^2}{a^p}\right)=(0,0,0)$, and
$\lim_{t\rightarrow\infty}\phi =+\infty$. 
\end{thm}
\textbf{Proof}.
As before, let $O^+(x_0)$ be the positive orbit passing at $t_0$ through a regular point $x_0\in
   \left\{(H, \rho_m, a, y, \phi)\in \Omega: H>0\right\}$. From equation  \eqref{consb}, it follows that the set  $\rho_m>0$ is invariant for the flow of \eqref{consb} with the restriction \eqref{Fried2bB} along the orbit $O^+(x_0)$; besides $\rho_m$ is different from zero if initially $\rho_m(t_0)$ it is so. This implies  $H$ is never zero because of equation \eqref{Fried2bB}, $3 H(t)^2\geq \rho_m(t)>0$ for all $t>t_0$.  Then
$H$ is always non-negative if initially it is non-negative. Furthermore, from equation \eqref{Rachd2}, it follows that  $H$ is decreasing and non-negative, then
$\exists \lim_{t\rightarrow\infty} H(t)=\eta\geq 0$ and 
$$ \int_{t_0}^\infty \left(\frac{1}{6} K^2 p a(s)^{-p}+\frac{1}{2}
   \gamma  \rho_m (s)+\frac{1}{2} y(s)^2\right) ds = H(t_0)-\eta<+\infty.$$   As in Theorem \ref{Proposition
I}, the total derivative of $f(t)=\left(\frac{1}{6} K^2 p a(t)^{-p}+\frac{1}{2}
   \gamma  \rho_m (t)+\frac{1}{2} y(t)^2\right)$ is bounded, and the improper integral $\int_{t_0}^\infty f(t) dt$ is convergent. Then, $\lim_{t\rightarrow +\infty} f(t)=0$, which, along with the non-negativeness of each term of $f(t)$, implies $$ \lim_{t\rightarrow \infty} \left(\rho_m, y, \frac{K^2}{a^p}\right)=(0,0,0).$$  
It can be proved that $\lim_{t\rightarrow\infty}\phi=+\infty$ in the same way as it was proved for Theorem \ref{Proposition II}. From equation \eqref{Fried2bB} we have $\lim_{t\rightarrow\infty} V(\phi)=3\eta^2-\Lambda$.  The function $V$ is strictly decreasing with respect to $\phi$, then $V(\phi)>\lim_{\phi\rightarrow\infty} V(\phi)$ for all $\phi$. Hence,  $\lim_{t\rightarrow\infty} V(\phi(t))\geq
\lim_{\phi\rightarrow\infty} V(\phi)$. Thus there are two cases to be considered: 
 \begin{enumerate}
   \item If $\lim_{t\rightarrow\infty} V(\phi(t))= \lim_{\phi\rightarrow\infty} V(\phi)$,
   by continuity of $V$ it follows
$\lim_{t\rightarrow\infty}\phi=+\infty$.
   \item If  $ \lim_{t\rightarrow\infty} V(\phi(t))> \lim_{\phi\rightarrow\infty} V(\phi)$, then,  since $V$ is continuous and strictly decreasing, it follows that it will be a unique $\bar{\phi}$ such as
   $$ \lim_{t\rightarrow\infty} V(\phi(t))=V(\bar{\phi}).$$    Because  $V$ is continuous, it follows that $$ \lim_{t\rightarrow\infty} \phi=\bar{\phi}.$$  
From equation \eqref{13EQ} we have
$$ \lim_{t\rightarrow\infty}\frac{d}{d
t}y=-V'(\bar{\phi})>0.$$   Hence,  there exists $t'$ such as
$\frac{d}{d t}y>-V'(\bar{\phi})/2$ for all $ t\geq t'$. Therefore, 
$$ y(t)- y(t')
>-\frac{V'(\bar{\phi})}{2}(t- t'),$$  which is impossible because
$\lim_{t\rightarrow\infty}y(t)=0$. Finally,
$\lim_{t\rightarrow\infty}\phi=+\infty$. $\blacksquare$
 \end{enumerate} 
 Additionally, if 
$\lim_{\phi\rightarrow\infty} V(\phi)=0$, then
$H\rightarrow \sqrt{\frac{\Lambda}{3}}$ as $t \rightarrow\infty$.

	\textbf{We will now show} how our result generalizes previous theorems:  
\begin{enumerate}
    \item[(A)] Setting $\chi (\phi) \equiv 1$ in Theorem \ref{Theorem5.3} we obtain the following: 
    \begin{cor}[Leon \& Franz-Silva, 2019]
\label{Proposition III}
Under the hypotheses (i), (ii), (iii) and (iv) of Theorem \ref{Theorem5.3}, and assuming  $\chi (\phi) \equiv 1$, then, 
$ \lim_{t\rightarrow \infty} \left( \rho_m, y, \frac{K^2}{a^p}\right)=(0,0,0)$, and $\lim_{t\rightarrow\infty}(a, \phi) =(+\infty,+\infty)$. 
\end{cor}
\item[(B)] Setting $\chi (\phi) \equiv 1$ (minimal coupling), $\Lambda=0$, and choosing $G_0(a)\equiv 0$ (flat FLRW universe), and $\Lambda=0$ in Theorem \ref{Theorem5.3}, we have the following: 
\begin{cor}[Miritzis 2003. Proposition 4, \cite{Miritzis:2003ym}]
\label{Prop4Miritzis} Under the hypotheses (i), (ii), and (iii) of Theorem \ref{Theorem5.3}, and setting $\chi (\phi) \equiv 1$, $\Lambda=0$, and $G_0(a)\equiv 0$, then,  $\lim_{t \rightarrow  + \infty} y(t)=\lim_{t \rightarrow  + \infty} \rho_m(t) =0$ and $\lim_{t \rightarrow  + \infty} \phi (t)= +\infty$.
 \end{cor}
\item[(C)]  Setting $\chi (\phi) \equiv 1$ (minimal coupling), $\Lambda=0$, $G_0(a)\equiv 0$ (flat FLRW universe), and $\rho_m=0$ (vacuum)  in Theorem \ref{Theorem5.3} we have the following:
 \begin{cor}[Corollary of Proposition 4 of \cite{Miritzis:2003ym}]\label{thm2.2} Under the hypotheses (i), and (ii) of Theorem \ref{Theorem5.3}, and setting  $\chi (\phi) \equiv 1$, $\Lambda=0$,  $G_0(a)\equiv 0$, and $\rho_m=0$, then $\lim_{t \rightarrow  + \infty} y(t) =0$ and $\lim_{t \rightarrow  + \infty} \phi (t)= +\infty$.
 \end{cor}
\end{enumerate}
Finally, we present the theorem: 
\begin{thm}[Leon \& Franz-Silva, 2019]\label{thmIIIFINAL} Assuming
\begin{enumerate}
\item[(i)] $V(\phi)\in C^2(\mathbb{R})$ such that the possibly empty set $\{\phi: V(\phi)<0\}$ is bounded; 
\item[(ii)] and the possibly empty set of singular points of  $V(\phi)$ is finite.
\item[(iii)] $\phi_*$ is a strict minimum, possibly degenerated, of $V(\phi)$ with non-negative critical value.
\item[(iv)] $\Lambda= 0$, and $G_0(a)\geq 0$ such as $\frac{a  G_{0}'(a)}{G_{0}(a)}\leq -p<0$, for all $a>0$. 
\end{enumerate}
 Then ${\bf
p}_*:=(\phi, y, \rho_m, H)=\left(\phi_*, 0, 0, \sqrt{\frac{V(\phi_*)}{3}}\right)$ is an asymptotically stable equilibrium point.
\end{thm}
\textbf{Proof}. Let us define 
\begin{equation}
\label{defW}
W(\phi, y, \rho_m, H)=H^2-\frac{1}{3}\left(\frac{1}{2}y^2+V(\phi)+\rho_m\right):= \frac{1}{3}G_0(a),
\end{equation}
which satisfies \footnote{Observe that in \cite{Giambo:2009byn}, where the case $G_0(a)=-\frac{3k}{a^2}$ was studied, leading to $\dot W=-2 H W$.}
\begin{equation}
\label{16Eq}
\dot W=H W \frac{a  G_{0}'(a)}{G_{0}(a)} <-p H W.
\end{equation}
Therefore, $W$ is a non-negative and decreasing function of $t$.

Let us define 
\begin{equation}
\label{17Eq}
\epsilon =\frac{1}{2}y^2 + V(\phi) + \rho_m, \quad \dot\epsilon =-3 H \left(\gamma  \rho_m +y^2\right).
\end{equation}
This implies that $\epsilon$ is decreasing too.

Firstly, it is assumed that $V(\phi_*)>0$.  Let 
$\tilde{V}>V(\phi_*)$ be a regular value of $V$ such as the connected component of $V^{-1}\left((-\infty,\tilde{V}]\right)$
that contains $\phi_*$ is a compact set in $\mathbb{R}$.  Let us denote this set by $A$  and define  $\Psi$ as
$$ \Psi=\left\{(\phi,y, \rho_m,H)\in\mathbb{R}^4: \phi\in A,
\epsilon \leq \tilde{V}, \rho_m\geq 0, W(\phi, y, \rho_m, H)\in[0, \bar{W}]\right\},$$  where $\bar{W}$ is positive. It can be proved that then $\Psi$ is a compact set as follows:

\begin{enumerate}
\item $\Psi$ is a closed set in $\mathbb{R}^4$.
\item $V(\phi_*)\leq V(\phi) \leq \tilde{V}$, for all $\phi\in A$.
\item Since $\frac{1}{2} y^2 + V(\phi_*)\leq \frac{1}{2} y^2 + V(\phi) + \rho_m= \epsilon \leq \tilde{V}$, it therefore follows that $y$ is bounded.
\item From $\rho_m\leq \tilde{V} - \frac{1}{2} y^2 - V(\phi) \leq \tilde{V} -  V(\phi_*)$, it is a consequence that $\rho_m$ is bounded. 
\item From \eqref{defW}, and due to the above facts, it follows that:
\begin{equation}
\frac{V(\phi_*)}{3}\leq\frac{V(\phi)}{3}\leq H^2=W+\frac{1}{3}\left(\frac{1}{2}y^2+V(\phi)+\rho_m\right) \leq \bar{W} + \frac{\tilde{V}}{3}.
\end{equation} That is, $H$ is also bounded. 
\end{enumerate}
Let us define $\Psi_+\subseteq \Psi$, the connected component of  $\Psi$
containing ${\bf p}_*$.  Following the same arguments as in 
\cite{Giambo:2009byn,Giambo:2008ck} we prove that $\Psi_+$ is positively invariant with respect to \eqref{Non_min2}. Let  $\mathbf{x}(t)$ being any solution starting at $\Psi_+$, and defining $\bar{t}=\sup\left\{t>0:
H(t)>0\right\}\in\mathbb{R}\cup\{+\infty\}$. 
When $t<\bar{t}$, equations \eqref{16Eq} and \eqref{17Eq} imply that both $W$ and $\epsilon$ decrease.  
Moreover, let us assume that there exists $t<\bar{t}$ such as $\phi(t)\notin A$. Hence, $V(\phi(t))>\tilde{V}$, however 
\begin{equation*}
\tilde{V}< V(\phi(t)) \leq \frac{1}{2}y(t)^2 + V(\phi(t)) + \rho_m(t)= \epsilon(t) \leq \tilde{V},
\end{equation*}
which is a contradiction. Therefore,  $\phi(t)\in A, \quad \forall t<\bar{t}$. But, $W\geq 0$ along the flow under \eqref{Non_min2}, due to  $G_0(a)\geq 0$, and so by hypothesis it follows: 
\begin{equation*}
H(t)^2 \geq \frac{1}{3}\left(\frac{1}{2}y(t)^2+V(\phi(t))+\rho_m(t)\right)\geq  \underbrace{\frac{V(\phi(t))}{3}\geq \frac{V(\phi_*)}{3}}_{\text{Because}\; \phi(t)\in A, \quad \forall t<\bar{t}}.
\end{equation*}
 That is, as long as $H$ remains positive, it is strictly bounded away from zero and thus $\bar{t}=+\infty$. Therefore, $\Psi_+$ satisfies the hypothesis of LaSalle's invariance Theorem \cite{LaSalle,wiggins}. Considering the monotonic functions  $\epsilon$ and $W$ defined on $\Psi_+$ it follows that any solution with initial state on $\Psi_+$ must be such that $H y^2 \rightarrow 0, H \gamma \rho_m \rightarrow 0$ as $t\rightarrow +\infty$.  Since $H$ is strictly bounded away from zero on  $\Psi_+$ it follows that 
$y\rightarrow 0$, $\rho_m\rightarrow 0$ (recall $1\leq \gamma \leq 2$) and 
$H^2-\frac{V(\phi)}{3}\rightarrow 0$ as $t\rightarrow +\infty$.  
However, from
hypotheses $G_{0}(a)\geq 0$ and  $\frac{a  G_{0}'(a)}{G_{0}(a)}\leq -p<0$, for all $a>0$, we have 
\begin{equation*}
\dot{H}=-\frac{1}{2}\left(\gamma \rho_m+y^2\right)+\frac{1}{6}a G_0'(a) \leq -\frac{1}{2}\left(\gamma \rho_m+y^2\right)-\frac{p}{6} G_0(a) \leq 0. 
\end{equation*} 
Due to $H$ being monotonically decreasing and it is bounded to approach zero, it must have a limit. This implies that 
$V(\phi)$ must also have a limit. This limit must be $V(\phi_*)$.  Otherwise,  $V^{\prime}(\phi)$ would tend to a non zero value, and so would be the right hand side of \eqref{13EQ}, which is a contradiction. Therefore, the solution tends to ${\bf p}_*$.

If $V(\phi_*)=0$, the set $\Psi$ is connected and we choose  $\Psi_{+}$ as the subset of $\Psi$ with $H\geq0$. The unique equilibrium point on  $\Psi_{+}$ with  $H=0$ is the equilibrium point
${\bf p}_*$, then, if $H(t)\rightarrow 0$ the solution is forced to tend to the equilibrium point due to $H$ is monotonic.  On the contrary, if $H(t)$ tends to a positive number, as before, we will have  $y\rightarrow 0, \rho_m\rightarrow 0, W \rightarrow 0,\,V(\phi)\rightarrow V(\phi_*)=0$, and hence $H$ will necessarily tend to zero. $\blacksquare$

\subsubsection{Case of positive curvature $k=1$.} 
\label{positive-k}
For $k=+1, W<0$, we cannot guarantee the monotony of $H$, so we have to adapt the previous arguments in exactly the same way as in \cite{Giambo:2008ck}.  That is, $\phi_*$ is a local minimum of $V(\phi)$ with $V(\phi_*)>0$.  The 
$\tilde{V}>V(\phi_*)$ is a regular value of $V$ so the connected  component of $V^{-1}\left((-\infty,\tilde{V}]\right)$
 contains $\phi_*$ as the only critical point of $V$, and is a compact set in $\mathbb{R}$. It is considered a solution $\mathbf{x}(t)=(\phi(t), y(t), \rho_m(t), H(t))$ such as $\frac{1}{2} y(0)^2 +V(\phi(0)) + \rho_m(0)\leq \tilde{V}$, and let $\bar{W}<0$ a value to determine, to act as a lower bound for $W$. 
Taking the initial condition near the equilibrium point ${\bf
p}_*$, then $H(0)>0$; since $W(0)>\bar{W}$, from the equations $\dot W= H W \frac{a  G_{0}'(a)}{G_{0}(a)}\geq 0$ and $\dot \epsilon =-3 H \left(\gamma  \rho_m +y^2\right)\leq 0$, we have $W\geq \bar{W}$ ($W$ now, it will be monotonic increasing and bounded by above by zero) and $\epsilon\leq \tilde{V}$. The last inequality implies that $V(\phi(t))\leq \tilde{V}$. This implies that $\phi(t)$ satisfies $V(\phi_*)\leq V(\phi(t))\leq \tilde{V}$. Then,
\begin{equation*}
H^2= \frac{\epsilon}{3}  + \frac{W}{3} \geq \frac{V(\phi_*)}{3}+ \frac{\bar{W}}{3}= H_*^2 +\frac{\bar{W}}{3} \implies
H\geq  \bar{H}:= \left(\sqrt{1+\frac{\bar{W}}{3 {H_*}^2}}\right) H_*, 
\end{equation*}
where $H_*=\sqrt{\frac{V(\phi_*)}{3}}$.
Choosing $\bar{W}$ small enough such as $H_*^2>-\frac{\bar{W}}{3}$, we have $H(0)>0 \implies H(t)\geq \bar{H} >0$.
That is, $H$ is bounded away zero which is combined with the monotony of $W$, $\epsilon$, and using the LaSalle's invariance Theorem as in the case $W\geq 0$ leads to $y\rightarrow 0, \rho_m\rightarrow 0, W \rightarrow 0$ as $t\rightarrow \infty$, and the equilibrium point ${\bf
p}_*$ is approached asymptotically.  

If $V(\phi_*)=0$, the equilibrium point satisfies $H_*=0$, and the nearby solutions may re-collapse as $H$ changes sign.  Collapsing models were exhaustively studied, e.g., in \cite{Giambo:2009zza,Giambo:2015tja,Giambo:2014jfa,Giambo:2013bya,Giambo:2009zz,Giambo:2008sa,Giambo:2008ya,Giambo:2005se,Giambo:2002tp,Giambo:2001wi} for a wide class of self-interacting, self-gravitating homogeneous scalar field models.

\section{Dynamical systems analysis for arbitrary $V(\phi)$ and $\chi(\phi)$}
\label{SECT:6}

\subsection{Local dynamical systems analysis}
\label{SECT:6.1}
In this section we provide a dynamical system analysis of the system \eqref{Non_min2} for an arbitrary $V(\phi)$ and arbitrary $\chi(\phi)$. 
Let be defined
\begin{align}
  & \lambda= -\frac{W'(\phi)}{W(\phi)}, \quad f=\frac{W''(\phi)}{W(\phi)}-\frac{W'(\phi)^2}{W(\phi)^2},  \quad  g= - \frac{\chi '(\phi)}{\chi (\phi)}, \label{parametrizationA}
\end{align}
where $W(\phi)=\Lambda + V(\phi)$, and we assume $W(\phi)\geq 0$. The idea is assume $f$ and $g$ can be explicitly written as functions of $\lambda$. Then, the conditions \eqref{parametrizationA} can be written alternatively as 
\begin{equation}
\frac{d \lambda}{d \phi}= -f(\lambda), \quad 
\frac{d W}{d \phi}= -\lambda W, \quad
\frac{d \chi}{d \phi}=-g(\lambda) \chi, 
\end{equation}
which can be integrated in quadrature as
\begin{equation}
 \phi (\lambda )=\phi (1)-\int_1^{\lambda } \frac{1}{f(s)} \, ds, \quad
 W (\lambda )=W(1) e^{\int_1^{\lambda } \frac{s}{f(s)} \, ds}, \quad
 \chi (\lambda )=\chi(1) e^{\int_1^{\lambda } \frac{g(s)}{f(s)} \, ds}.
\end{equation}
The last equations can be used to generate the potentials and couplings by giving the functions $f(\lambda)$ and $g(\lambda)$ as input. 

We define the new variables
\begin{equation}
 x= \frac{\dot\phi}{\sqrt{6}H}, \quad \Omega_m= \frac{\rho_m}{3 H^2}, \quad \Omega_0= \frac{G_0(a)}{3 H^2}, \quad \lambda=-\frac{V'(\phi)}{\Lambda +V(\phi)},
\end{equation}
where $ G_0(a)=q a^{-p}, p\geq 0$, and then we obtain the dynamical system:
\begin{subequations}
\label{systH1}
\begin{align}
&x^{\prime}=\frac{1}{2} x \left(3 \gamma  \Omega_m+p \Omega_0+6
   x^2-6\right)-\sqrt{\frac{3}{2}} \lambda  \left(x^2+\Omega_0+\Omega_m-1\right)  +\frac{\sqrt{6}}{4} (3 \gamma -4) \Omega_m g(\lambda
   ),\\
&\Omega_m^{\prime}= \frac{1}{2} \Omega_m \left(\sqrt{6} (4-3
   \gamma ) x g(\lambda )+2 \left(3 \gamma  (\Omega_m-1)+p
   \Omega_0+6 x^2\right)\right),\\
&\Omega_0^{\prime}= \Omega_0 \left(3 \gamma\Omega_m+p (\Omega_0-1)+6 x^2\right),\\
&\lambda^{\prime}= -\sqrt{6} x f(\lambda ),   
\end{align}
\end{subequations}
where the prime means derivative with respect to $\tau=\ln a$, with the restriction
\begin{equation}
    x^2+\Omega _{m}+\Omega_{0}=1-\frac{W(\phi)}{3H^2} \leq 1, 
\end{equation}
and now the free functions are $f(\lambda), g(\lambda)$. We impose the condition $\gamma \in [1,2]$.

 The equilibrium points of the system \eqref{systH1} are the following: 
\begin{enumerate}
    \item[$A_1(\hat{\lambda})$:] $(x, \Omega_m, \Omega_0, \lambda)=\left(\frac{(4-3 \gamma ) g(\hat{\lambda})}{\sqrt{6} (\gamma -2)},
   1-\frac{(4-3 \gamma )^2 g(\hat{\lambda})^2}{6 (\gamma
   -2)^2}, 0, \hat{\lambda}\right)$  represents a matter - kinetic scaling solution. 
For $1\leq \gamma< 2$, we deduce that $A_1(\hat{\lambda})$ is a sink under one of the conditions (i)- (x) in  \ref{AppA}. If exists, it will be never a source. 
   
   \item[$A_2(\hat{\lambda})$:] $(x, \Omega_m, \Omega_0, \lambda)=\left(\frac{\sqrt{\frac{2}{3}} (p-3 \gamma )}{(3 \gamma -4) g\left(\hat{\lambda
   }\right)}, \frac{2 (6-p) (p-3 \gamma )}{3 (4-3 \gamma )^2
   g(\hat{\lambda})^2}, \frac{2 (p-3 \gamma ) (\gamma
   -2)}{(4-3 \gamma )^2 g(\hat{\lambda})^2}+1, \hat{\lambda}\right)$ represents a matter- scalar field - ``geometric'' fluid scaling solution. For $1\leq \gamma< 2$, we deduce that
$A_2(\hat{\lambda})$ is a sink under one of the conditions (i) - (viii) in  \ref{AppA}.
It is nonhyperbolic for $p=6$, saddle for $p=2$. If exists, it will be never a source. 
   
   \item[$A_3(\hat{\lambda})$:] $(x, \Omega_m, \Omega_0, \lambda)=\left(\frac{\sqrt{6} \gamma }{(4-3 \gamma ) g(\hat{\lambda})+2
   \hat{\lambda }}, \frac{2 (4-3 \gamma ) g(\hat{\lambda})
   \hat{\lambda }+4 \left(\hat{\lambda }^2-3 \gamma \right)}{\left((3
   \gamma -4) g(\hat{\lambda})-2 \hat{\lambda }\right)^2}, 0
   , \hat{\lambda}\right)$ represents a matter - scalar field scaling solution. 
In this case, we can proceed semi-analytically, that is, for non-minimal coupling $g\equiv 0$, and assuming $1\leq \gamma<2$, $A_3(\hat{\lambda})$ is a sink for 
 \begin{enumerate}
     \item $1\leq \gamma <2, p>3 \gamma, -\frac{2 \sqrt{6} \gamma }{\sqrt{9 \gamma -2}}\leq \hat{\lambda }<-\sqrt{3} \sqrt{\gamma }, f'(\hat{\lambda })<0$, or 
     \item $1\leq \gamma <2, p>3 \gamma,
   \sqrt{3} \sqrt{\gamma }<\hat{\lambda }\leq \frac{2 \sqrt{6} \gamma }{\sqrt{9 \gamma -2}}, f'(\hat{\lambda })>0$. 
 \end{enumerate}
  Otherwise, it is a saddle.  For non-minimal coupling the analysis has to be done numerically. 
  
   \item[$A_4(\hat{\lambda})$:] $(x, \Omega_m, \Omega_0, \lambda)=\left(-1, 0, 0,\hat{\lambda}\right)$ is a kinetic dominated solution representing an stiff fluid. 
   $A_4(\hat{\lambda})$ is a source for 
   \begin{enumerate}
       \item $1\leq \gamma <\frac{4}{3},  0\leq p<6,  g(\hat{\lambda })<\frac{\sqrt{6} (\gamma -2)}{3 \gamma -4},  f'(\hat{\lambda })>0,  \hat{\lambda }>-\sqrt{6}$, or 
       \item $\gamma =\frac{4}{3},   0\leq p<6,  f'(\hat{\lambda })>0,  \hat{\lambda }>-\sqrt{6}$, or 
   \item $\frac{4}{3}<\gamma <2,  0\leq p<6,  g(\hat{\lambda })>\frac{\sqrt{6} (\gamma -2)}{3 \gamma -4}, 
   f'(\hat{\lambda })>0,  \hat{\lambda }>-\sqrt{6}$.
   \end{enumerate}

   $A_4(\hat{\lambda})$ is a sink for
   \begin{enumerate}
       \item $1\leq \gamma <\frac{4}{3}, p>6, g(\hat{\lambda })>\frac{\sqrt{6} (\gamma -2)}{3 \gamma -4}, f'(\hat{\lambda })<0, \hat{\lambda }<-\sqrt{6}$, or 
       \item $\frac{4}{3}<\gamma <2, p>6, g(\hat{\lambda })<\frac{\sqrt{6} (\gamma -2)}{3 \gamma -4}, f'(\hat{\lambda })<0, \hat{\lambda }<-\sqrt{6}$.
   \end{enumerate}
   
   \item[$A_5(\hat{\lambda})$:] $(x, \Omega_m, \Omega_0, \lambda)=\left(0, 0, 1, \hat{\lambda}\right)$ is a solution dominated by the effective energy density of $G_0(a)$. It is nonhyperbolic with a 3D unstable manifold for $p>6$.
   
   \item[$A_6(\hat{\lambda})$:] $(x, \Omega_m, \Omega_0, \lambda)=\left(1, 0, 0, \hat{\lambda}\right)$  is a kinetic dominated solution representing a stiff fluid. 
   $A_6(\hat{\lambda})$ is a source for
   \begin{enumerate}
       \item $1\leq \gamma <\frac{4}{3},  p<6,  g(\hat{\lambda })>-\frac{\sqrt{6} (\gamma -2)}{3 \gamma -4},  f'(\hat{\lambda })<0,  \hat{\lambda }<\sqrt{6}$, or 
       \item $\gamma =\frac{4}{3},  p<6,    f'(\hat{\lambda })<0,  \hat{\lambda }<\sqrt{6}$, or 
   \item $\frac{4}{3}<\gamma<2,  p<6,  g(\hat{\lambda })<-\frac{\sqrt{6} (\gamma -2)}{3 \gamma -4},  f'(\hat{\lambda })<0, 
   \hat{\lambda }<\sqrt{6}$.
   \end{enumerate}
      $A_6(\hat{\lambda})$ is a  sink for 
    \begin{enumerate}
        \item $1\leq \gamma <\frac{4}{3},  p>6,  g(\hat{\lambda })<-\frac{\sqrt{6} (\gamma -2)}{3 \gamma -4},  f'(\hat{\lambda })>0,  \hat{\lambda }>\sqrt{6}$, or 
        \item $1\leq \gamma >\frac{4}{3},  p>6,   g(\hat{\lambda })>-\frac{\sqrt{6} (\gamma -2)}{3 \gamma -4},  f'(\hat{\lambda })>0,  \hat{\lambda }>\sqrt{6}$.
    \end{enumerate}
   
   \item[$A_7(\hat{\lambda})$:] $(x, \Omega_m, \Omega_0, \lambda)=\left(\frac{p}{\sqrt{6} \hat{\lambda }}, 0, 1-\frac{p}{\hat{\lambda }^2} , \hat{\lambda}\right)$ is a scaling solution where the energy density of the scalar field and the effective energy density from $G_0(a)$ scale with the same order of magnitude. 
   
   For $1\leq \gamma\leq 2$, $A_7(\hat{\lambda})$, it is a sink for one of the conditions (i) - (xliv) in  \ref{AppA}.  It is never a source.
   
   \item[$A_8(\hat{\lambda})$:] $(x, \Omega_m, \Omega_0, \lambda)=\left(\frac{\hat{\lambda }}{\sqrt{6}}, 0, 0, \hat{\lambda}\right)$  represents the typical quintessence scalar field dominated solution. Assuming $1\leq \gamma \leq 2$,  $A_8(\hat{\lambda})$ is a sink for 
   \begin{enumerate}
       \item $1\leq \gamma <\frac{4}{3},  p>\hat{\lambda }^2,  f'(\hat{\lambda })>0,  0<\hat{\lambda }<\sqrt{6},  g(\hat{\lambda })<\frac{6 \gamma -2 \hat{\lambda }^2}{4 \hat{\lambda }-3 \gamma  \hat{\lambda}}$, or 
   \item $1\leq \gamma <\frac{4}{3},  p>\hat{\lambda }^2,  g(\hat{\lambda })>\frac{6 \gamma -2 \hat{\lambda }^2}{4 \hat{\lambda }-3 \gamma  \hat{\lambda }},  -\sqrt{6}<\hat{\lambda }<0, 
   f'(\hat{\lambda })<0$, or 
   \item $\gamma=\frac{4}{3},  p>\hat{\lambda }^2,  -2<\hat{\lambda }<0,  f'(\hat{\lambda })<0$, or 
   \item $\gamma=\frac{4}{3},  p>\hat{\lambda }^2,  f'\hat{\lambda})>0,  0<\hat{\lambda }<2$, or 
   \item $\frac{4}{3}<\gamma <2,  p>\hat{\lambda }^2,  g(\hat{\lambda })>\frac{6 \gamma -2 \hat{\lambda }^2}{4 \hat{\lambda }-3 \gamma  \hat{\lambda }}, 
   f'(\hat{\lambda })>0,  0<\hat{\lambda }<\sqrt{6}$, or 
   \item $\frac{4}{3}<\gamma <2,  p>\hat{\lambda }^2,  -\sqrt{6}<\hat{\lambda }<0,  g(\hat{\lambda })<\frac{6 \gamma -2 \hat{\lambda
   }^2}{4 \hat{\lambda }-3 \gamma  \hat{\lambda }},  f'(\hat{\lambda })<0$.
   \end{enumerate}
 It is never a source.
   
   \item[$A_9$:] $(x, \Omega_m, \Omega_0, \lambda)=\left(0, 0, 0, 0\right)$ represents the vacuum de Sitter solution associated to the minimum of the potential. It is a sink for $p>0, \gamma>0, f(0)>0$. Otherwise, it is a saddle. 
   
   \item[$A_{10}(\tilde{\lambda})$:] $(x, \Omega_m, \Omega_0, \lambda)=\left(0, 1, 0, \tilde{\lambda}\right)$, where we denote by  $\tilde{\lambda}$, the values of $\lambda$ for which  $g(\lambda)=0$. It represents a non-interacting matter dominated solution. It is a saddle. 
\end{enumerate}
\subsection{Global dynamical analysis}
\label{SECT:3.3}
In Section \ref{SECT:6.1}, we have investigated the stability of the equilibrium points using Hubble- normalized equations, which is essentially based on the Copeland, Liddle \& Wands's approach \cite{Copeland:1997et}. It is well-known that this procedure is well-suited to investigate local stability features of the equilibrium points. However, it does not provide a global description of the phase space when generically $\phi$ diverges or when $H\rightarrow 0$; in which case the method fails. For this reason, we present a new global systems analysis for arbitrary $V(\phi)$ and $\chi(\phi)$ motivated by the approach by Uggla \& Alho \cite{Alho:2014fha}. For simplicity we set $\Lambda=0$. 

The new functions are $W_{V}(\phi), W_{\chi}(\phi)$, such as 
\begin{equation}
  W_{V}(\phi)=  \frac{V'(\phi)}{V(\phi)} -N, \quad W_{\chi}(\phi)= \frac{\chi'(\phi)}{\chi(\phi)} - M, 
\end{equation}
which are assumed to be well-defined, where the constants $N$ and $M$ are defined by:
\begin{equation}
N=  \lim_{\phi \rightarrow + \infty}  \frac{V'(\phi)}{V(\phi)}, \quad M=  \lim_{\phi \rightarrow + \infty}  \frac{\chi'(\phi)}{\chi(\phi)},
\end{equation}
and they are assumed finite, such as
\begin{equation}
  \lim_{\phi \rightarrow + \infty}   W_{V}(\phi)=0, \quad  \lim_{\phi \rightarrow + \infty}   W_{\chi}(\phi)=0.
\end{equation}
We set  $ G_0(a)=q a^{-p}, p\geq 0$ as before. 
\\
Defining the variables
\begin{align}
&T=\frac{m}{m+H}, \quad \theta=  \tan ^{-1} \left(\frac{\dot \phi}{\sqrt{2 V(\phi)+ 2 \rho_m + 2 G_0(a)}}\right), \nonumber \\
& \Omega_m= \frac{\rho_m}{3 H^2}, \quad \Omega_0= \frac{G_0(a)}{3 H^2}, \quad m>0, \end{align}
such that
\begin{align}
  & V(\phi)=  \frac{3 m^2 (1-T)^2 (\cos (2 \theta )-2 \Omega_{0}-2 \Omega_{m}+1)}{2 T^2}, \nonumber \\
  & H = m \left(\frac{1-T}{T}\right),\quad \dot{\phi}= \frac{\sqrt{6} m (1-T) \sin (\theta )}{T},\quad
\rho_{m}=\frac{3 m^2(1-T)^2 \Omega_{m}}{T^2},
\end{align}
and the time variable  $\tau= \ln a$,
to obtain the unconstrained dynamical system:
\begin{subequations}
\label{23-syst}
\begin{align}
   & \frac{d T}{d\tau}= \frac{1}{2} (1-T) T \left(3 \gamma  \Omega_m+6 \sin ^2(\theta )+p \Omega_0\right),\\
   & \frac{d \theta}{d\tau}= -\frac{\sec (\theta )}{4 \sqrt{6}} \Bigg\{3 \sqrt{6} \sin (3 \theta )+6 \left(\cos (2 \theta ) - 2
   \Omega_0+1\right) (N+W_V(\phi)) \nonumber \\
   & +\Omega_m \left(-6 \sqrt{6} \gamma  \sin (\theta )-12 (2 M+N+W_V(\phi)+2 W_{\chi}(\phi))+18 \gamma  (M+W_{\chi}(\phi))\right)\nonumber \\
   & +\sqrt{6} \sin (\theta ) (3-2 p \Omega_0)\Bigg\},\\
   & \frac{d \Omega_m}{d\tau}= -\frac{1}{2}  \Omega_m \Bigg\{6 \cos (2 \theta )-\sqrt{6} (3 \gamma -4) \sin (\theta ) (M+W_{\chi}(\phi))\nonumber \\
   & -2 (3 \gamma 
   (\Omega_m-1)+p \Omega_0+3)\Bigg\},\\
   & \frac{d \Omega_0}{d\tau}= - \Omega_0 (-3 \gamma  \Omega_m+3 \cos (2 \theta )-p \Omega_0+p-3),\\
   & \frac{d \phi}{d\tau}=\sqrt{6} \sin (\theta ).
\end{align}
\end{subequations}

\subsubsection{Asymptotic analysis as $\phi\rightarrow \infty$.}
\label{SECT:3.4}
In this section we discuss the stability of the equilibrium points of \eqref{23-syst} as $\phi\rightarrow \infty$ for functions $V$ and $\chi$  well-behaved at infinity of exponential orders $N$ and $M$, respectively.

\begin{defn}[Definition 1 \cite{Foster:1998sk}] 
\label{kWBI1}
Let $V:\mathbb{R}\longrightarrow \mathbb{R}$ a $C^2$ function. Let there exists some $\phi_0> 0$ for which $V(\phi)>0$ for all $\phi>\phi_0$.  If exists 
$N<\infty$, such that the function
\begin{equation} W_V:   [\phi_0, \infty) \longrightarrow \mathbb{R}, \quad
\phi \longrightarrow \frac{V^{\prime}(\phi)}{V(\phi)}-N 
\end{equation}
is well-defined, and satisfies 
\begin{equation}
\lim_{\phi\rightarrow \infty} W_V(\phi)=0.
\end{equation}
Then, we say that $V$ is  well-behaved at Infinity (WBI) of exponential order $N$.
\end{defn}

\begin{defn}[Definition 2 \cite{Foster:1998sk}]
\label{kWBI2}
A $C^k$ function $V(\phi)$ is a class k WBI function if it is WBI of exponential order $N$, and there are
 $\phi_0> 0$, and a coordinate transformation $\varphi=h(\phi)$ which maps the interval
$[\phi_0,\infty)$  onto $(0, \epsilon]$, where $\epsilon=h(\phi_0)$, satisfying $\lim_{\phi\rightarrow +\infty}h(\phi)=0$, and with the following additional
properties:
\begin{enumerate}
\item $h$ is $C^{k+1}$ and strictly decreasing.
\item The functions 
\begin{equation}
 \bar{W}_V=\left\{\begin{array}{cc}
\frac{V'(h^{-1}(\varphi))}{V(h^{-1}(\varphi))}-N, & \varphi>0,\\
0, & \varphi=0 \end{array}\right.
\end{equation} and 
\begin{equation}
\bar{h'}(\varphi)=\left\{\begin{array}{cc}
h'(h^{-1}(\varphi)), & \varphi>0,\\
\lim_{\phi\rightarrow \infty} h'(\phi), & \varphi=0 \end{array}\right.
\end{equation} are $C^k$ on the closed interval $[0, \epsilon]$; and
\item \begin{equation}
\frac{d \bar{W}_V}{d \varphi}(0)=\frac{d \bar{h'}}{d \varphi}(0)=0.
\end{equation}
\end{enumerate}
\end{defn}
The last hypotheses are equivalent to
$$\lim_{\varphi \to 0} \, \frac{W_{\chi}'\left(h^{-1}(\varphi )\right)}{h'\left(h^{-1}(\varphi )\right)},$$
and
$$\lim_{\varphi \to 0} \, \frac{h''\left(h^{-1}(\varphi )\right)}{h'\left(h^{-1}(\varphi )\right)}=0.$$
Assuming the $C^2$ functions $V(\phi)$ and  $\chi(\phi)$ are class 2 WBI, we obtain the unconstrained dynamical system:
\begin{subequations}
\label{31-syst} 
\begin{align}
   & \frac{d T}{d{\tau}}= \frac{1}{2} (1-T)T \left(3 \gamma  \Omega_m+6 \sin ^2(\theta )+p \Omega_0\right),\\
   & \frac{d \theta}{d\tau}= -\frac{\sec (\theta )}{4 \sqrt{6}} \Bigg\{3 \sqrt{6} \sin (3 \theta )+6 \left(\cos (2 \theta ) - 2
   \Omega_0+1\right) (N+\bar{W}_V(\varphi)) \nonumber \\
   & +\Omega_m \left(-6 \sqrt{6} \gamma  \sin (\theta )-12 (2 M+N+\bar{W}_V(\varphi)+2 \bar{W}_{\chi}(\varphi))+18 \gamma  (M+\bar{W}_{\chi}(\varphi))\right)\nonumber \\
   & +\sqrt{6} \sin (\theta ) (3-2 p \Omega_0)\Bigg\},\\
   & \frac{d \Omega_m}{d{\tau}}= -\frac{1}{2} \Omega_m \Bigg\{6 \cos (2 \theta )-\sqrt{6} (3 \gamma -4) \sin (\theta ) (M+\bar{W}_{\chi}(\varphi))\nonumber \\
   & -2 (3 \gamma (\Omega_m-1)+p \Omega_0+3)\Bigg\},\\
   & \frac{d \Omega_0}{d{\tau}}= - \Omega_0 (-3 \gamma  \Omega_m+3 \cos (2 \theta )-p \Omega_0+p-3),\\
   & \frac{d \varphi}{d{\tau}}=\sqrt{6} \bar{h'}(\varphi)\sin (\theta ).
\end{align}
\end{subequations}
defined on the phase space 
\begin{align}
   & \Big\{(T, \theta, \Omega_m, \Omega_0, \varphi)\in\mathbb{R}^5: 0\leq T\leq 1, -\frac{\pi}{2}\leq \theta \leq \frac{\pi}{2}, 0\leq \varphi\leq h({\phi_0}), \nonumber\\
    & {2 T^2} V(h^{-1}(\varphi))= 3 m^2 (1-T)^2 (\cos (2 \theta )-2 \Omega_{0}-2 \Omega_{m}+1)\Big\}.
\end{align} $\theta$ is unique modulo $2\pi$. It has been chosen such that $\cos \theta\geq 0$. In the following list	$\tan^{-1}[x,y]$
gives the arc tangent of $y/x$, taking into account on which quadrant the point $(x,y)$ is in.  When $x^2+y^2=1$, $\tan^{-1}[x,y]$ gives the number $\theta$  such as $x=\cos\theta$ and $y=\sin\theta$.
\\
The equilibrium points of system \eqref{31-syst} with $\varphi=0$  (i.e., corresponding to $\phi\rightarrow \infty$) are the following:
\begin{enumerate}
    \item[$B_1$:] $\left(0,\tan ^{-1}\left[\sqrt{1-\frac{N^2}{6}},-\frac{N}{\sqrt{6}}\right]+2 \pi  c_1,0,0,0\right), c_1\in \mathbb{Z}$, represents a scalar field dominated solution, satisfying $H\rightarrow \infty$. It is always a nonhyperbolic saddle. 
    
    \item[$B_2$:] $\left(1,\tan ^{-1}\left[\sqrt{1-\frac{N^2}{6}},-\frac{N}{\sqrt{6}}\right]+2 \pi  c_1,0,0,0\right), c_1\in \mathbb{Z}$, represents an scalar field dominated solution, satisfying $H\rightarrow 0$.  The case of physical interest is when it is nonhyperbolic with a 4D stable manifold under one of the conditions (i)- (x) of \ref{AppB}. 
   
    \item[$B_3$:] $\left(0,2 \pi  c_1,0,1,0\right), c_1\in \mathbb{Z}$, with eigenvalues  $\left\{0,\frac{p-6}{2},\frac{p}{2},p,p-3 \gamma \right\}$. It represents a ``geometric fluid'' dominated solution with $H\rightarrow \infty$. The physical interesting situation is when it is nonhyperbolic with a 4D unstable manifold for $p>6, 1\leq \gamma \leq 2$. It is a nonhyperbolic saddle otherwise. 
        
    \item[$B_4$:] $\left(1,2 \pi  c_1,0,1,0\right), c_1\in \mathbb{Z}$, represents a ``geometric fluid'' dominated solution with $H\rightarrow 0$. It is a nonhyperbolic saddle. 
    
    \item[$B_5$:] $\left(0,\tan ^{-1}\left[\sqrt{1-\frac{p^2}{6 N^2}},-\frac{p}{\sqrt{6} N}\right]+2 \pi  c_1,0,1-\frac{p}{N^2},0\right), c_1\in \mathbb{Z}$ represents a scaling solution where neither the energy density of the ``geometric fluid'', nor the energy density of the scalar field completely dominates, that satisfies $H\rightarrow \infty$. It is a nonhyperbolic saddle. 
    
        \item[$B_6$:] $\left(1,\tan ^{-1}\left[\sqrt{1-\frac{p^2}{6 N^2}},-\frac{p}{\sqrt{6} N}\right]+2 \pi  c_1,0,1-\frac{p}{N^2},0\right), c_1\in \mathbb{Z}$, represents a scaling solution where neither the energy density of ``geometric fluid'' nor the energy density of the scalar field completely dominates, that satisfies $H\rightarrow 0$. The situation of physical interest is when it is nohyperbolic with a 4D stable manifold under one of the conditions (i)- (xl) in \ref{AppB}. 
    
    \item[$B_7$:] $\left(0,\tan ^{-1}\left[\sqrt{1-\frac{2 (p-3 \gamma )^2}{3 M^2(4-3 \gamma )^2}},\frac{2(p-3 \gamma)}{\sqrt{6}M(4 -3 \gamma)}\right]+2 \pi  c_1,\frac{2 (6-p) (p-3 \gamma )}{3 (4-3 \gamma )^2 M^2},\frac{(4-3 \gamma )^2
   M^2+2 (\gamma -2) (p-3 \gamma )}{(4-3 \gamma )^2 M^2},0\right), c_1\in \mathbb{Z}$, represents a matter- scalar field - ``geometric'' fluid scaling solution with $H\rightarrow \infty$. It is a nonhyperbolic saddle.

    \item[$B_8$:] $\left(1,\tan ^{-1}\left[\sqrt{1-\frac{2 (p-3 \gamma )^2}{3 M^2(4-3 \gamma )^2}},\frac{2(p-3 \gamma)}{\sqrt{6}M(4 -3 \gamma)}\right]+2 \pi  c_1,\frac{2 (6-p) (p-3 \gamma )}{3 (4-3 \gamma )^2 M^2},\frac{(4-3 \gamma )^2
   M^2+2 (\gamma -2) (p-3 \gamma )}{(4-3 \gamma )^2 M^2},0\right), c_1\in \mathbb{Z}$ represents a matter- scalar field - ``geometric'' fluid scaling solution with $H\rightarrow 0$. The situation of physical interest is when it is nohyperbolic with a 4D stable manifold  under the conditions (i) - (xi) of \ref{AppB}. 
   
      \item[$B_9$:] $\left(0,\tan ^{-1}\left[ \frac{\sqrt{6 (2-\gamma )^2-(4-3 \gamma )^2 M^2}}{\sqrt{6}(2-\gamma )},\frac{(4-3 \gamma ) M}{\sqrt{6}(2-\gamma)}\right]+2 \pi  c_1,\frac{6 (2-\gamma )^2-(4-3 \gamma )^2 M^2}{6 (2-\gamma )^2},0,0\right), c_1\in \mathbb{Z}$ represents a matter- scalar field scaling solution with $H\rightarrow \infty$. It is a nonhyperbolic saddle. 
   
      \item[$B_{10}$:] $\left(1,\tan ^{-1}\left[ \frac{\sqrt{6 (2-\gamma )^2-(4-3 \gamma )^2 M^2}}{\sqrt{6}(2-\gamma )},\frac{(4-3 \gamma ) M}{\sqrt{6}(2-\gamma)}\right]+2 \pi  c_1,\frac{6 (2-\gamma )^2-(4-3 \gamma )^2 M^2}{6 (2-\gamma )^2},0,0\right), c_1\in \mathbb{Z}$ represents a matter- scalar field scaling solution with $H\rightarrow 0$.
  The situation of physical interest is when it is nohyperbolic with a 4D stable manifold for 
   \begin{enumerate}
    \item $1\leq \gamma <\frac{4}{3}, \;  N<-\sqrt{6}, \;  \frac{N}{3 \gamma -4}-\sqrt{\frac{6 \gamma ^2-12 \gamma +N^2}{(3 \gamma -4)^2}}<M<\frac{\sqrt{6} \gamma -2 \sqrt{6}}{3 \gamma -4}, \;  p>\frac{6 \gamma ^2-12 \gamma -9 \gamma
   ^2 M^2+24 \gamma  M^2-16 M^2}{2 \gamma -4}$, or 
   \item $\frac{4}{3}<\gamma <2, \;  N<-\sqrt{6}, \;  \frac{\sqrt{6} \gamma -2 \sqrt{6}}{3 \gamma -4}<M<\sqrt{\frac{6 \gamma ^2-12 \gamma +N^2}{(3 \gamma -4)^2}}+\frac{N}{3
   \gamma -4}, \;  p>\frac{6 \gamma ^2-12 \gamma -9 \gamma ^2 M^2+24 \gamma  M^2-16 M^2}{2 \gamma -4}$, or 
   \item $1\leq \gamma <\frac{4}{3}, \;  N>\sqrt{6}, \;  \frac{2 \sqrt{6}-\sqrt{6} \gamma }{3 \gamma -4}<M<\sqrt{\frac{6
   \gamma ^2-12 \gamma +N^2}{(3 \gamma -4)^2}}+\frac{N}{3 \gamma -4}, \;  p>\frac{6 \gamma ^2-12 \gamma -9 \gamma ^2 M^2+24 \gamma  M^2-16 M^2}{2 \gamma -4}$, or 
   \item $\frac{4}{3}<\gamma <2, \;  N>\sqrt{6}, \;  \frac{N}{3
   \gamma -4}-\sqrt{\frac{6 \gamma ^2-12 \gamma +N^2}{(3 \gamma -4)^2}}<M<\frac{2 \sqrt{6}-\sqrt{6} \gamma }{3 \gamma -4}, \;  p>\frac{6 \gamma ^2-12 \gamma -9 \gamma ^2 M^2+24 \gamma  M^2-16 M^2}{2 \gamma -4}$.
   \end{enumerate}
     
      \item[$B_{11}$:] $\left(0,\tan^{-1}\left[\sqrt{1-\frac{6 \gamma ^2}{\left(2 N+M (4-3 \gamma )\right)^2}},-\frac{\sqrt{6}
\gamma }{2 N+M (4-3 \gamma )}\right]+2 \pi  c_1, \frac{4 N (2 M+N)-6 (2+M N) \gamma }{(2 N+M (4-3 \gamma ))^2},0,0\right), c_1\in \mathbb{Z}$ represents a matter- scalar field scaling solution with $H\rightarrow \infty$.
It is a hyperbolic saddle. 
     \item[$B_{12}$:] $\left(1,\tan^{-1}\left[\sqrt{1-\frac{6 \gamma ^2}{\left(2 N+M (4-3 \gamma )\right)^2}},-\frac{\sqrt{6}
\gamma }{2 N+M (4-3 \gamma )}\right]+2 \pi  c_1,
\frac{4 N (2 M+N)-6 (2+M N) \gamma }{(2 N+M (4-3 \gamma ))^2},0,0\right), c_1\in \mathbb{Z}$ represents a matter- scalar field scaling solution with $H\rightarrow 0$.
The situation of physical interest is when it is nohyperbolic with a 4D stable manifold under one of the conditions (i) - (lviii) in \ref{AppB}. 
\end{enumerate}
Summarizing, we have provided a dynamical system analysis of the system \eqref{Non_min2} for arbitrary $V(\phi)$ and arbitrary $\chi(\phi)$. We have exhaustively examined the equilibrium points in a phase space, as well as for $\phi\rightarrow +\infty$, obtaining equilibrium points that represent some solutions of cosmological interest, such as several types of scaling solutions, a kinetic dominated solution representing a stiff fluid, a solution dominated by an effective energy density of geometric origin, a quintessence scalar field dominated solution, the vacuum de Sitter solution associated to the minimum of the potential, and a non-interacting matter dominated solution. All such revelations demonstrate very rich cosmological phenomenologies.

Now, we proceed to study some particular realizations of the above model with emphasis in dynamics as $\phi \rightarrow \infty$. 

\subsubsection{Example: an scalar field model with potential $V(\phi)= V_0 e^{-\lambda \phi}$ in vacuum.}
\label{Section2.3}

In this section, we consider a scalar field model with potential $V(\phi)= V_0 e^{-\lambda \phi}$ in vacuum for the flat FLRW metric. That is, $N=-\lambda, M=0, W_{V}\equiv 0, W_{\chi}\equiv 0$, $\Omega_m\equiv 0, \Omega_0\equiv 0, \rho_m\equiv 0, G_0 \equiv 0$ and $\chi\equiv 1$, and using the time variable $\tau= \ln a$, we obtain the unconstrained dynamical system:
\begin{align}\label{syst3A}
&\frac{d T}{d{\tau}}=3 (1-T) T \sin ^2(\theta ), \quad 
\frac{d \theta}{d{\tau}}= \frac{1}{2} \cos (\theta ) \left(\sqrt{6} \lambda -6 \sin (\theta )\right),
\end{align}
defined in the finite cylinder $\mathbf{S}$ with boundaries $T=0$ and $T=1$.
\begin{table}[t]
\begin{center}
\begin{tabular}{|c|c|c|c|}
\hline
Label & $\left(T, \theta \right)$ & Existence &  Stability\\ \hline \hline
$P_1$ &  $\left(0, -\frac{\pi }{2} + 2 c_1 \pi \right)$ & $\forall \lambda$ & Saddle for $\lambda < -\sqrt{6}$. \\ 
&&& Nonhyperbolic for $\lambda = -\sqrt{6}$. \\ 
&&& Source for $\lambda > -\sqrt{6}$. \\ \hline
$P_2$ & $\left(0, \frac{\pi }{2} + 2 c_1 \pi \right)$ & $\forall \lambda$  & Saddle for $\lambda >\sqrt{6}$. \\
&&& Nonhyperbolic  for $\lambda =\sqrt{6}$.\\ 
&&& Source for $\lambda <\sqrt{6}$. \\ \hline 
$P_3$ &   $\left(0, \arcsin\left(\frac{\lambda }{\sqrt{6}}\right)\right)$ & $-\sqrt{6} \leq \lambda \leq \sqrt{6}$  & Nonhyperbolic  for  $\lambda \in \{-\sqrt{6},0, \sqrt{6}\}$. \\ 
&&& Saddle for  $-\sqrt{6}< \lambda <0$ or \\
&&& $0<\lambda <\sqrt{6}$. \\ \hline 
$P_4$ &  $\left(1,  -\frac{\pi }{2} + 2 c_1 \pi \right)$  & $\forall \lambda$ & Sink for $\lambda < -\sqrt{6}$. \\ 
&&& Nonhyperbolic  for $\lambda = -\sqrt{6}$. \\ 
&&& Saddle for $\lambda > -\sqrt{6}$. \\ \hline
$P_5$ &  $\left(1, \frac{\pi }{2}+ 2 c_1 \pi \right)$ & $\forall \lambda$  & Sink for $\lambda >\sqrt{6}$. \\
&&& Nonhyperbolic  for $\lambda =\sqrt{6}$.\\ 
&&& Saddle for $\lambda <\sqrt{6}$. \\ \hline 
$P_6$ &  $ \left( 1, \arcsin\left(\frac{\lambda }{\sqrt{6}}\right) \right)$ & $-\sqrt{6} \leq \lambda \leq \sqrt{6}$ & Nonhyperbolic  for $\lambda \in \{-\sqrt{6},0, \sqrt{6}\}$. \\ 
&&& Sink for $-\sqrt{6}< \lambda <0$ or \\
&&& $0<\lambda <\sqrt{6}$. \\ \hline 
\end{tabular}
\end{center}\caption{\label{critsyst3A} Existence conditions and stability conditions of the equilibrium points of equations  \eqref{syst3A}, $c_1\in \mathbb{Z}$.}
\end{table}
\begin{figure}[t]
\begin{center}
\subfigure[]{\includegraphics[scale=0.5]{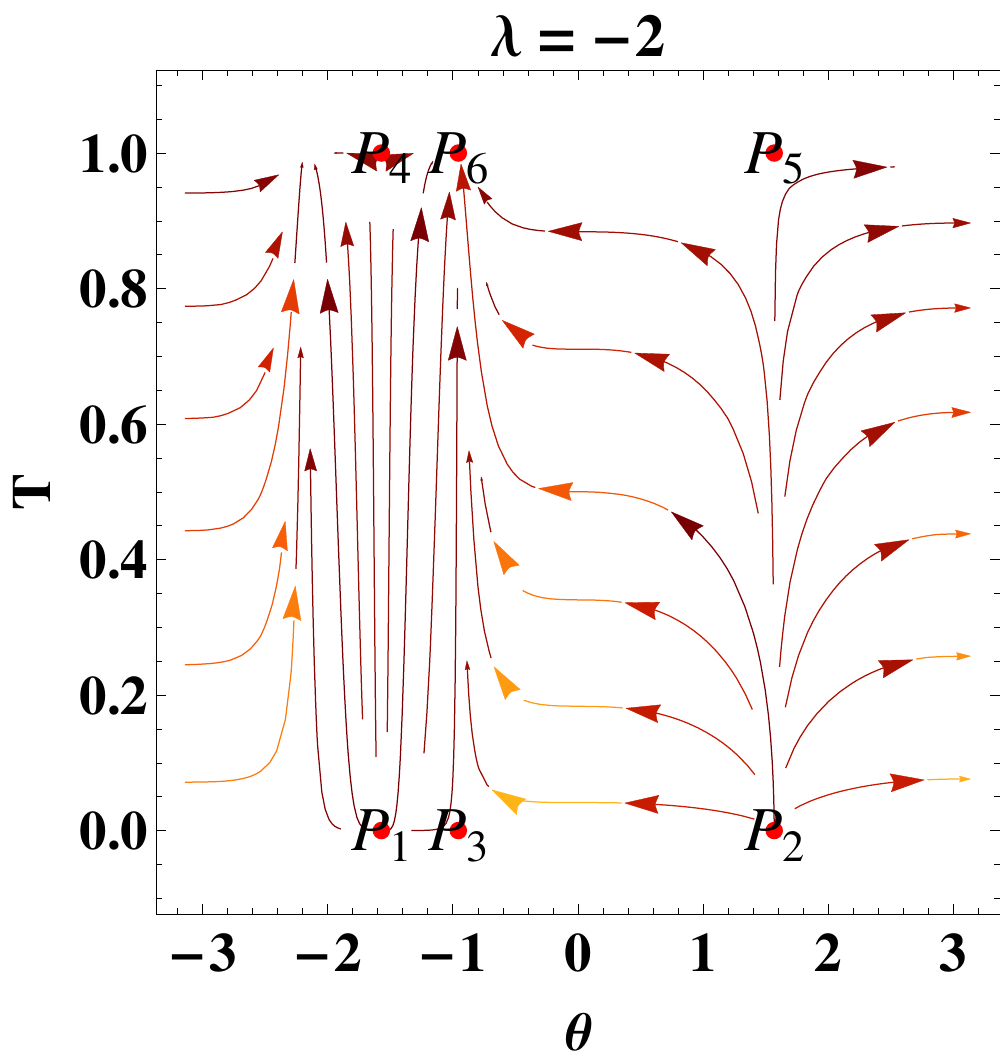} }\hspace{2cm}
\subfigure[]{\includegraphics[scale=0.5]{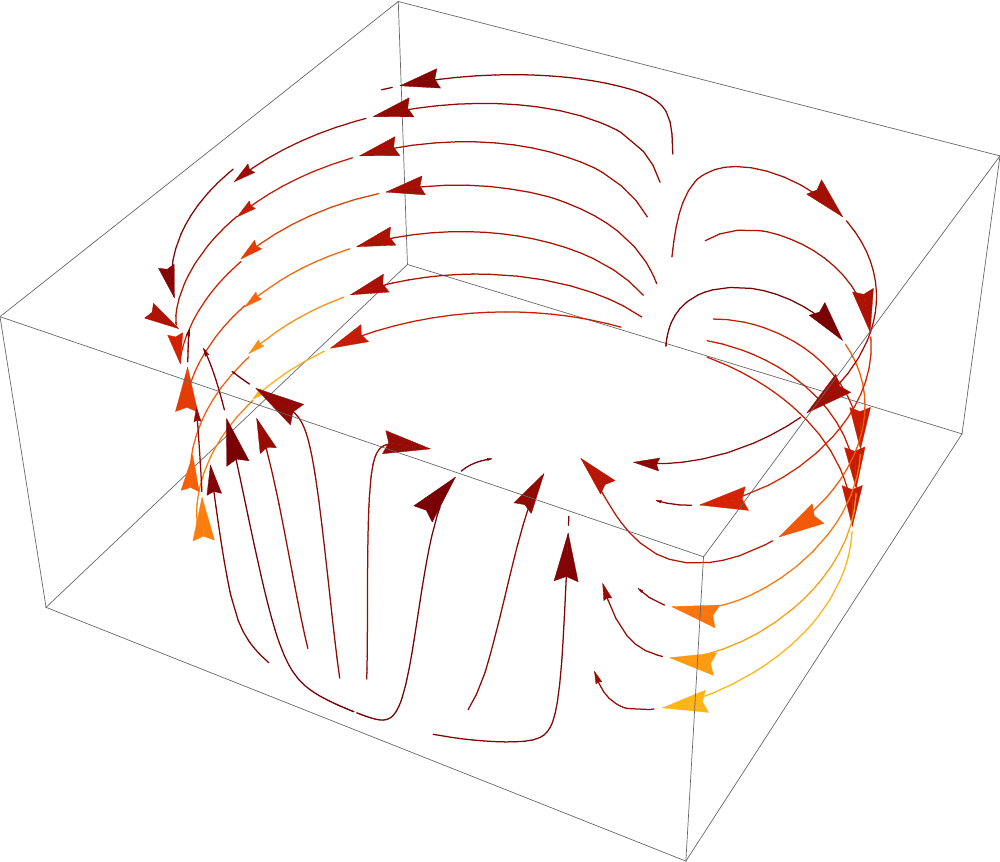}}
\subfigure[]{\includegraphics[scale=0.5]{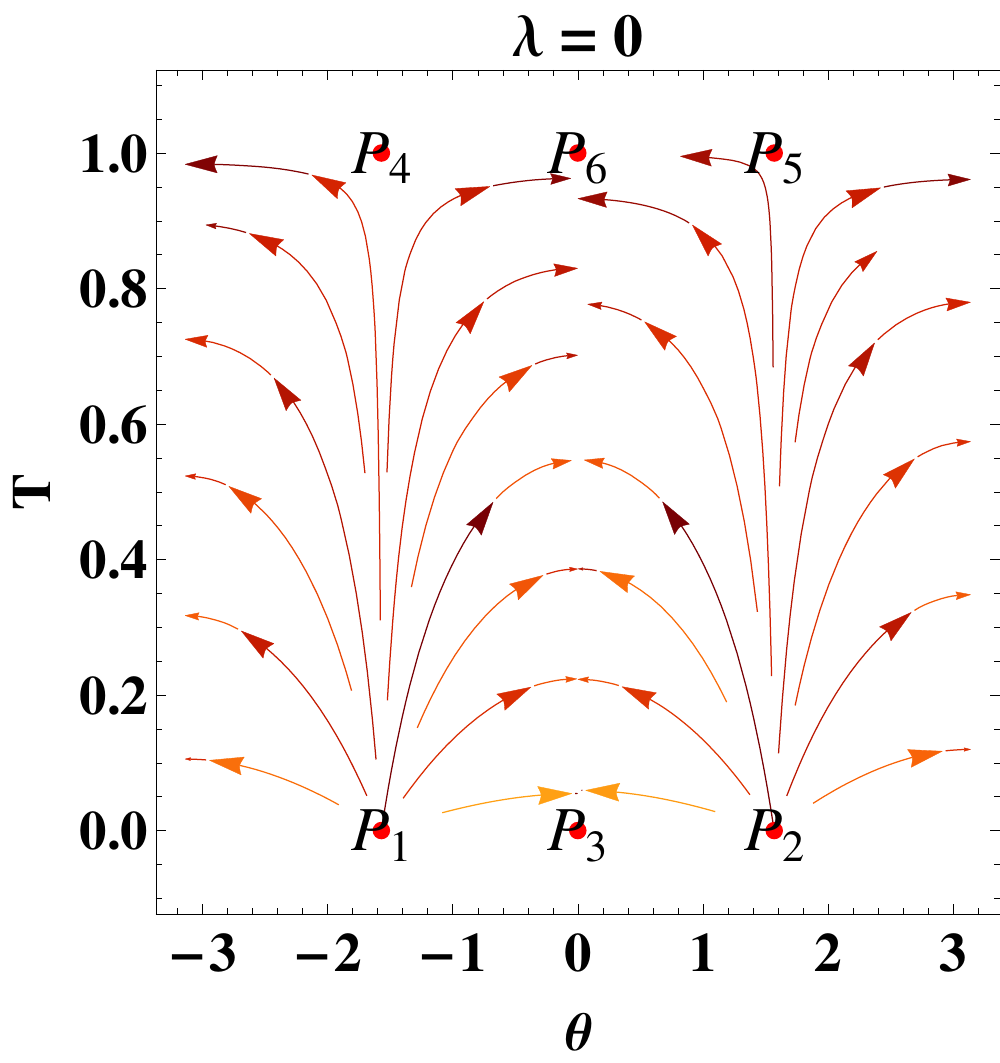} }\hspace{2cm}
\subfigure[]{\includegraphics[scale=0.5]{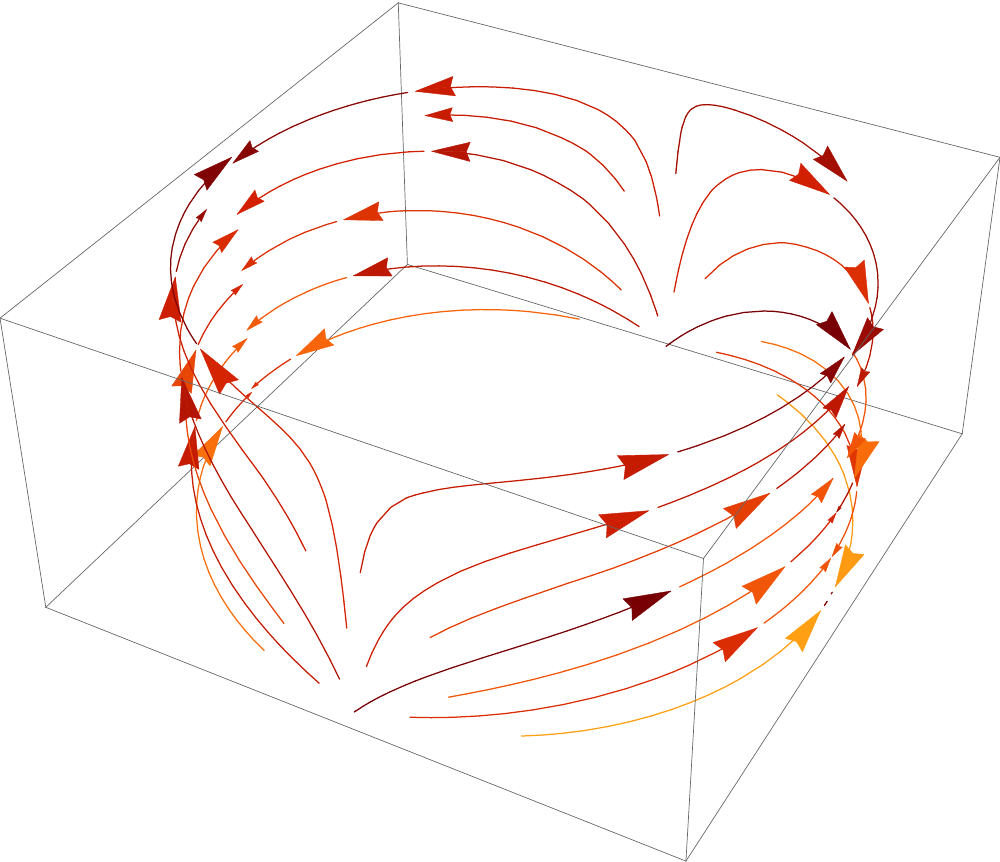}}
\end{center}\caption{\label{Fig2.7} Phase portrait of equations   \eqref{syst3A} (left panel). Projection over the cylinder $\mathbf{S}$ (right panel) for different values of $\lambda$.}
\end{figure}
\begin{figure}[t]
\begin{center}
\subfigure[]{\includegraphics[scale=0.5]{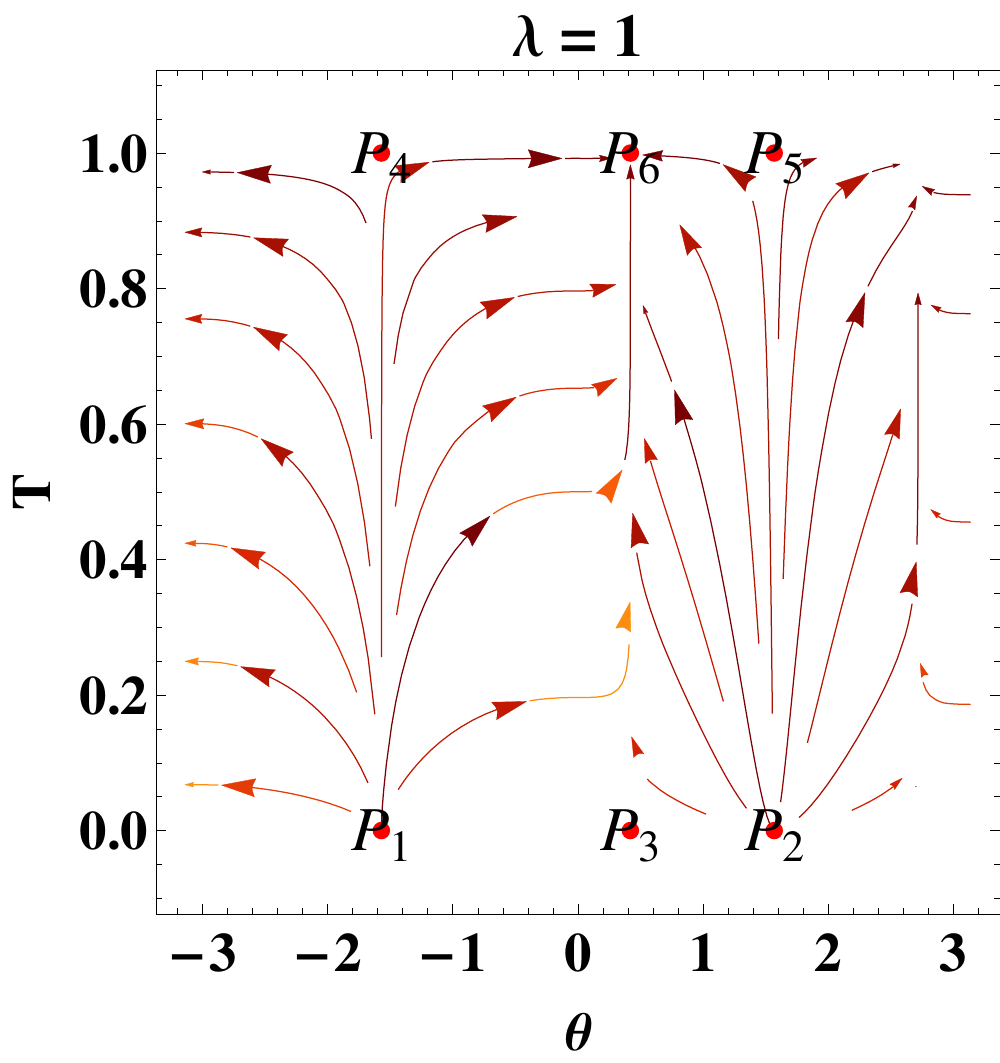} }\hspace{2cm}
\subfigure[]{\includegraphics[scale=0.5]{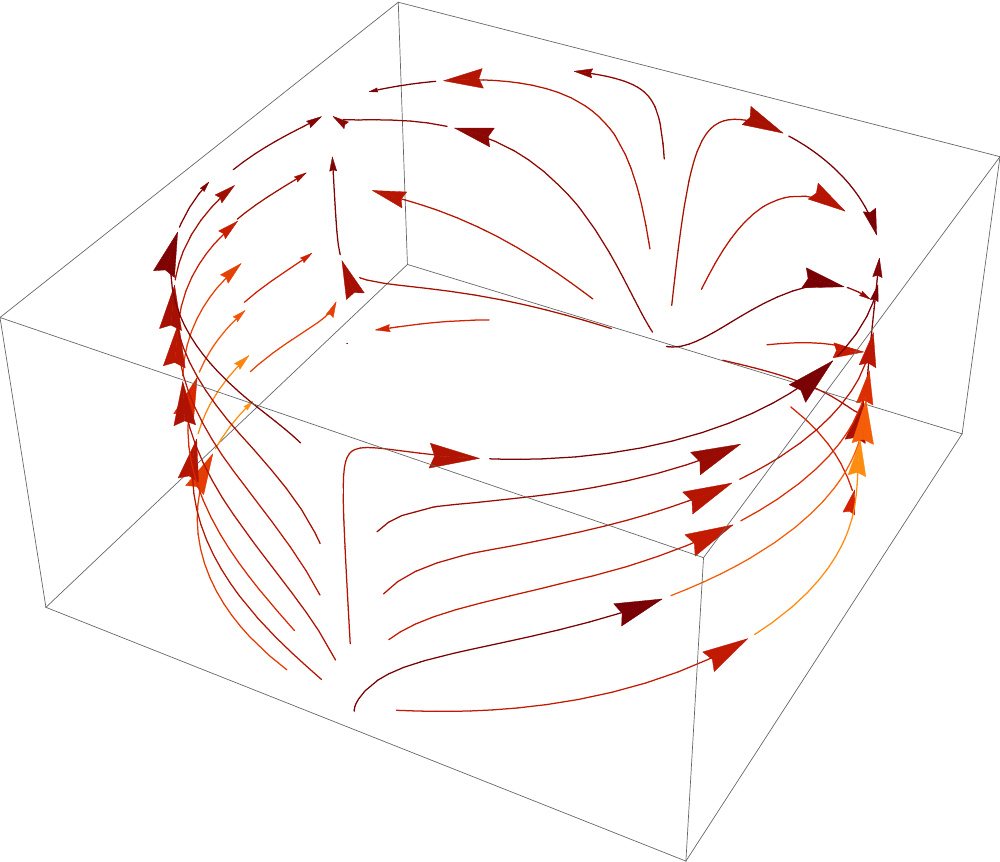}}
\subfigure[]{\includegraphics[scale=0.5]{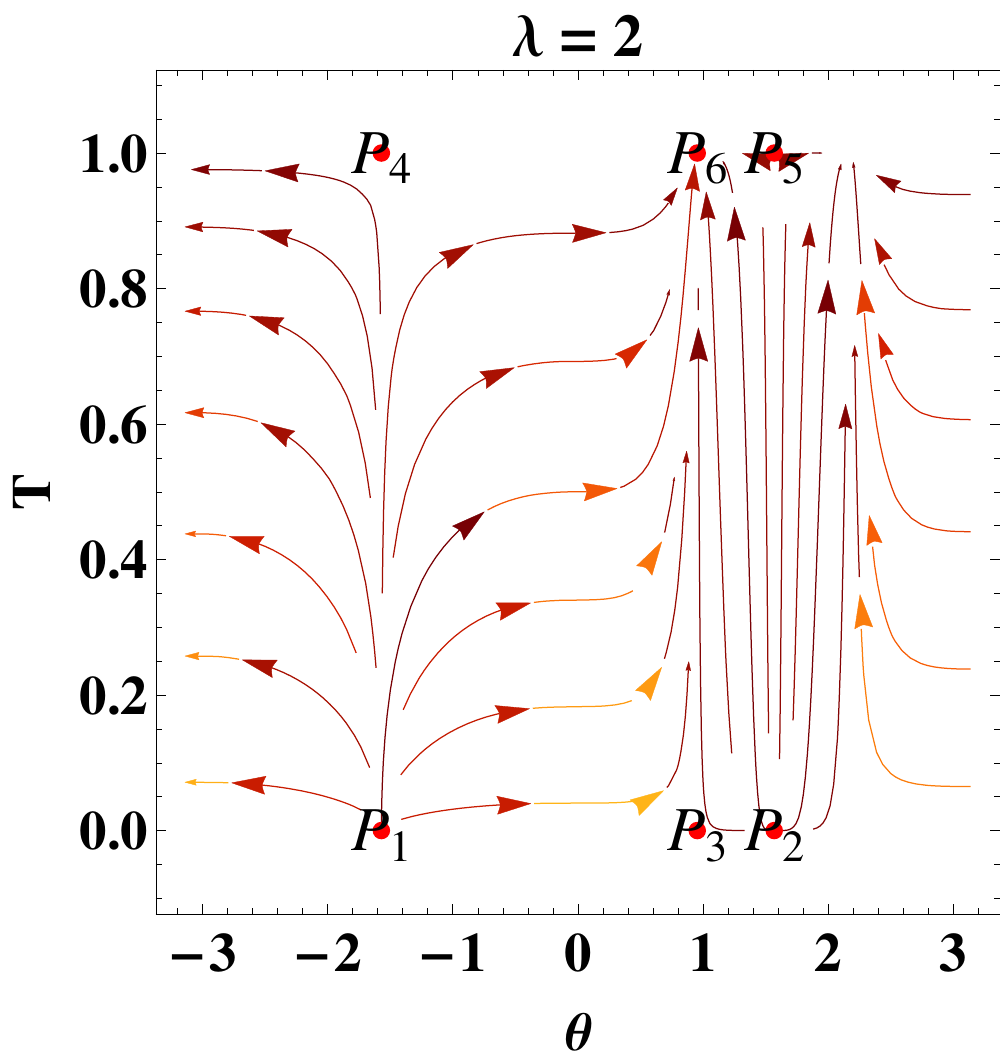} }\hspace{2cm}
\subfigure[]{\includegraphics[scale=0.5]{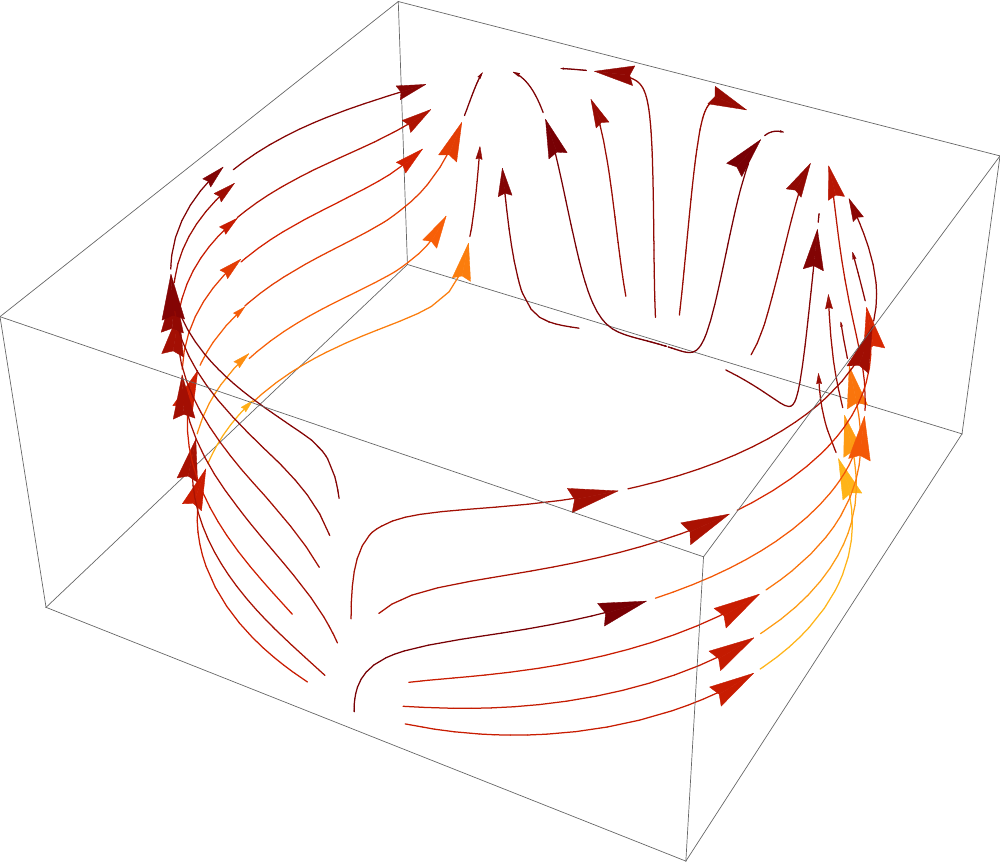}}
\end{center}\caption{\label{Fig2.7b} Phase portrait of equations   \eqref{syst3A} (left panel). Projection over the cylinder $\mathbf{S}$ (right panel) for different values of $\lambda$.}
\end{figure}
The variable $T$ is suitable for global analysis \cite{Alho:2014fha}, due to 
\begin{equation}
\label{3monotony}
\frac{d T}{d \tau}\Big|_{\sin \theta=0}=0, \quad \frac{d^2 T}{d {\tau}^2}\Big|_{\sin \theta=0}=0, \quad \quad \frac{d^3 T}{d {\tau}^3}\Big|_{\sin \theta=0}=9 \lambda ^2 (1-T) T.
\end{equation}
From the first equation of \eqref{syst3A} and equation \eqref{3monotony}, $T$ is a monotonically increasing function on $\mathbf{S}$. As a consequence, all orbits originate from the invariant subset $T=0$ (which contains the $\alpha$-limit), which is classically related to the initial singularity with $H \rightarrow \infty$, and ends on the invariant boundary subset $T=1$, which corresponds asymptotically to $H=0$. 

The equilibrium points of equations   \eqref{syst3A} are the following: 
\begin{enumerate}
\item  $P_1: (T,\theta)=\left(0, -\frac{\pi }{2} + 2 c_1 \pi \right),  c_1\in \mathbb{Z}$,  with eigenvalues $\left\{3,\sqrt{\frac{3}{2}} \lambda +3\right\}$. It represents a kinetic dominated solution with $H\rightarrow \infty$. The stability conditions of $P_1$ are the following:
        \begin{enumerate}
        \item Saddle for $\lambda < -\sqrt{6}$.
        \item Nonhyperbolic for $\lambda = -\sqrt{6}$.
        \item Source for $\lambda > -\sqrt{6}$.
        \end{enumerate} 
    
\item  $P_2: (T,\theta)= \left(0, \frac{\pi }{2}  + 2 c_1 \pi \right),  c_1\in \mathbb{Z}$, with eigenvalues $\left\{3,3-\sqrt{\frac{3}{2}} \lambda \right\}$. It represents a kinetic dominated solution with $H\rightarrow \infty$. The stability conditions of $P_2$ are the following: 
        \begin{enumerate}
        \item Source for $\lambda <\sqrt{6}$.
        \item Nonhyperbolic for $\lambda =\sqrt{6}$.
        \item Saddle for $\lambda >\sqrt{6}$.
        \end{enumerate}

 \item  $P_3: (T,\theta)=\left(0, \arcsin\left(\frac{\lambda }{\sqrt{6}}\right)\right)$, with eigenvalues $\left\{\frac{\lambda ^2}{2},\frac{1}{2} \left(\lambda
   ^2-6\right)\right\}$. It represents a scalar field dominated solution with $H\rightarrow \infty$. The stability conditions of $P_3$ are the following:
        \begin{enumerate} 
        \item Nonhyperbolic  for $\lambda \in \{-\sqrt{6},0, \sqrt{6}\}$.
        \item Saddle for  $-\sqrt{6}< \lambda <0$ or  $0<\lambda <\sqrt{6}$. 
        \end{enumerate}
\item  $P_4: (T,\theta)=\left(1,  -\frac{\pi }{2}  + 2 c_1 \pi \right),  c_1\in \mathbb{Z}$, with eigenvalues 
  $\left\{-3,\sqrt{\frac{3}{2}} \lambda +3\right\}$. It represents a kinetic dominated solution with $H\rightarrow 0$. The stability conditions of $P_4$ are the following: 
       \begin{enumerate}
       \item Sink for $\lambda < -\sqrt{6}$. 
       \item Nonhyperbolic  for $\lambda = -\sqrt{6}$. 
       \item Saddle for $\lambda > -\sqrt{6}$. 
       \end{enumerate}
  
\item  $P_5: (T,\theta)= \left(1, \frac{\pi }{2}  + 2 c_1 \pi \right),  c_1\in \mathbb{Z}$, with eigenvalues $\left\{-3,3-\sqrt{\frac{3}{2}} \lambda \right\}$.  It represents a kinetic dominated solution with $H\rightarrow 0$. The stability conditions of $P_5$ are the following:          \begin{enumerate}
       \item It is a sink for $\lambda >\sqrt{6}$.
       \item Nonhyperbolic  for $\lambda =\sqrt{6}$.
       \item Saddle for $\lambda <\sqrt{6}$. 
       \end{enumerate}
 
\item  $P_6: (T,\theta)=\left( 1, \arcsin\left(\frac{\lambda }{\sqrt{6}}\right)\right)$ with eigenvalues $\left\{-\frac{\lambda ^2}{2},\frac{1}{2} \left(\lambda
   ^2-6\right)\right\}$. It represents a scalar field dominated solution with $H\rightarrow 0$. The stability conditions of $P_6$ are the following:
        \begin{enumerate}
        \item Nonhyperbolic for $\lambda \in \{-\sqrt{6},0, \sqrt{6}\}$.
        \item Sink for $-\sqrt{6}< \lambda <0$ or  $0<\lambda <\sqrt{6}$.
        \end{enumerate}
\end{enumerate}
In table \ref{critsyst3A} are summarized the existence conditions and stability conditions of the equilibrium points of equations   \eqref{syst3A}.
 
In the Figures \ref{Fig2.7} and \ref{Fig2.7b} are presented some orbits of the flow of equations   \eqref{syst3A} (left panel); and a projection over the cylinder $\mathbf{S}$ (right panel) for different values of $\lambda$.   

\subsubsection{Example: a scalar-field cosmology with generalized harmonic potential $
V(\phi)= \mu^3 \left[\frac{\phi^2}{\mu} + b f \cos\left(\delta + \frac{\phi}{f}\right)\right]$, $b\neq 0$ in vacuum.}
\label{Sect:2.4.1}

In this section we proceed to the asymptotic analysis as $\phi\rightarrow \infty$ of a scalar-field cosmology with generalized harmonic potential 
$V(\phi)= \mu^3 \left[\frac{\phi^2}{\mu} + b f \cos\left(\delta + \frac{\phi}{f}\right)\right]$, $b\neq 0$ in vacuum, for a flat FLRW model.  We set $N=0, M=0, W_{\chi}\equiv 0$, $\Omega_m\equiv 0, \Omega_0\equiv 0, \rho_m\equiv 0, G_0 \equiv 0$ and $\chi\equiv 1$. \\
Furthermore, 
 \begin{equation}
\label{36-syst}
{W}_V(\phi)=
\frac{2 \phi -b \mu  \sin \left(\delta +\frac{\phi }{f}\right)}{b f \mu 
   \cos \left(\delta +\frac{\phi }{f}\right)+\phi ^2}.
\end{equation}

\begin{figure}[t]
\begin{center}
	\subfigure[$(b, f, \delta, \mu)= (0.1, 0.33, 0, 0.9)$]{\includegraphics[scale=0.5]{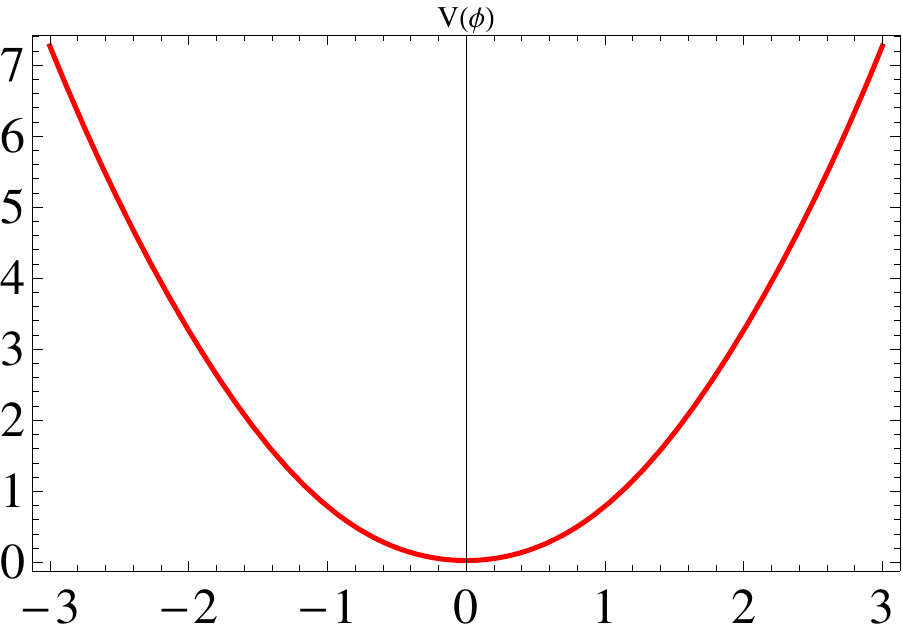} } \hspace{2cm}
	\subfigure[$(b, f, \delta, \mu)= (0.1, 0.33, 0, 0.9)$]{\includegraphics[scale=0.5]{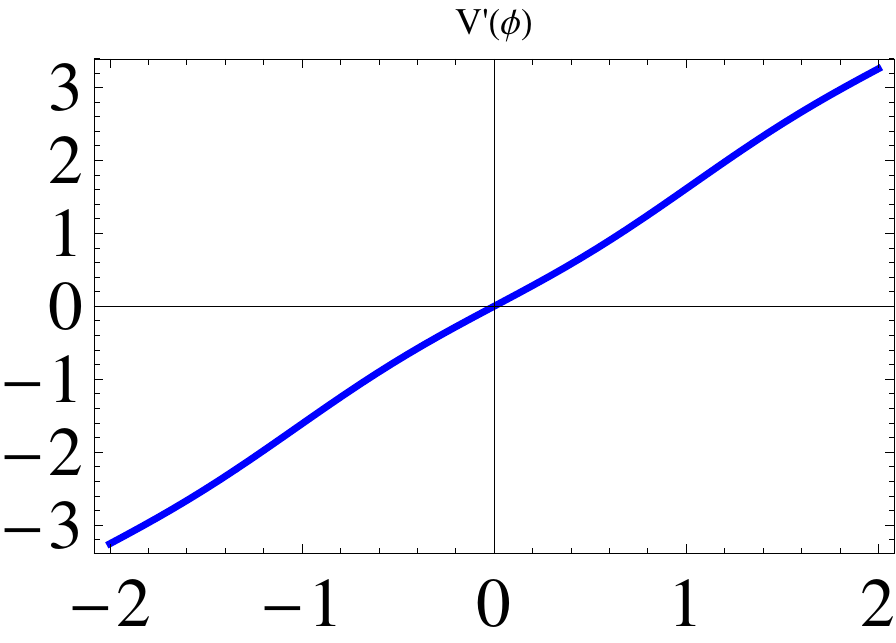} }
	\subfigure[$(b, 
   f, \delta, \mu)= (0.99, 0.09, 0, 0.9)$]{\includegraphics[scale=0.5]{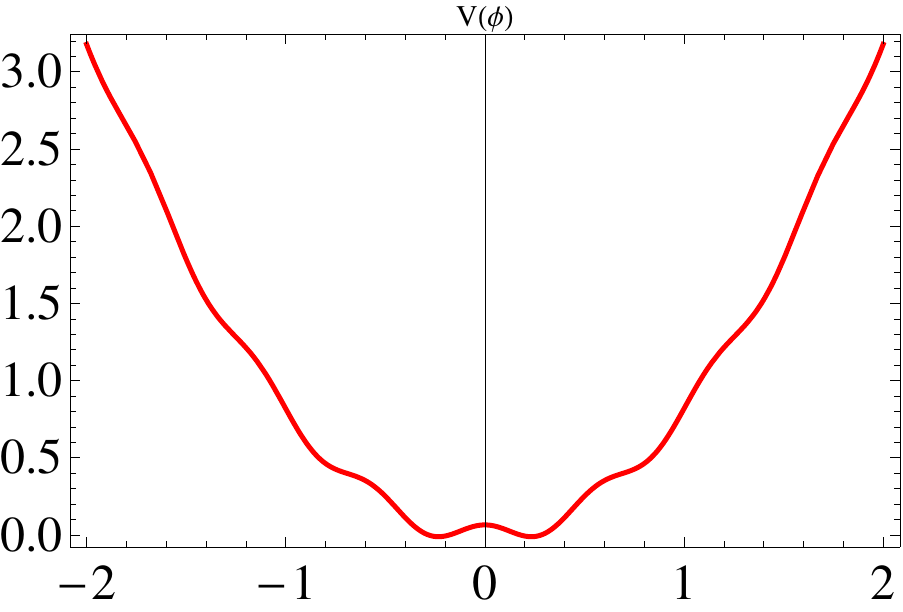}} \hspace{2cm}
	\subfigure[$(b, 
   f, \delta, \mu)= (0.99, 0.09, 0, 0.9)$]{\includegraphics[scale=0.5]{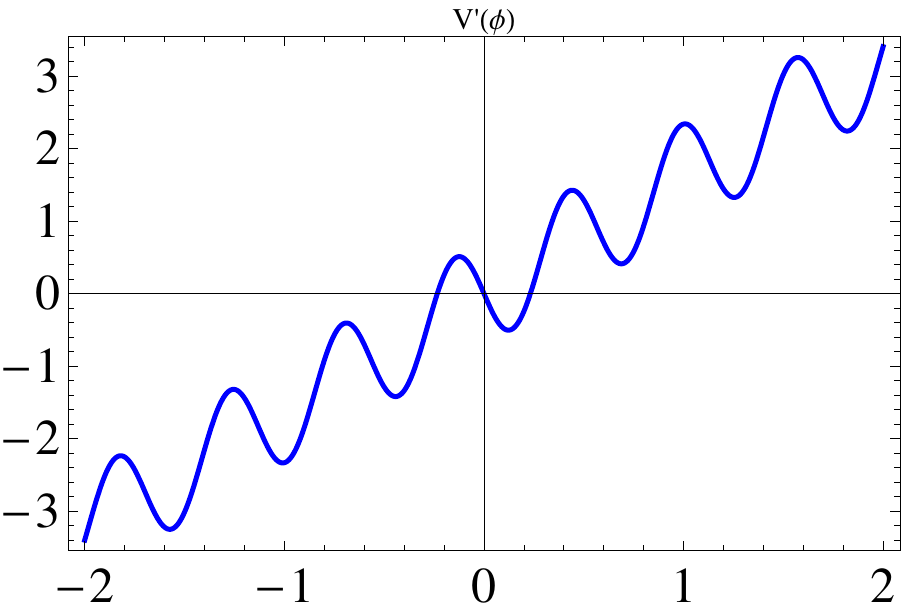} }
\end{center}\caption{\label{PlotMonodromy-Potential} with the generalized harmonic potential $
V(\phi)= \mu^3 \left[\frac{\phi^2}{\mu} + b f \cos\left(\delta + \frac{\phi}{f}\right)\right]$ and its derivative.} 
\end{figure}

In figure \ref{PlotMonodromy-Potential} the potential $V(\phi)$ and its derivative $V^{\prime}(\phi)$ are represented for some values of the parameters $(b, f, \delta, \mu)$. The condition for the existence of a local minimum at the origin is $\delta=0, \mu ^3 \left(\frac{2}{\mu }-\frac{b}{f}\right)>0$; with $V(0)=b f \mu^3$.   
The condition for the existence of a local maximum at the origin is $\delta=0, \mu ^3 \left(\frac{2}{\mu }-\frac{b}{f}\right)<0$; with $V(0)=b f \mu^3$. For $\delta= 0,\mu= \frac{2 f}{b},\phi= 0$, where the origin is a degenerated local minimum of order two with $V(0)=\frac{8 f^4}{b^2}$. 

For the dynamical system analysis of the system when $|\phi|<\infty$, we remit the reader to Section  \ref{Sect2.4}. There, we analyze the dynamics as $\phi\rightarrow \infty$. 
\\
Defining the transformation
\begin{equation}
\label{transform34}
\varphi=h(\phi)=\varphi=\left(\delta +\frac{\phi}{f}\right)^{-\frac{1}{4}},
\end{equation}
we deduce that $V(\phi)$ is 2 WBI with exponential order $N=0$.  
 \begin{equation}
\bar{W}_V(\varphi)=\left\{\begin{array}{cc}
\frac{-b \mu  \varphi^8 \sin \left(\frac{1}{\varphi^4}\right)-2 \delta  f \varphi^8+2 f \varphi^4}{f \left(b \mu  \varphi^8 \cos
   \left(\frac{1}{\varphi^4}\right)+f \left(\delta  \varphi^4-1\right)^2\right)}, & \varphi>0,\\
0, & \varphi=0 \end{array}\right.,
\end{equation}
\begin{equation}
\bar{h'}(\varphi)=\left\{\begin{array}{cc}
-\frac{\varphi^5}{4 f}, & \varphi>0,\\
0,  & \varphi=0 \end{array}\right.,
\end{equation}
that satisfy the conditions (ii) and (iii) of Definition \ref{kWBI2}.
Hence, we have the dynamical system 
\begin{subequations}
\label{model1Alternative1-5syst3}
\begin{align}
& \frac{d T}{d {\tau}}=3 (1-T) T \sin ^2(\theta), \label{4syst3}\\
& \frac{d \theta}{d{\tau}}=-\frac{1}{2}  \cos (\theta ) \left(6 \sin (\theta )+\sqrt{6} \bar{W}_V(\varphi)\right),\\
& \frac{d \varphi}{d {\tau}}=\sqrt{6} \bar{h'}(\varphi) \sin (\theta ),
\end{align}
\end{subequations}
where we have used the new time variable $\tau= \ln a$,  
defined on a phase space which consists of the vector product $\mathbf{S}\times J$ of finite cylinder $\mathbf{S}$ with boundaries $T=0$ and $T=1$ with the interval 
$J=\left[0,  \left(\delta +\frac{\phi_0}{f}\right)^{-\frac{1}{4}}\right]$. 
The variable $T$ is suitable for global analysis \cite{Alho:2014fha}, due to 
\begin{align}
\label{4monotony}
& \frac{d T}{d \tau}\Big|_{\sin \theta=0}=0, \quad \frac{d^2 T}{d \tau^2}\Big|_{\sin \theta=0}=0, \nonumber \\
& \frac{d^3 T}{d \tau^3}\Big|_{\sin \theta=0}=\frac{9 (1-T) T v^8 \left(b \mu  v^4 \sin \left(\frac{1}{v^4}\right)+2 f \left(\delta  v^4-1\right)\right)^2}{f^2 \left(b \mu
    v^8 \cos \left(\frac{1}{v^4}\right)+f \left(\delta  v^4-1\right)^2\right)^2}. 
\end{align}
From equation \eqref{4syst3} and equation \eqref{4monotony}, $T$ is a monotonically increasing function on $\mathbf{S}\times J$. As a consequence, all orbits originate from the invariant subset $T=0$ (which contains the $\alpha$-limit), which is classically related to the initial singularity with $H \rightarrow \infty$, and ends on the invariant boundary subset $T=1$, which corresponds to asymptotically $H=0$. 
\\
The (curves of) equilibrium points of \eqref{model1Alternative1-5syst3} are the following:
\begin{enumerate} 
\item $({T}, \theta, \varphi)= \left(T_c, 2 n \pi, 0\right)$, with eigenvalues $\{-3,0,0\}$. It is nonhyperbolic.
\item $({T}, \theta, \varphi)= \left(T_c, \pi+ 2 n \pi, 0\right)$, with eigenvalues $\{-3,0,0\}$. It is nonhyperbolic.
\item $({T}, \theta, \varphi)= \left(0, -\frac{\pi}{2}+2 n \pi, 0\right)$, with eigenvalues $\{3,3,0\}$. It is nonhyperbolic.
\item $({T}, \theta, \varphi)= \left(0, \frac{\pi}{2}+2 n \pi, 0\right)$, with eigenvalues $\{3,3,0\}$. It is nonhyperbolic.
\item $({T}, \theta, \varphi)= \left(1, -\frac{\pi}{2}+2 n \pi, 0\right)$, with eigenvalues $\{-3,3,0\}$. Behaves as saddle. 
\item $({T}, \theta, \varphi)= \left(1, \frac{\pi}{2}+2 n \pi, 0\right)$, with eigenvalues $\{-3,3,0\}$. Behaves as saddle. 
\end{enumerate}

\begin{figure}
    \centering
    \includegraphics[scale=0.7]{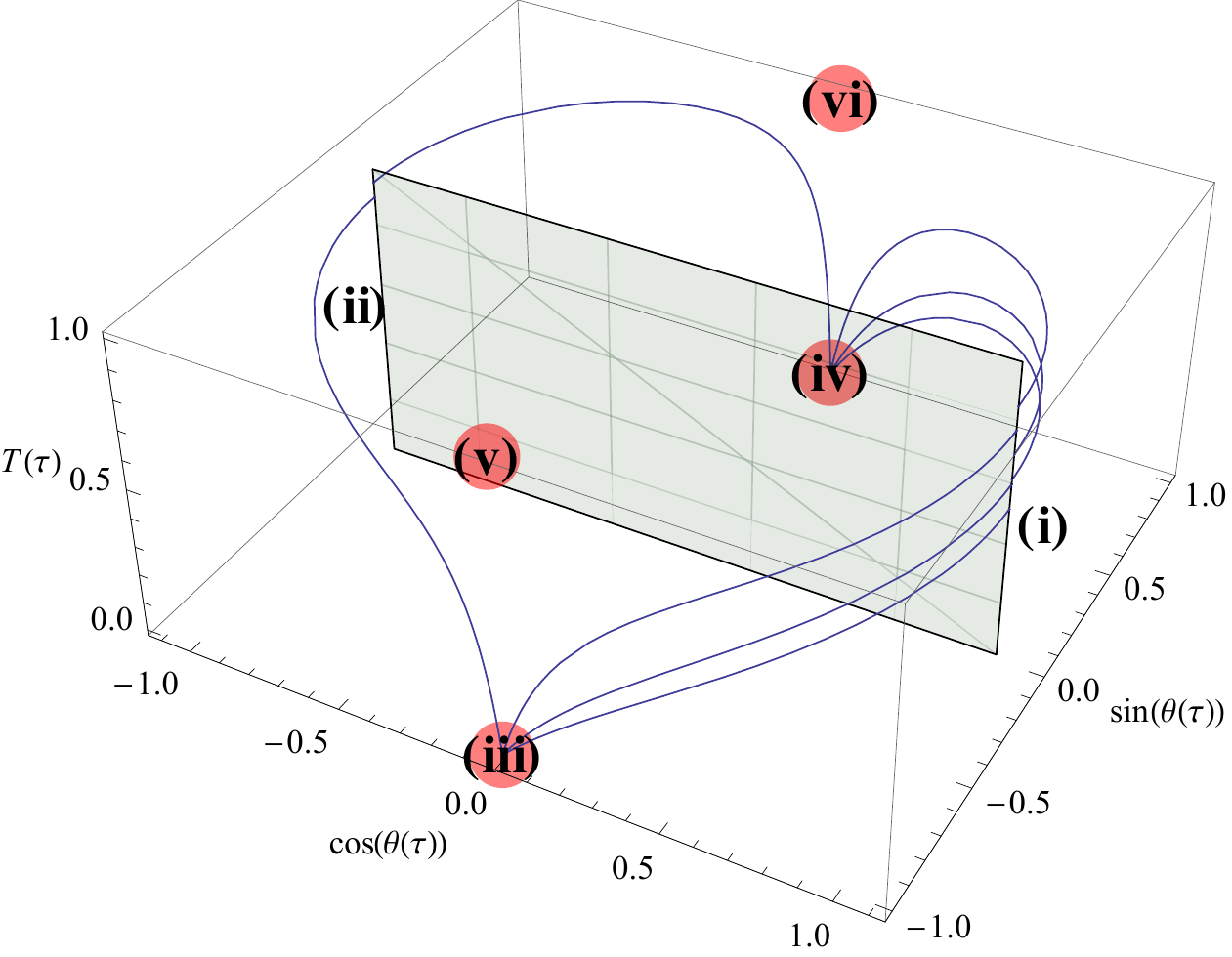}
    \caption{Evaluation of $(\cos \theta, \sin\theta, T)$ at some orbits of the system \eqref{model1Alternative1-5syst3} for $(b, f, \delta, \mu)= (0.1, 0.33, 0, 1)$. In the plot are represented the points (iii) and (iv) are sources; the points (v) and (vi) are saddles. The vertical plane represents the invariant set $\{(T,\theta, \varphi):\varphi=0, \sin(\theta)=0\}$. The vertical sides (i) and (ii) of this rectangle are the local sinks. }
    \label{fig:36}
\end{figure}
In the Figure \ref{fig:36}, it is evaluated $(\cos \theta, \sin\theta, T)$ at some orbits of the system \eqref{model1Alternative1-5syst3} for $(b, f, \delta, \mu)= (0.1, 0.33, 0, 1)$. In the plot are represented the points (iii) and (iv), which are sources; the points (v) and (vi), which are saddles. The vertical plane represents the invariant set $\{(T,\theta, \varphi):\varphi=0, \sin(\theta)=0\}$. The vertical sides (i) and (ii) of this rectangle are the local sinks.
 
\subsubsection{Example: a scalar-field cosmology with generalized harmonic potential $
V(\phi)= \mu ^3 \left[b f \left(\cos (\delta )-\cos \left(\delta +\frac{\phi }{f}\right)\right)+\frac{\phi ^2}{\mu}\right]
$, $b\neq 0$, in vacuum.}
\label{Sect:2.5.1}

In this section, we proceed to the asymptotic analysis as $\phi\rightarrow \infty$ of a scalar-field cosmology with generalized harmonic potential  $V(\phi)= \mu ^3 \left[b f \left(\cos (\delta )-\cos \left(\delta +\frac{\phi }{f}\right)\right)+\frac{\phi ^2}{\mu}\right]
$, $b\neq 0$ in vacuum, for a flat FLRW model.  We set $N=0, M=0, W_{\chi}\equiv 0$, $\Omega_m\equiv 0, \Omega_0\equiv 0, \rho_m\equiv 0, G_0 \equiv 0$ and $\chi\equiv 1$. \\
Furthermore,
 \begin{equation}
 \label{43-syst}
{W}_V(\phi)=\frac{b \mu  \sin \left(\delta +\frac{\phi }{f}\right)+2 \phi }{b f \mu \left(\cos (\delta )-\cos \left(\delta +\frac{\phi
   }{f}\right)\right)+\phi ^2}.
\end{equation}

\begin{figure}[t]
\begin{center}
	\subfigure[$(b, f, \delta, \mu)= (0.1, 0.33, 0, 0.9)$]{\includegraphics[scale=0.5]{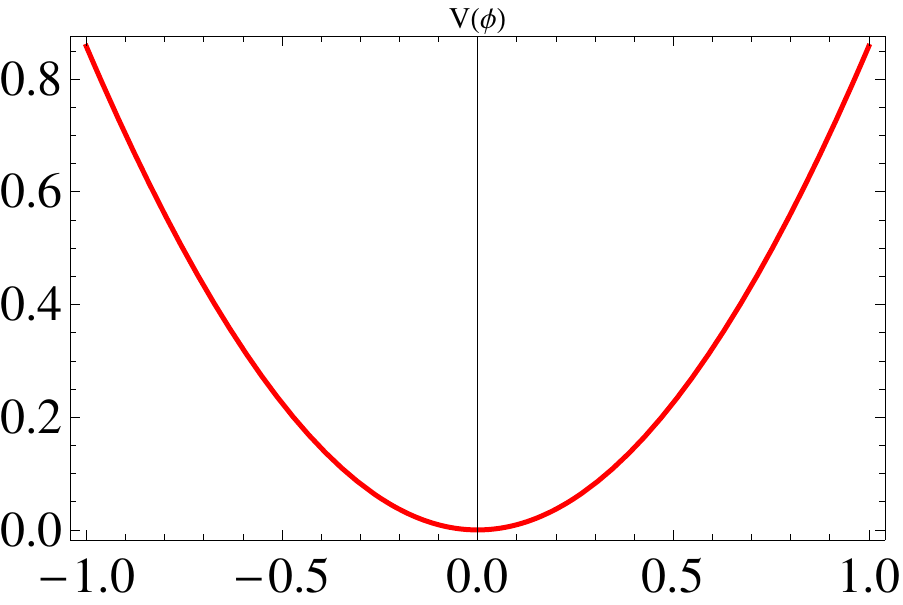} }\hspace{2cm}
	\subfigure[$(b, f, \delta, \mu)= (0.1, 0.33, 0, 0.9)$]{\includegraphics[scale=0.5]{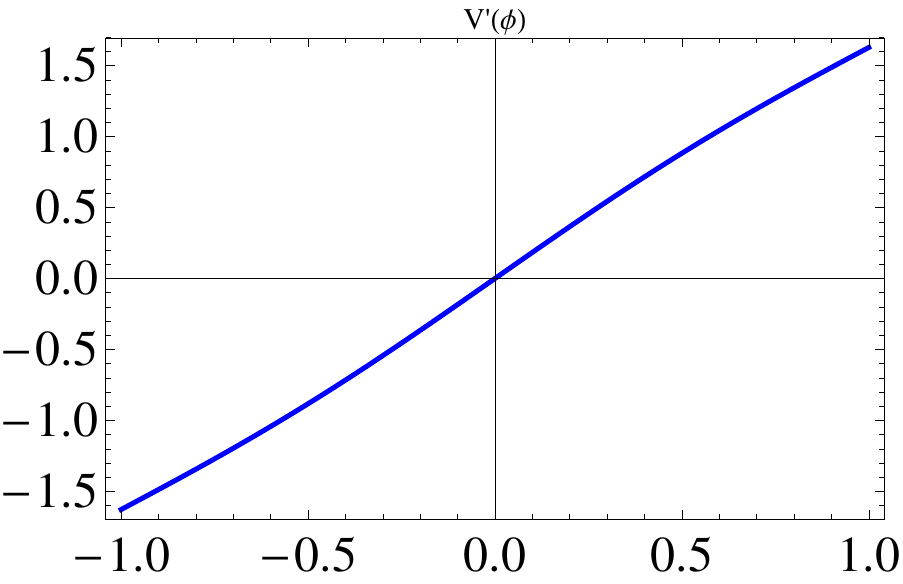} }
			\subfigure[$(b, f, \delta, \mu)= (0.99, 0.09, 0, 0.9)$]{\includegraphics[scale=0.5]{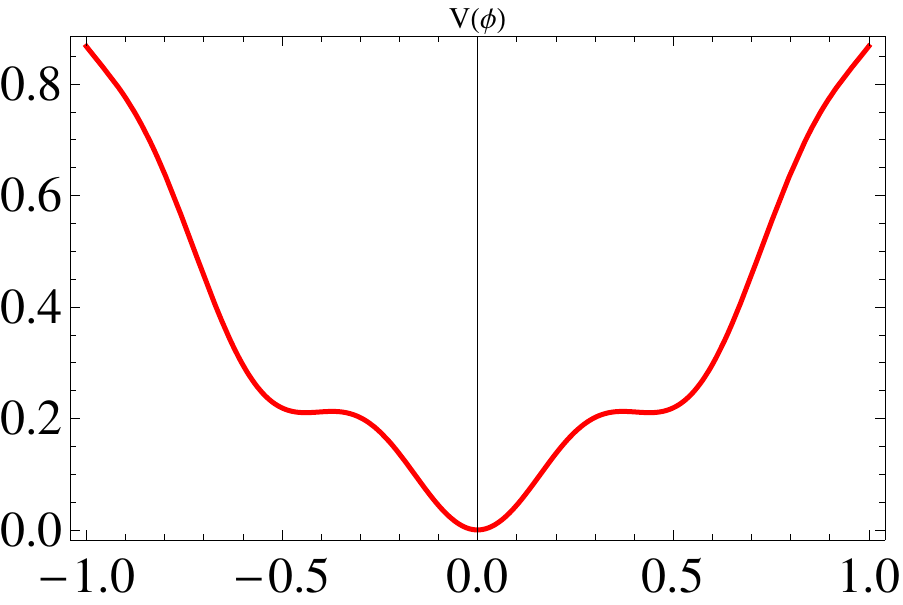}}\hspace{2cm}
	\subfigure[$(b, f, \delta, \mu)= (0.99, 0.09, 0, 0.9)$]{\includegraphics[scale=0.5]{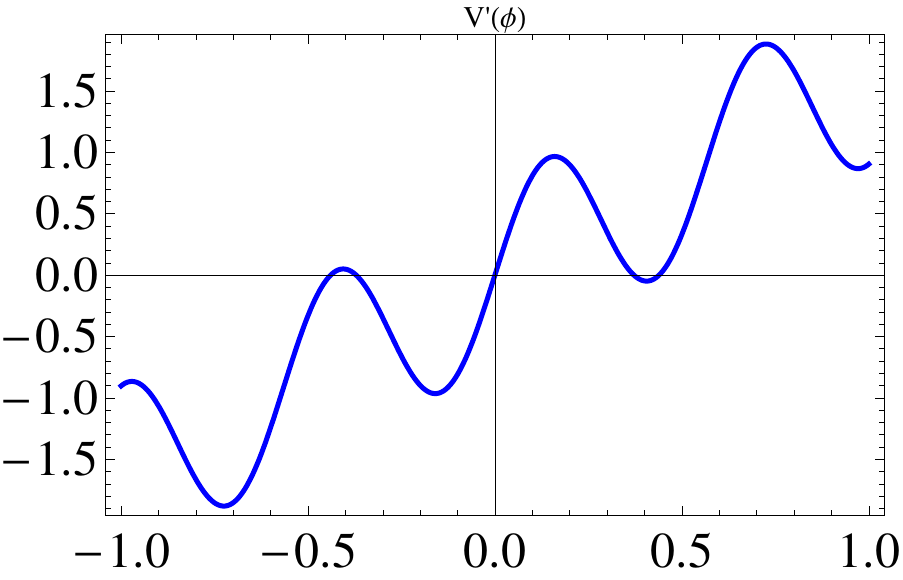} }
\end{center}\caption{\label{2PlotMonodromy-Potential} Generalized harmonic potential $
V(\phi)= \mu ^3 \left[b f \left(\cos (\delta )-\cos \left(\delta +\frac{\phi }{f}\right)\right)+\frac{\phi ^2}{\mu}\right]$ and its derivative.} 
\end{figure}

In the figure \ref{2PlotMonodromy-Potential}, the generalized harmonic potential  $V(\phi)$ and its derivative $V^{\prime}(\phi)$ for some values of the parameters $(b, f, \delta, \mu)$ is represented.

For the dynamical system analysis of the system when $|\phi|<\infty$, we remit the reader to Section  \ref{Sect.2.5}. There, we analyze the dynamics as $\phi\rightarrow \infty$. 
\\
Using again the transformation \eqref{transform34}, we deduce that $V(\phi)$ is 2 WBI with exponential order $N=0$.  
 \begin{equation}
\bar{W}_V(\varphi)=\left\{\begin{array}{cc}
\frac{b \mu  \varphi^8 \sin \left(\frac{1}{\varphi^4}\right)-2 \delta  f \varphi^8+2 f \varphi^4}{f \left(b \mu  \varphi^8 \left(\cos (\delta )-\cos
   \left(\frac{1}{\varphi^4}\right)\right)+f \left(\delta  \varphi^4-1\right)^2\right)}, & \varphi>0,\\
0, & \varphi=0 \end{array}\right.,
\end{equation}
\begin{equation}
\bar{h'}(\varphi)=\left\{\begin{array}{cc}
-\frac{\varphi^5}{4 f}, & \varphi>0,\\
0,  & \varphi=0 \end{array}\right.,
\end{equation}
that satisfy the conditions (ii) and (iii) of Definition \ref{kWBI2}.
Hence, we have the dynamical system 
\begin{subequations}
\label{Alternative1-5syst3}
\begin{align}
& \frac{d T}{d {\tau}}=3 (1-T) T \sin ^2(\theta), \label{5syst3}\\
& \frac{d \theta}{d{\tau}}=-\frac{1}{2}  \cos (\theta ) \left(6 \sin (\theta )+\sqrt{6} \bar{W}_V(\varphi)\right),\\
& \frac{d \varphi}{d {\tau}}=\sqrt{6} \bar{h'}(\varphi) \sin (\theta ),
\end{align}
\end{subequations}
where we have used the new time variable $\tau= \ln a$ defined on a phase space which consists of the vector product $\mathbf{S}\times J$ of the finite cylinder $\mathbf{S}$ with boundaries $T=0$ and $T=1$ with the interval 
$J=\left[0,  \left(\delta +\frac{\phi_0}{f}\right)^{-\frac{1}{4}}\right]$. 
The variable $T$ is suitable for global analysis \cite{Alho:2014fha}, due to 
\begin{align}
\label{5monotony}
&\frac{d T}{d \tau}\Big|_{\sin \theta=0}=0, \quad \frac{d^2 T}{d \tau^2}\Big|_{\sin \theta=0}=0, \nonumber\\
& \frac{d^3 T}{d \tau^3}\Big|_{\sin \theta=0}= \frac{9 (1-T) T v^8 \left(b \mu  v^4 \sin \left(\frac{1}{v^4}\right)+2 f \left(\delta  v^4-1\right)\right)^2}{f^2 \left(b \mu
    v^8 \cos \left(\frac{1}{v^4}\right)+f \left(\delta  v^4-1\right)^2\right)^2}. 
\end{align}
From the equation \eqref{5syst3} and equation \eqref{5monotony}, $T$ is a monotonically increasing function on $\mathbf{S}\times J$. As a consequence, all orbits are originated from the invariant subset $T=0$ (which contains the $\alpha$-limit), which is classically related to the initial singularity with $H \rightarrow \infty$, and ends on the invariant boundary subset $T=1$, which corresponds to asymptotically $H=0$. 
\\
The (curves of) equilibrium points of \eqref{Alternative1-5syst3} are the following: 
\begin{enumerate} 
\item $({T}, \theta, \varphi)= \left(T_c, 2 n \pi, 0\right)$, with eigenvalues $\{-3,0,0\}$. It is nonhyperbolic.
\item $({T}, \theta, \varphi)= \left(T_c, \pi+ 2 n \pi, 0\right)$, with eigenvalues $\{-3,0,0\}$. It is nonhyperbolic.
\item $({T}, \theta, \varphi)= \left(0, -\frac{\pi}{2}+2 n \pi, 0\right)$, with eigenvalues $\{3,3,0\}$. It is nonhyperbolic.
\item $({T}, \theta, \varphi)= \left(0, \frac{\pi}{2}+2 n \pi, 0\right)$, with eigenvalues $\{3,3,0\}$. It is nonhyperbolic.
\item $({T}, \theta, \varphi)= \left(1, -\frac{\pi}{2}+2 n \pi, 0\right)$, with eigenvalues $\{-3,3,0\}$. Behaves as saddle. 
\item $({T}, \theta, \varphi)= \left(1, \frac{\pi}{2}+2 n \pi, 0\right)$, with eigenvalues $\{-3,3,0\}$. Behaves as saddle. 
\end{enumerate}
\begin{figure}
    \centering
    \includegraphics[scale=0.7]{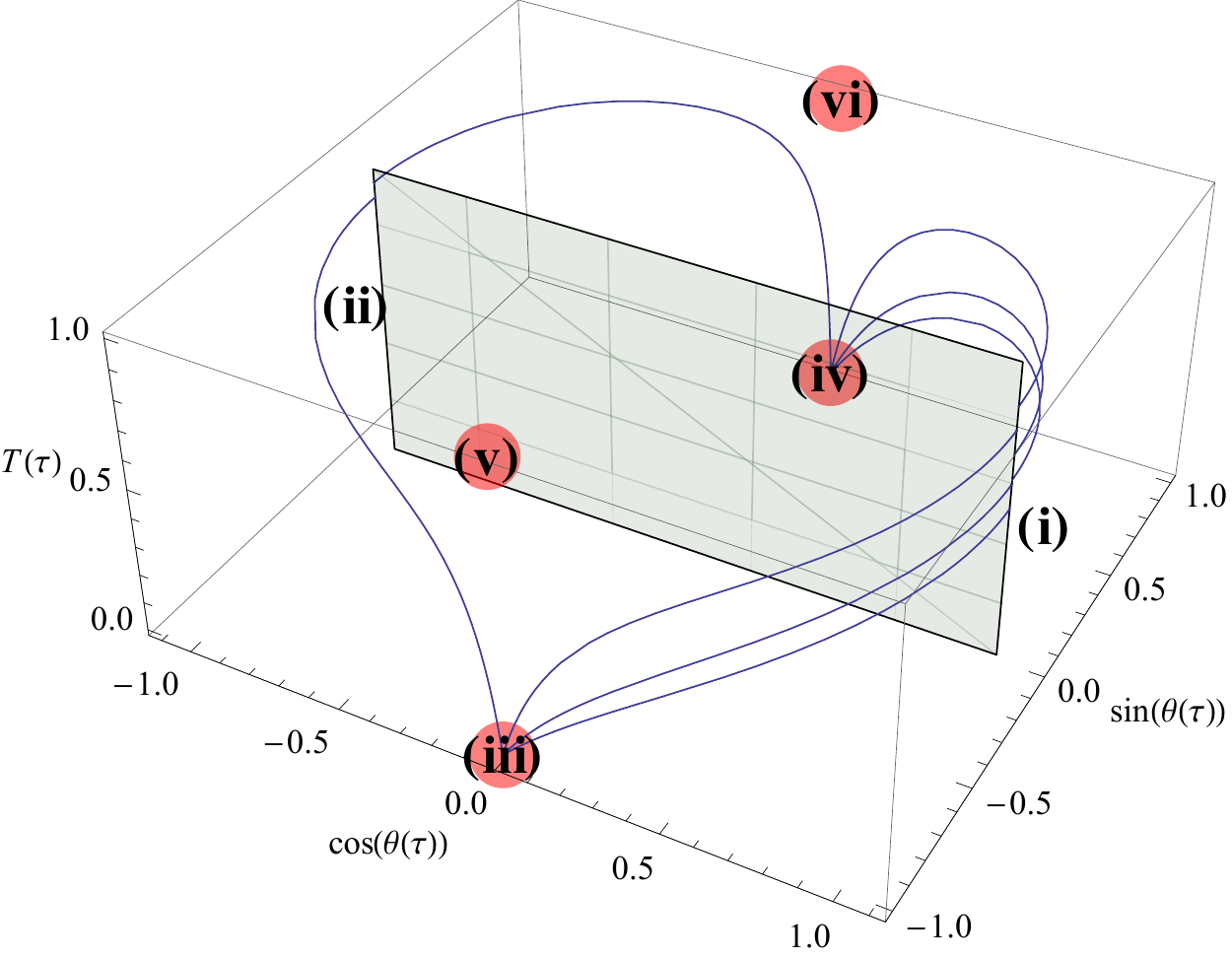}
    \caption{Evaluation of $(\cos \theta, \sin\theta, T)$ at some orbits of the system \eqref{Alternative1-5syst3} for $(b, f, \delta, \mu)= (0.1, 0.33, 0, 1)$. In the plot are represented the points (iii) and (iv), they are sources; the points (v) and (vi) are saddles. The vertical plane represents the invariant set $\{(T,\theta, \varphi):\varphi=0, \sin(\theta)=0\}$. The vertical sides (i) and (ii) of this rectangle are the local sinks. }
    \label{fig:41}
\end{figure}
In the Figure \ref{fig:41},  it is evaluated $(\cos \theta, \sin\theta, T)$ at some orbits of the system \eqref{Alternative1-5syst3} for $(b, f, \delta, \mu)= (0.1, 0.33, 0, 1)$. In the plot are represented the points (iii) and (iv), which are sources; the points (v) and (vi), which are saddles. The vertical plane represents the invariant set $\{(T,\theta, \varphi):\varphi=0, \sin(\theta)=0\}$. The vertical sides (i) and (ii) of this rectangle are the local sinks.
\section{Alternative dynamical systems formulations for a scalar-field cosmology with generalized harmonic potentials in vacuum.}
\label{SECT4}
In general, although the functions $W_{V}$ defined in \eqref{36-syst} or \eqref{43-syst} are smooth as $\phi\rightarrow +\infty$, they do not behave reasonably as the extreme points of $V$ are reached. Therefore, in this section we provide an alternative dynamical systems formulation for an scalar-field cosmology with generalized harmonic potentials in a vacuum.  The general scope of the section is to present some examples that violate one or more hypotheses of the Theorems proved in Section \ref{SECT:Non_min}, obtaining some counterexamples.  
 
\subsection{Scalar-field cosmology with generalized harmonic potential $
V(\phi)= \mu^3 \left[\frac{\phi^2}{\mu} + b f \cos\left(\delta + \frac{\phi}{f}\right)\right]$, $b\neq 0$ in vacuum.}
\label{Sect2.4}

In this section, we proceed with the qualitative analysis  of a scalar-field cosmology with a generalized harmonic potential $V(\phi)= \mu^3 \left[\frac{\phi^2}{\mu} + b f \cos\left(\delta + \frac{\phi}{f}\right)\right]$, $b\neq 0$ in a vacuum using an alternative dynamical systems formulation. 

Introducing the compact variables 
\begin{equation}
    u=\frac{\dot\phi}{\sqrt{2 \rho_c}}, \quad v= \frac{\phi}{f}, 
\end{equation}
we can rewrite the Friedmann equation as 
\begin{equation}
  \frac{3 H^2}{\rho_c}=\left[u^2+ \frac{\mu^3 f}{\rho_c}\left(\frac{f v^2}{\mu} + b \cos\left(\delta + v\right)\right)\right].
    \end{equation}
As we are interested in an expanding universe we choose the positive solution for $H$ of the previous equation. Hence, by introducing $\tau =\frac{\sqrt{2 \rho_c}}{f} t$, and
redefining constants $\rho_{c}= \frac{1}{2} b f \mu ^3, \quad k=\frac{2 f}{b \mu}$,
we obtain the equations:   
\begin{align}
&\frac{d u}{d \tau}=-\frac{\sqrt{6}}{4} b k \mu  u \sqrt{k v^2+u^2+2 \cos (\delta +v)}-k \mu  v+\sin (\delta +v)), \quad\frac{d v}{d\tau}=u.  \label{0monodronomyeqs1ab}
\end{align}

The origin $(u,v)=(0,0)$ is an equilibrium point if $\delta=0$. Then, the eigenvalues of the linearization are \\ 
$\left\{\frac{1}{4} \left(-\sqrt{k \mu  \left(3 b^2 k \mu -16\right)+16}-\sqrt{3} b k \mu \right),\frac{1}{4} \left(\sqrt{k \mu  \left(3
   b^2 k \mu -16\right)+16}-\sqrt{3} b k \mu \right)\right\}$.

	\begin{figure}[t]
\begin{center}
\subfigure[\label{monodromy-potential-1-b} $(b, f, \delta, \mu)= (0.1, 0.33, 0, 0.9)$]{\includegraphics[scale=0.4]{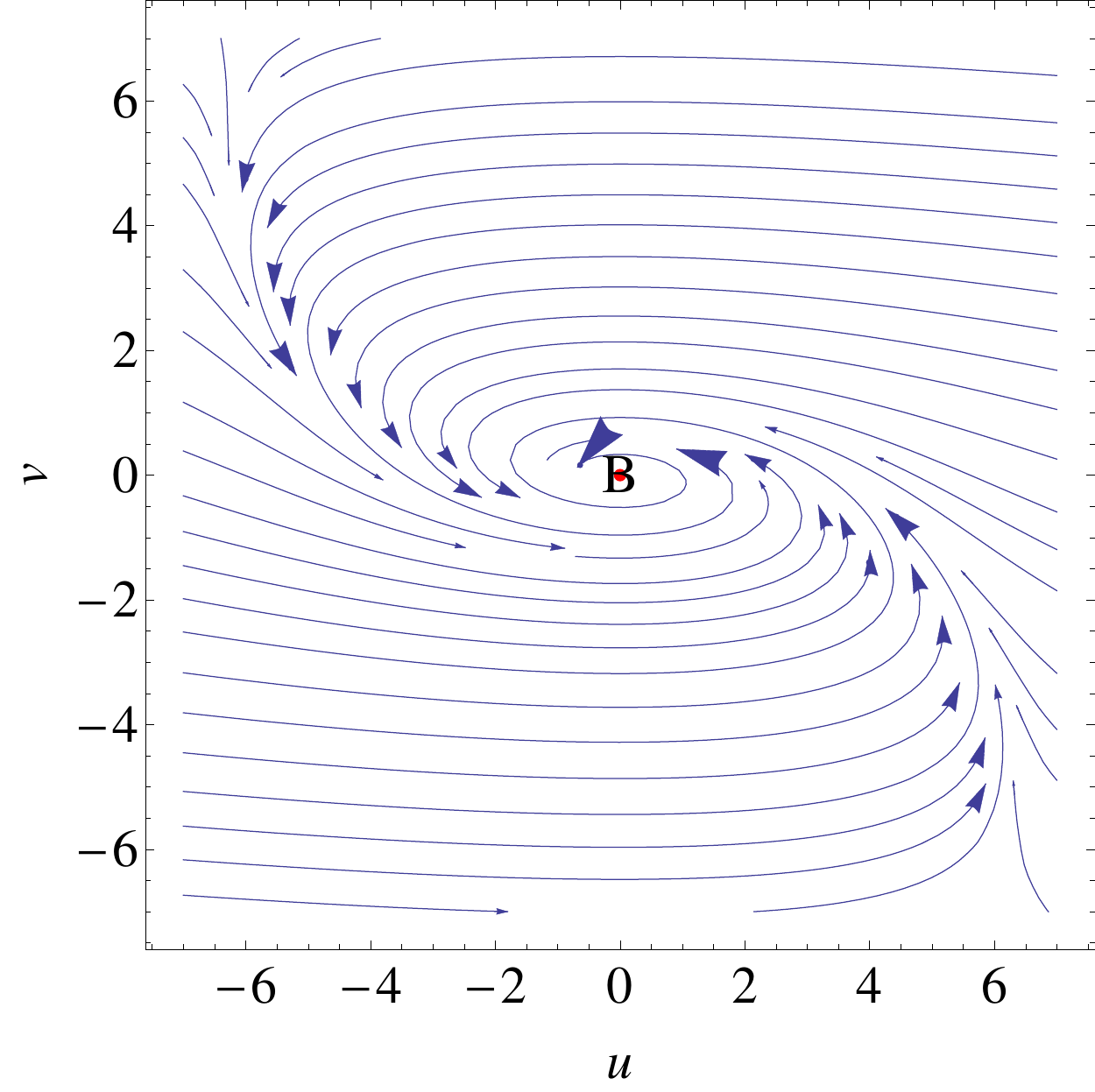}}\hspace{2cm}
\subfigure[\label{monodromy-potential-1-a}  $(b, f, \delta, \mu)= (0.99, 0.09, 0, 0.9)$]{\includegraphics[scale=0.4]{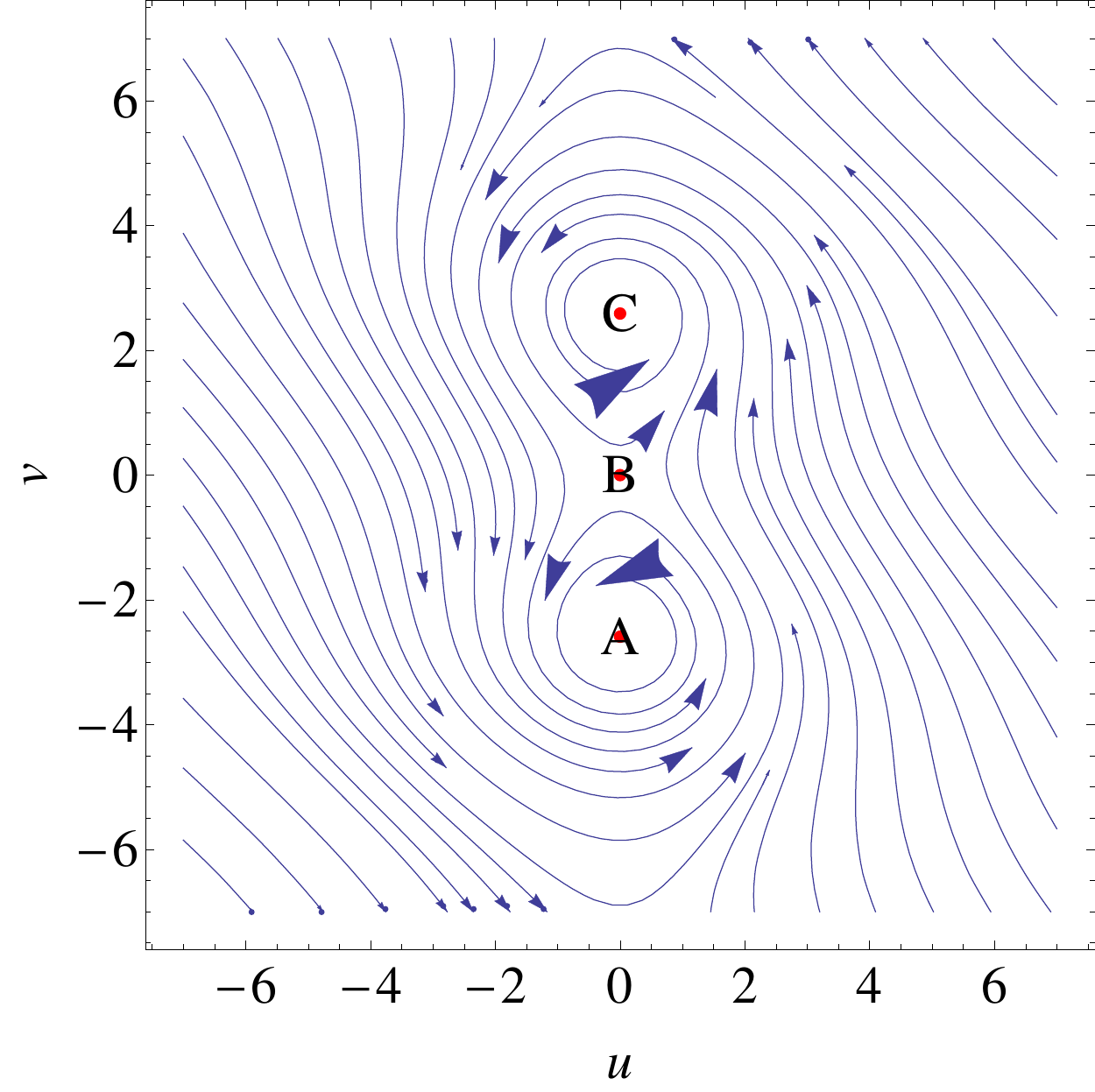}}
\caption{Phase portrait of equations    \eqref{0monodronomyeqs1ab} for some values of parameters $(b, f, \delta, \mu)$.}
\end{center}
\end{figure}
The origin is a sink for
\begin{enumerate}
    \item $\mu <0, k<\frac{1}{\mu }, b\geq \frac{4}{3} \sqrt{\frac{3 k \mu -3}{k^2 \mu ^2}}$, or 
    \item $\mu >0, k>\frac{1}{\mu }, b\geq \frac{4}{3} \sqrt{\frac{3 k \mu -3}{k^2 \mu ^2}}$. 
\end{enumerate}
It is a source if
\begin{enumerate}
\item $\mu <0, k<\frac{1}{\mu }, b\leq -\frac{4}{3} \sqrt{\frac{3 k \mu -3}{k^2 \mu ^2}}$, o
\item $\mu >0, k>\frac{1}{\mu }, b\leq -\frac{4}{3} \sqrt{\frac{3 k \mu -3}{k^2 \mu ^2}}$. 
\end{enumerate}
Finally, when $\delta=0$ and $\mu =0$, or $k \mu <1$ the origin is a saddle. 

Now, for $k\neq 0$ and $|k \mu v_c|\leq 1$, we have the equilibrium points $(u,v)=(0,v_c)$ such as  
$\sin (\delta +v)-k \mu  v=0$.  
Obtaining a real valued linearization matrix. It is additionally required that
$\frac{3 v \sin (\delta +v)}{2 \mu }+3 \cos (\delta +v)\geq 0$. 

If $\delta=0$ and $|k \mu v|>1$, there are no equilibrium points apart of the origin.  

In general, for $\delta \neq 0$ the system \eqref{0monodronomyeqs1ab} admits no equilibrium points $(u,v)=(0,v_c)$, apart from the origin, for $|k \mu v_c|>1$. 

If $|k \mu v_c|\leq 1$, we have the equilibrium points $(u,v)=(0,v_c)$ where $v_c$ are the roots of the transcendental equation $\sin (\delta +v)-k \mu  v=0$.  

For $\delta\neq 0$ and $|k \mu v_c|\leq 1$, we have
    $\delta = -v_c+ \arcsin\left(k \mu v_c\right)$, and we obtain the eigenvalues \\
$\Big\{-\frac{\sqrt{k \mu  \left(3 b^2 k \mu  \left(2 \sqrt{1-k^2 \mu ^2 v_c^2}+k v_c^2\right)-32\right)+32 \sqrt{1-k^2 \mu ^2 v_c^2}}+b k \mu 
   \sqrt{6 \sqrt{1-k^2 \mu ^2 v_c^2}+3 k v_c^2}}{4 \sqrt{2}}$,\\
	$\frac{\sqrt{k \mu  \left(3 b^2 k \mu  \left(2 \sqrt{1-k^2 \mu ^2 v_c^2}+k
   v_c^2\right)-32\right)+32 \sqrt{1-k^2 \mu ^2 v_c^2}}-b k \mu  \sqrt{6 \sqrt{1-k^2 \mu ^2 v_c^2}+3 k v_c^2}}{4 \sqrt{2}}\Big\}$. 

For the choice of parameters $(b, f, \delta, \mu)= (0.1, 0.33, 0, 0.9)$ we have 
   $\rho_c=\frac{24057}{2000000}\approx 0.0120285, k=\frac{22}{3}\approx 7.33333$. 
The only equilibrium point is the origin with eigenvalues $\{-0.285788+2.34911 i,-0.285788-2.34911 i\}$, which a stable spiral. In figure \ref{monodromy-potential-1-b} we present some orbits of the flow of \eqref{0monodronomyeqs1ab} for the choice of parameters $(b, 
   f, \delta, \mu)=(0.1, 0.33, 0, 0.9)$.
For this choice of parameters we verified the hypotheses and the results of Theorems \ref{tm} and \ref{thm2.1} ($\lim_{t\rightarrow \infty } \dot\phi=0$ and $\lim_{t\rightarrow \infty } \phi=0$).

Substituting the values $(b, 
   f, \delta, \mu)= (0.99, 0.09, 0, 0.9)$, we obtain
   $\rho_c=\frac{649539}{20000000}\approx 0.032477, k=\frac{20}{99}\approx 0.20202$. 
The transcendental equation is $\frac{2 v}{11}-\sin (v)=0$. 
Therefore, there are three equilibrium points
\begin{enumerate}
    \item $A:=(u,v)=(0, -2.64078)$. The linearization matrix is complex-valued with eigenvalues $\{ +0.997194 i,  -1.06199 i\}$. 
    \item $B:=(u,v)=(0,0)$, with eigenvalues $\{-0.985828,0.829944\}$. It is a saddle. 
       \item $C:=(u,v)=(0, 2.64078)$ The linearization matrix is complex-valued  with eigenvalues $\{0.\, +0.997194 i,0.\, -1.06199 i\}$. 
\end{enumerate}
In this case, the value of the potential has negative values at the stable equilibrium points . It is well-known that a  negative potential constant generates an equilibrium state which is just the Anti - de Sitter (AdS) equilibrium solution. 

In figure \ref{monodromy-potential-1-a}, we present some orbits of the flow of \eqref{0monodronomyeqs1ab} for  $(b, f, \delta, \mu)= (0.99, 0.09, 0, 0.9)$. For these choices of parameters the hypotheses 
{\it{$V(\phi)\geq 0$ and $V(\phi)=0$, if and only if $\phi=0$}} of Theorem \ref{tm} are violated, but the result $\lim_{t\rightarrow +\infty} \dot \phi =0$ holds. The hypotheses {\it{$V(\phi)\geq 0$ and $V(\phi)=0$, if and only if $\phi=0$}} and {\it{$V^{\prime}(\phi)<0$ for $\phi<0$ and $V^{\prime}(\phi)>0$ for $\phi>0$}} of Theorem \ref{thm2.1} are violated, and $\lim_{t\rightarrow +\infty}\phi$ can be finite (rather than zero or infinity). The recall of this Theorem relies on the last hypothesis. Finally, when the hypotheses {\it{$V(\phi)\geq 0$ and $V^{\prime}(\phi)<0\quad \forall \phi\in\mathbb{R}$}} of \ref{thm2.2} are violated, and $\lim_{t\rightarrow +\infty}\dot\phi=0, \lim_{t\rightarrow +\infty}\phi<\infty$.

\subsubsection{Oscillating regime.}
\label{Sect:2.4.2}
In the reference \cite{Rendall:2006cq}, oscillating scalar field models with potential $\frac{1}{2}\phi^2$ and potentials $\frac{1}{2}\phi^2+W(\phi)$ with $W$ smooth and $W(\phi)=o(\phi^3)$ were studied.
There were derived improved asymptotic expansions for the solution in homogeneous and isotropic spaces. Various generalizations were obtained for non-linear massive scalar fields, $k$- essence models and $f(R)$-gravity. 
In this section we investigate the potential $
V(\phi)= \mu^3 \left[\frac{\phi^2}{\mu} + b f \cos\left(\delta + \frac{\phi}{f}\right)\right]$, $b\neq 0$ looking for oscillatory behavior, as expected from the numerical investigations. We derive asymptotic expansions as well. Observing that the cosine corrections are $O(b f \mu^3)$, they therefore do not fall in the potential class studied by  \cite{Rendall:2006cq}. 

The pair
\begin{equation}
\left(\frac{\sqrt{2 \mu}\phi}{\sqrt{{\dot \phi}^2+2 \mu \phi^2}}, \quad \frac{\dot{\phi}}{\sqrt{{\dot{\phi}}^2+2 \mu \phi^2}}\right)
\end{equation}
define
a function of $t$ with values in the unit circle. Therefore, we define the angular function 
$\vartheta (t)$ that is unique under identification module $2\pi$, defined by 
\begin{equation}
\vartheta=  \tan ^{-1} \left(\frac{\dot \phi}{\sqrt{2 }\mu \phi}\right),
\end{equation}
together with
\begin{equation}
r=\sqrt{{\dot \phi}^2+2 \mu^2 \phi^2},
\end{equation} 
with inverse
\begin{equation}
\phi = \frac{r \cos (\vartheta )}{\sqrt{2} \mu },\quad \dot\phi= r \sin (\vartheta ). 
\end{equation}
They satisfy
\begin{equation}
-b f \mu ^3 \cos \left(\delta +\frac{r \cos (\theta )}{\sqrt{2} f \mu }\right)+3 H^2-\frac{r^2}{2}=0.
\end{equation}
For expanding universes ($H>0$), we obtain the equations: 
\begin{subequations}
\begin{align}
& \dot r=-\sqrt{\frac{3}{2}} r \sin ^2(\vartheta) \sqrt{2 b f \mu ^3 \cos \left(\delta +\frac{r \cos (\vartheta)}{\sqrt{2} f \mu
   }\right)+r^2} +  b \mu ^3 \sin (\vartheta) \sin \left(\delta +\frac{r \cos (\vartheta)}{\sqrt{2} f \mu }\right),\\
& \dot \vartheta=-\sqrt{2} \mu -\sqrt{\frac{3}{2}} \sin (\vartheta) \cos (\vartheta) \sqrt{2 b f \mu ^3 \cos \left(\delta +\frac{r \cos (\vartheta)}{\sqrt{2} f \mu }\right)+r^2} \nonumber \\
& +\frac{b \mu ^3 \cos (\theta ) \sin \left(\delta +\frac{r \cos (\theta
   )}{\sqrt{2} f \mu }\right)}{r}.
\end{align}
\end{subequations}
Observing that for $b\rightarrow 0$, we obtain the equations 
\begin{align}
&\dot r=-\sqrt{\frac{3}{2}} r^2 \sin ^2(\vartheta ), \quad
 \dot \vartheta=-\sqrt{2} \mu -\sqrt{\frac{3}{2}} r \sin (\vartheta ) \cos (\vartheta ).
\end{align}
The solutions of the limiting equation admit the asymptotic expansions \cite{Rendall:2006cq}
\begin{equation}
\vartheta(t)= -\sqrt{2} \mu  t + O(\ln t), \quad r(t)= \frac{4}{\sqrt{6} t}+ O(t^{-2}\ln t). 
\end{equation}
Hence, 
\begin{equation}
\phi(t)=\frac{4 \cos t}{\sqrt{6} t}+ O(t^{-2}\ln t), \quad \dot{\phi}(t)=\frac{4 \sin t}{\sqrt{6} t}+ O(t^{-2}\ln t).
\end{equation}
These expansions can be improved to the order $O(t^{-3}\ln t)$ as in \cite{Rendall:2006cq}. 
Instead, in order to obtain more accuracy in the asymptotic solution of the limiting problem as $b \rightarrow 0$, we use a similar argument as in \cite{Rendall:2006cq} to derive asymptotic expansions of the full problem ($b\neq 0$).
\\
Note that 
\begin{subequations}
\label{syst71}
\begin{align}
& \dot r= b \mu ^3 \sin (\delta ) \sin (\theta )+\left(\frac{b \mu ^2 \cos (\delta ) \cos (\theta ) \sin (\theta )}{\sqrt{2} f}-\sqrt{3} \sqrt{b f \mu ^3 \cos (\delta )} \sin ^2(\theta )\right) r+O\left(r^2\right),\\
& \dot \vartheta +\sqrt{2} \mu = \frac{b \mu ^3 \sin (\delta ) \cos (\vartheta )}{r}+\left(\frac{b \mu ^2 \cos (\delta ) \cos ^2(\vartheta )}{\sqrt{2} f}-\sqrt{3} \sin (\vartheta ) \cos (\vartheta ) \sqrt{b f \mu ^3 \cos (\delta )}\right)\nonumber \\
& \frac{r \left(\sqrt{6} f \tan (\delta ) \sin (\theta ) \cos ^2(\theta ) \sqrt{b f \mu ^3 \cos (\delta )}-b \mu ^2 \sin (\delta ) \cos ^3(\theta )\right)}{4 f^2 \mu } +O\left(r^2\right).
\end{align}
\end{subequations}

\begin{figure}
\begin{center}
\subfigure[]{\includegraphics[scale=0.5]{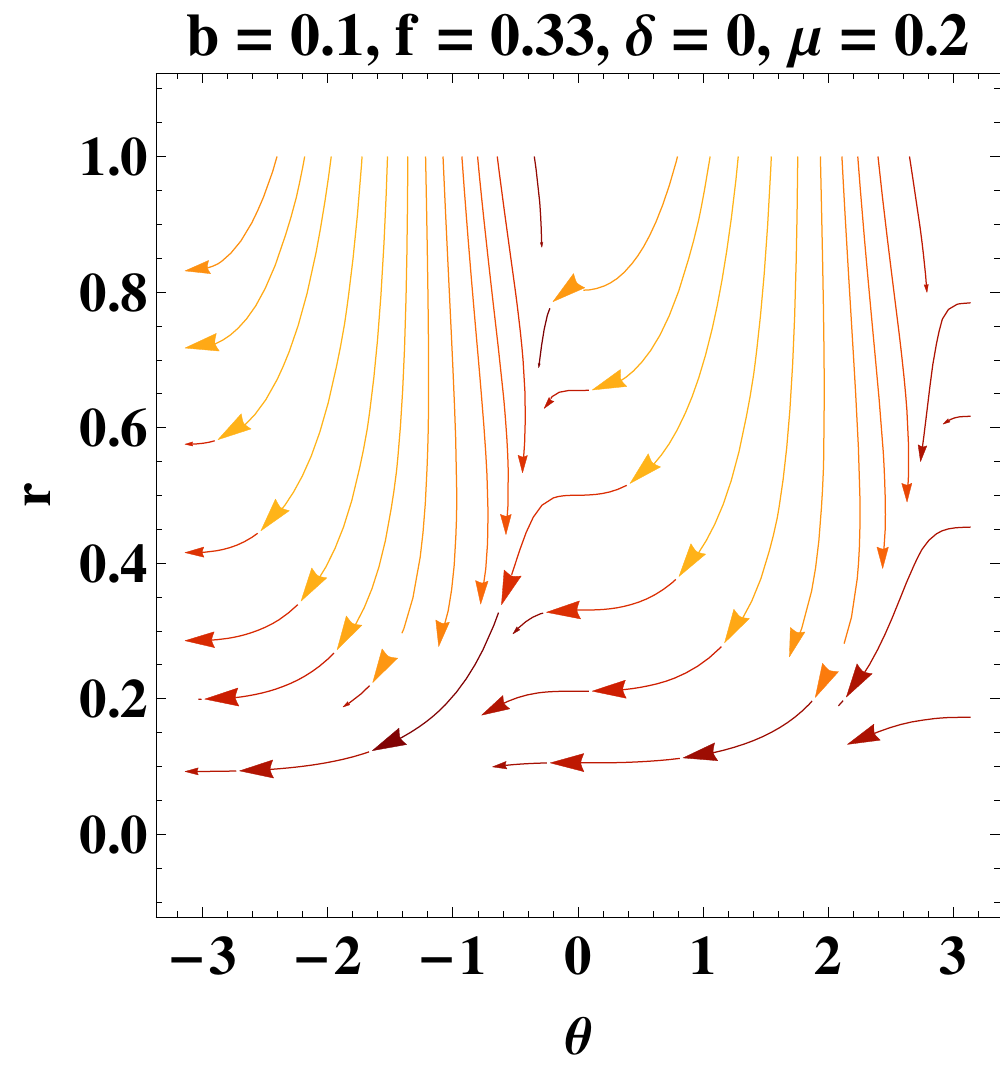} }\hspace{2cm}
\subfigure[]{\includegraphics[scale=0.5]{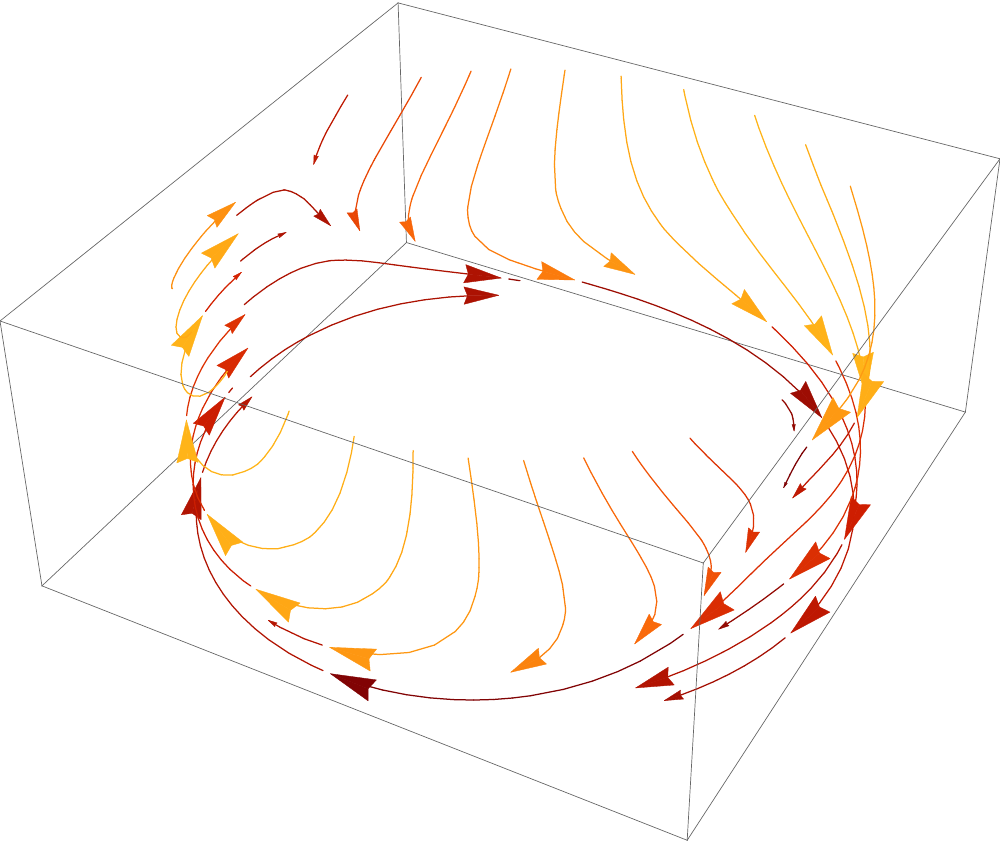}}
\subfigure[]{\includegraphics[scale=0.5]{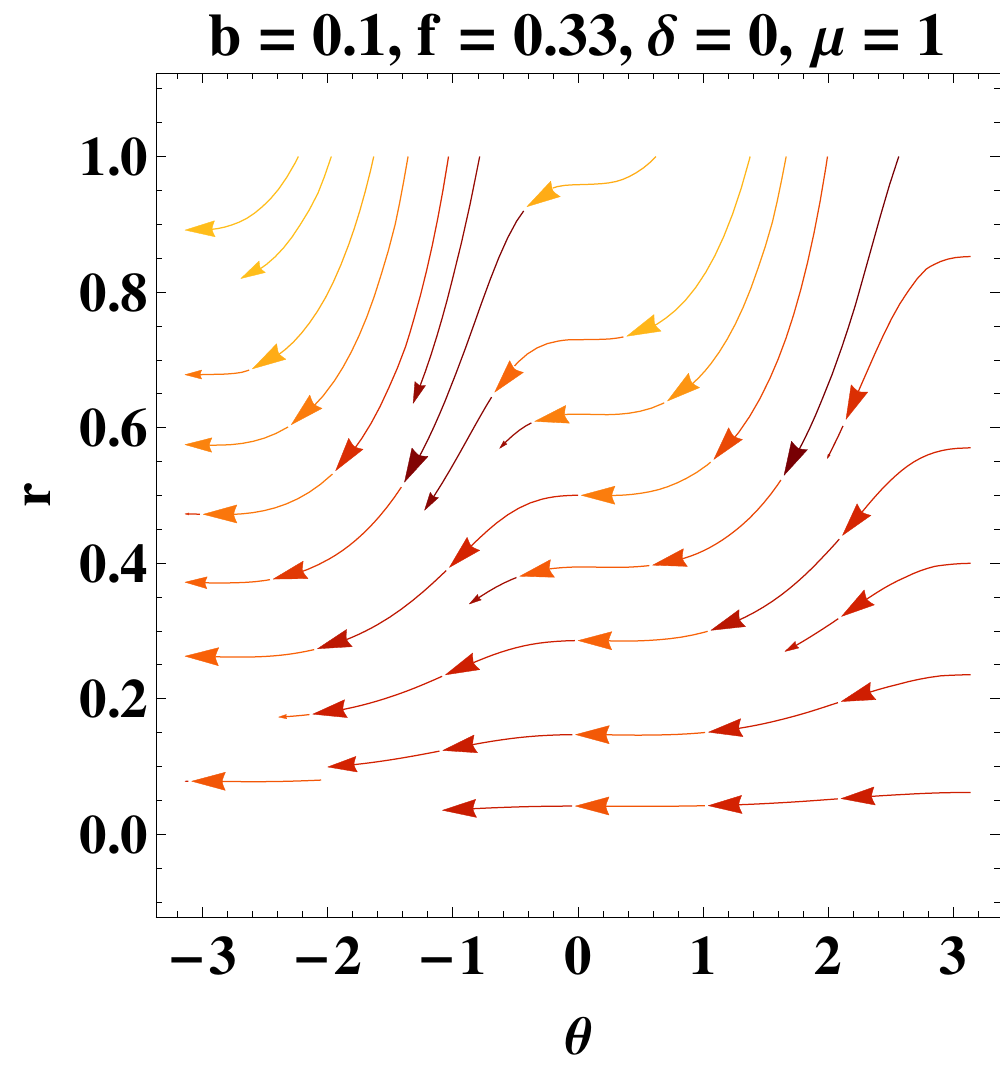} }\hspace{2cm}
\subfigure[]{\includegraphics[scale=0.5]{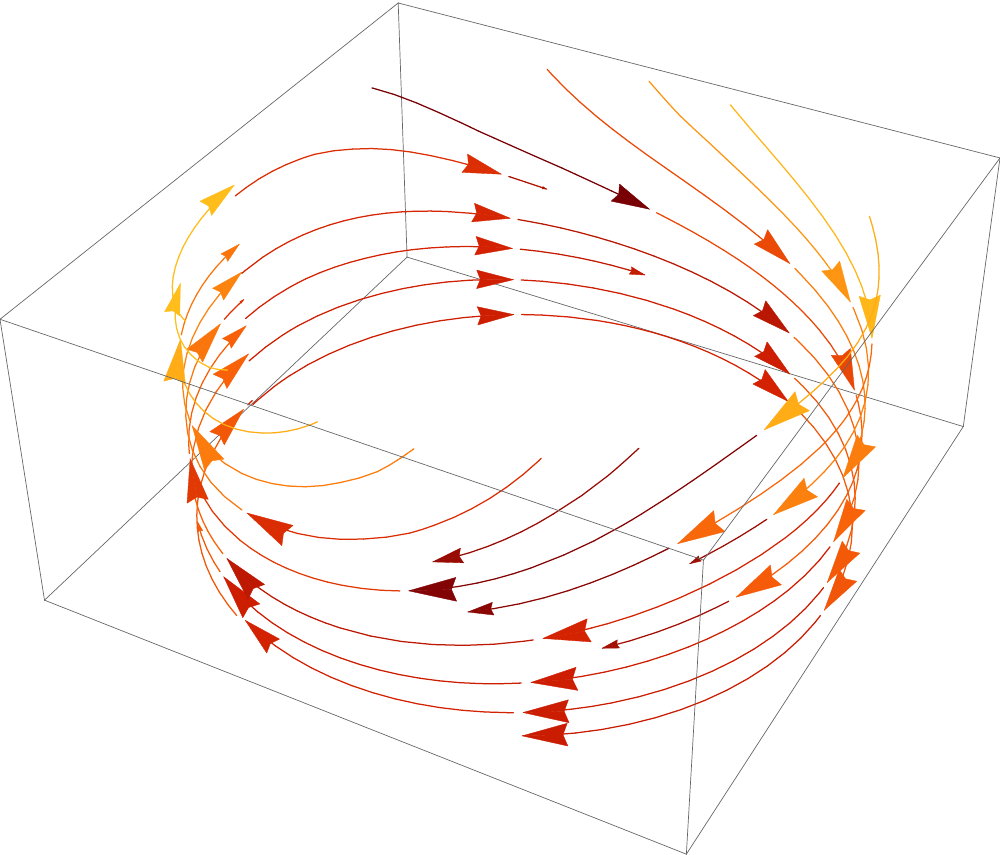}}
\end{center}\caption{\label{Fig6} Phase portrait of equations   \eqref{syst71} (left panel). Projection over the cylinder $\mathbf{S}$ (right panel) for $(b, f, \delta)= (0.1, 0.33, 0)$ and different values of $\mu$.}
\end{figure}

\begin{figure}
\begin{center}
\subfigure[]{\includegraphics[scale=0.5]{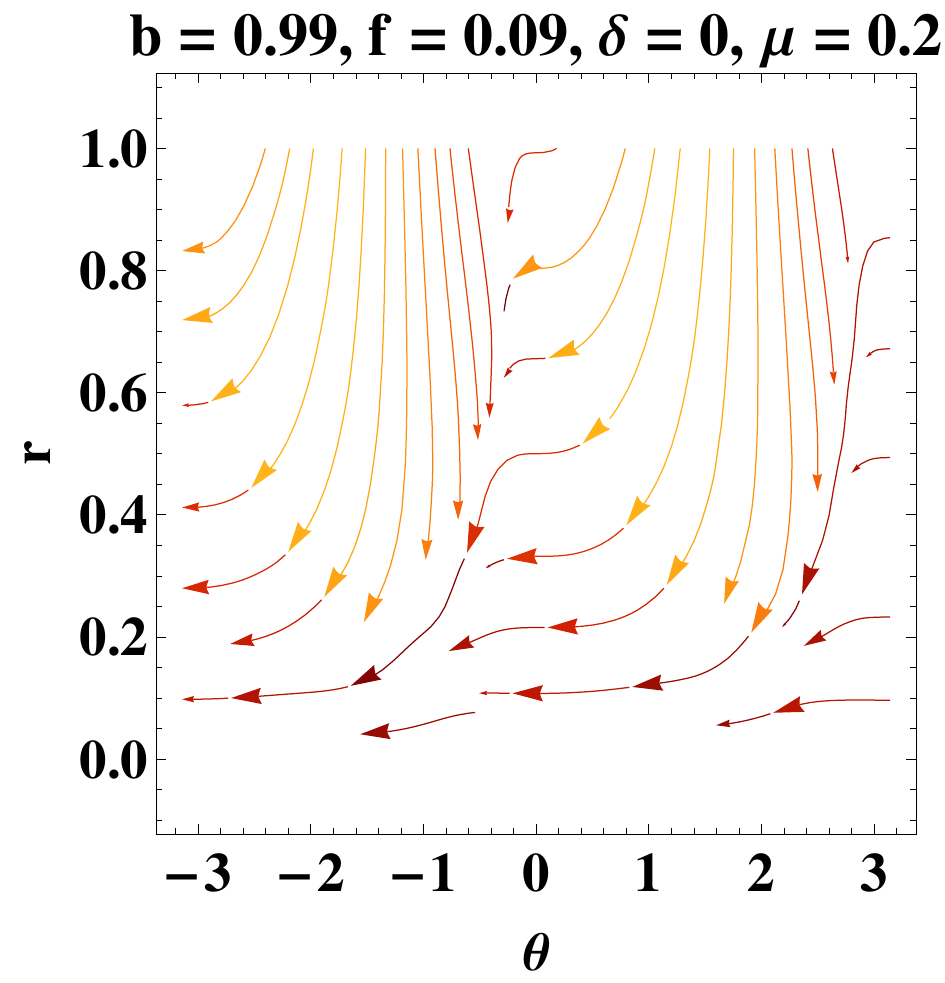} }\hspace{2cm}
\subfigure[]{\includegraphics[scale=0.5]{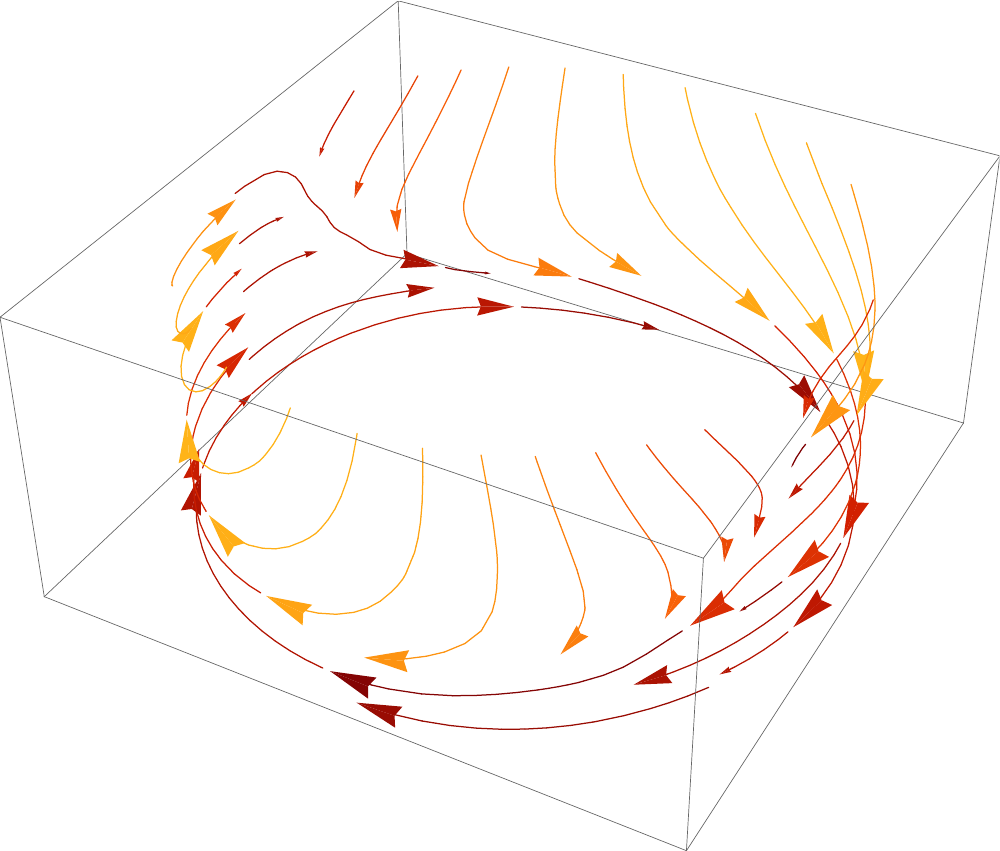}}
\subfigure[]{\includegraphics[scale=0.5]{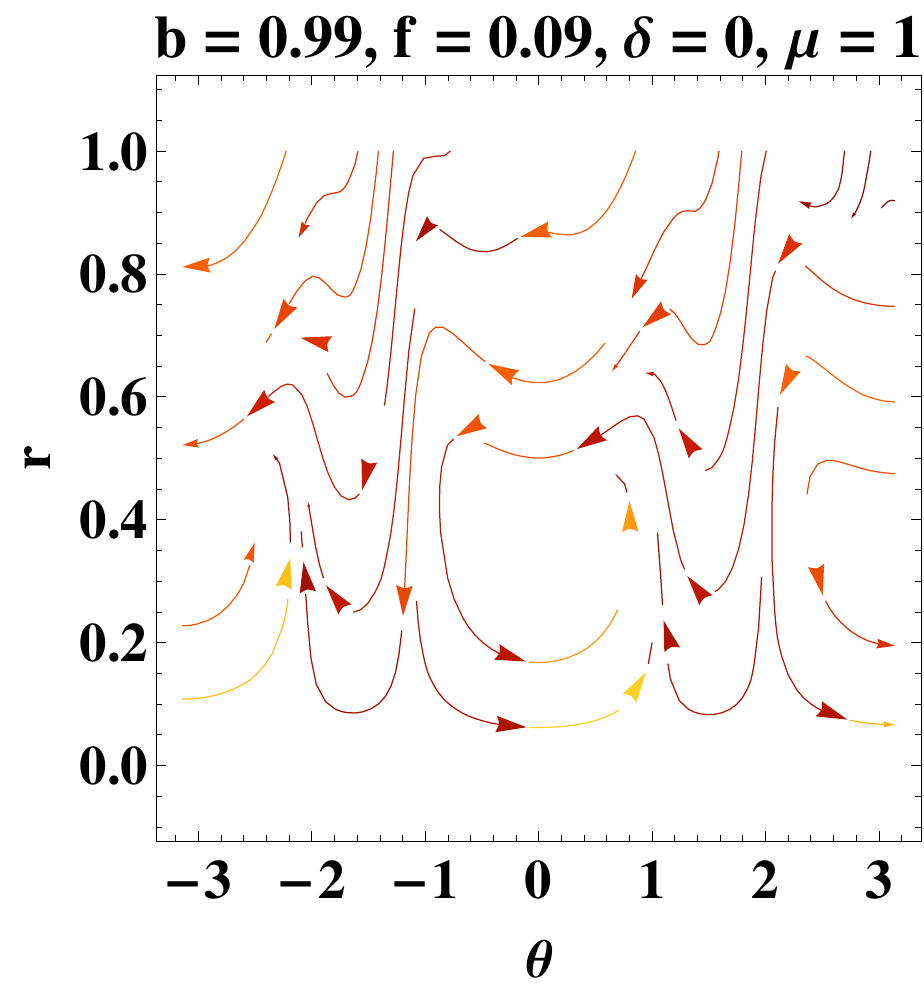} }\hspace{2cm}
\subfigure[]{\includegraphics[scale=0.5]{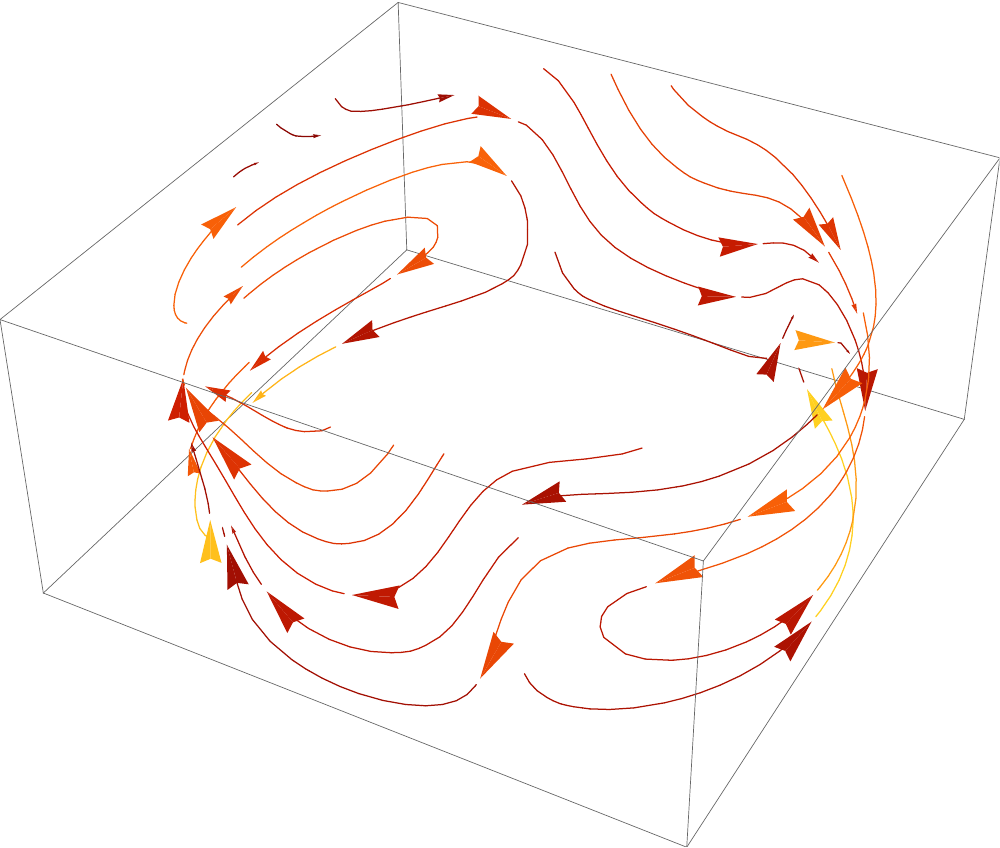}}
\end{center}\caption{\label{Fig7} Phase portrait of equations   \eqref{syst71} (left panel). Projection over the cylinder $\mathbf{S}$ (right panel) for $(b, f, \delta)= (0.99, 0.09, 0)$ and different values of $\mu$.}
\end{figure}

Obtaining an approximated solution near the oscillatory regime we can take the average with respect to $\vartheta$ over any orbit of period $2 \pi$, given by 
\begin{equation}
\langle f \rangle = \frac{1}{2\pi}\int_{c}^{c +2 \pi} f(\vartheta) d \vartheta, \quad \cos(c)\geq 0,
\end{equation}
to $f=(\dot r, \dot \vartheta)$, leading to 
\begin{equation}
\dot r= -\frac{1}{2} \sqrt{3} r \sqrt{b f \mu ^3 \cos (\delta )}, \quad \dot \vartheta=\frac{b \mu ^2 \cos (\delta )}{2 \sqrt{2} f}-\sqrt{2} \mu. 
\end{equation}
The averaged equations have solutions 
\begin{equation}
r(t)= r_0 e^{-\frac{1}{2} \sqrt{3} t \sqrt{b f \mu ^3 \cos (\delta )}}, \quad \vartheta (t)= \left(\frac{b \mu ^2 \cos (\delta )}{2 \sqrt{2} f}-\sqrt{2} \mu\right)t+\theta_0. 
\end{equation}

Introducing along with $\vartheta$, and $r$, the new variable 
\begin{equation}
\varepsilon=\frac{H}{\mu+H}, 
\end{equation} 
with inverse
\begin{equation}
 H= \frac{\mu  \varepsilon}{1-\varepsilon},
\end{equation}
satisfying
\begin{equation}
-b f \mu ^3 (\varepsilon -1)^2 \cos \left(\delta +\frac{r \cos (\vartheta )}{\sqrt{2} f \mu }\right)-\frac{1}{2} r^2 (\varepsilon -1)^2+3 \mu ^2 \varepsilon ^2=0,
\end{equation}
along with the time derivative  $\hat{\tau}$ given by 
\begin{equation}
\frac{d \hat{\tau}}{d t}:= \mu+ H,
\end{equation}
we obtain the equations 
\begin{subequations}
\begin{align}
& \varepsilon'=-\frac{r^2 (1-\varepsilon)^3 \sin ^2(\vartheta )}{2 \mu ^2},\\
& r'=-3 r \varepsilon  \sin ^2(\vartheta ) +b \mu ^2 (1-\varepsilon) \sin (\vartheta ) \sin \left(\delta +\frac{r \cos (\vartheta )}{\sqrt{2} f \mu }\right)\\
&\vartheta'=-\sqrt{2} (1-\varepsilon)-3 \varepsilon  \sin (\vartheta ) \cos (\vartheta )+\frac{b \mu ^2 (1-\varepsilon )
   \cos (\vartheta ) \sin \left(\delta +\frac{r \cos (\vartheta )}{\sqrt{2} f \mu }\right)}{r}.
\end{align}
\end{subequations}
Obtaining an approximated solution near the oscillatory regime we take the average with respect to $\vartheta$ over any orbit of period $2 \pi$, 
$\langle (\varepsilon', r', \vartheta') \rangle$, leading to
\begin{subequations}
\begin{align}
& \varepsilon'=\frac{r^2 (\varepsilon -1)^3}{4 \mu ^2},\\
& r'=-\frac{3 r \epsilon }{2},\\
& \vartheta'= \sqrt{2} (\varepsilon -1)-\Big\langle \frac{b \mu ^2 (\varepsilon -1) \cos (\theta ) \sin \left(\delta +\frac{r \cos (\theta )}{\sqrt{2} f \mu
   }\right)}{r}\Big\rangle. 
\end{align}
\end{subequations}
But, as $r\rightarrow 0$,
\begin{align}
& \frac{1}{2 \pi }\int_c^{c+2 \pi } \frac{b \mu ^2 (\varepsilon -1) \cos (\vartheta ) \sin \left(\delta +\frac{r \cos (\vartheta )}{\sqrt{2} f \mu
   }\right)}{r} \, d\theta\nonumber \\
	&  \sim \frac{b \mu  (\varepsilon -1) \cos (\delta )}{2 \sqrt{2} f}-\frac{r^2 (b (\varepsilon -1) \cos (\delta ))}{32 \left(\sqrt{2} f^3 \mu \right)}+O\left(r^3\right).
\end{align}
Finally, we have 
\begin{subequations}
\begin{align}
& \varepsilon'=-\frac{r^2 (1-\varepsilon)^3}{4 \mu ^2},\\
& r'=-\frac{3 r \varepsilon }{2},\\
& \vartheta'= \left[-\sqrt{2} + \frac{b \mu \cos (\delta )}{2 \sqrt{2} f}-\frac{r^2 (b \cos (\delta ))}{32 \left(\sqrt{2} f^3 \mu \right)}\right](1-\varepsilon), 
\end{align}
\end{subequations}
with the averaged constraint
\begin{equation}
\mu ^2 \left(3 \varepsilon ^2-b f \mu  (1-\varepsilon)^2 \cos (\delta )\right)+\frac{r^2 (1-\varepsilon)^2 (b \mu  \cos (\delta )-4 f)}{8 f}=0,
\end{equation}
as $r\rightarrow 0$. 
\\
The above system is integrable yielding
\begin{subequations}
\begin{align}
& r(\varepsilon )=\frac{\sqrt{2} \sqrt{c_1 (\varepsilon -1)^2+\mu ^2 (6 \varepsilon -3)}}{1-\varepsilon },\\
& \vartheta (\varepsilon )=\frac{\mu  \tanh ^{-1}\left(\frac{c_1 (\varepsilon -1)+3 \mu ^2}{\mu  \sqrt{9 \mu ^2-3 c_1}}\right) (b \mu 
   \cos (\delta )-4 f)}{f \sqrt{18 \mu ^2-6 c_1}}+\frac{b \mu  \cos (\delta )}{8 \sqrt{2} f^3 (1-\varepsilon )}+c_2, 
\end{align}
and 
\begin{align}
& 3(t-t_0)= \ln \left(\frac{(1-\varepsilon )^2 \left(\frac{3 \mu  \sqrt{3 \mu ^2-c_1}+\sqrt{3} c_1 \varepsilon -\sqrt{3} c_1+3 \sqrt{3} \mu ^2}{3 \mu  \sqrt{3 \mu ^2-c_1}-\sqrt{3} c_1 \varepsilon +\sqrt{3} c_1-3 \sqrt{3} \mu ^2}\right){}^{\frac{\mu
   }{\sqrt{\mu ^2-\frac{c_1}{3}}}}}{c_1 (\varepsilon -1)^2+\mu ^2 (6 \varepsilon -3)}\right) \nonumber \\
	& \sim \ln \left(\frac{\left(\frac{2 \mu  \left(\sqrt{9 \mu ^2-3 c_1}+3 \mu \right)-c_1}{c_1}\right){}^{\frac{\mu }{\sqrt{\mu ^2-\frac{c_1}{3}}}}}{c_1-3 \mu ^2}\right)-\frac{6 \mu ^2 \varepsilon }{c_1-3 \mu ^2}+O\left(\varepsilon ^2\right),
\end{align}
\end{subequations}
as $\varepsilon\rightarrow 0$. 

In the figure \ref{Fig6}, we present the phase portrait of equations   \eqref{syst71} (left panel) and the projection over the cylinder $\mathbf{S}$ (right panel) for $(b, f, \delta)= (0.1, 0.33, 0)$ and different values of $\mu$. Figure \ref{Fig7} presents the phase portrait of equations  \eqref{syst71} (left panel) and the projection over the cylinder $\mathbf{S}$ (right panel) for $(b, f, \delta)= (0.99, 0.09, 0)$ and different values of $\mu$. The plots show the oscillatory behavior of the solutions. 

\subsection{Scalar-field cosmology with generalized harmonic potential $
V(\phi)= \mu ^3 \left[b f \left(\cos (\delta )-\cos \left(\delta +\frac{\phi }{f}\right)\right)+\frac{\phi ^2}{\mu}\right]
$, $b\neq 0$, in vacuum.}
\label{Sect.2.5}
In this section, we proceed with the qualitative analysis study of a scalar-field cosmology with generalized harmonic potential $V(\phi)= \mu ^3 \left[b f \left(\cos (\delta )-\cos \left(\delta +\frac{\phi }{f}\right)\right)+\frac{\phi ^2}{\mu}\right]
$, $b\neq 0$ in a vacuum, using the following alternative formulation. 

We introduce the compact variables 
\begin{equation}
    u=\frac{\dot\phi}{\sqrt{2 \rho_c}}, \quad v= \frac{\phi}{f}, 
\end{equation}
satisfying
    \begin{equation}
        3 H^2=\mu ^3 \left(\frac{f^2 v^2}{\mu }-b f \cos (\delta +v)\right)+b f \mu ^3 \cos (\delta )+\rho_c u^2.
    \end{equation}
As we are interested in an expanding universe we choose the positive solution for $H$ of the previous equation. Hence, by introducing $\tau =\frac{\sqrt{2 \rho_c}}{f} t$, and
redefining constants $\rho_{c}= \frac{1}{2} b f \mu ^3, \quad k=\frac{2 f}{b \mu}$, we obtain the equations   
\begin{align}
\label{monodronomyeqs1ab} 
&\frac{d u}{d \tau}=-\frac{\sqrt{6}}{4} b k \mu  u \sqrt{2 \cos (\delta )+k v^2+u^2-2 \cos (\delta +v)}-k v-\sin (\delta +v), \quad \frac{d v}{d\tau}=u.  
\end{align}

	\begin{figure}[t]
\begin{center}
\subfigure[\label{2monodromy-potential-1-b} $(b, f, \delta, \mu)= (0.1, 0.33, 0, 0.9)$]{\includegraphics[scale=0.48]{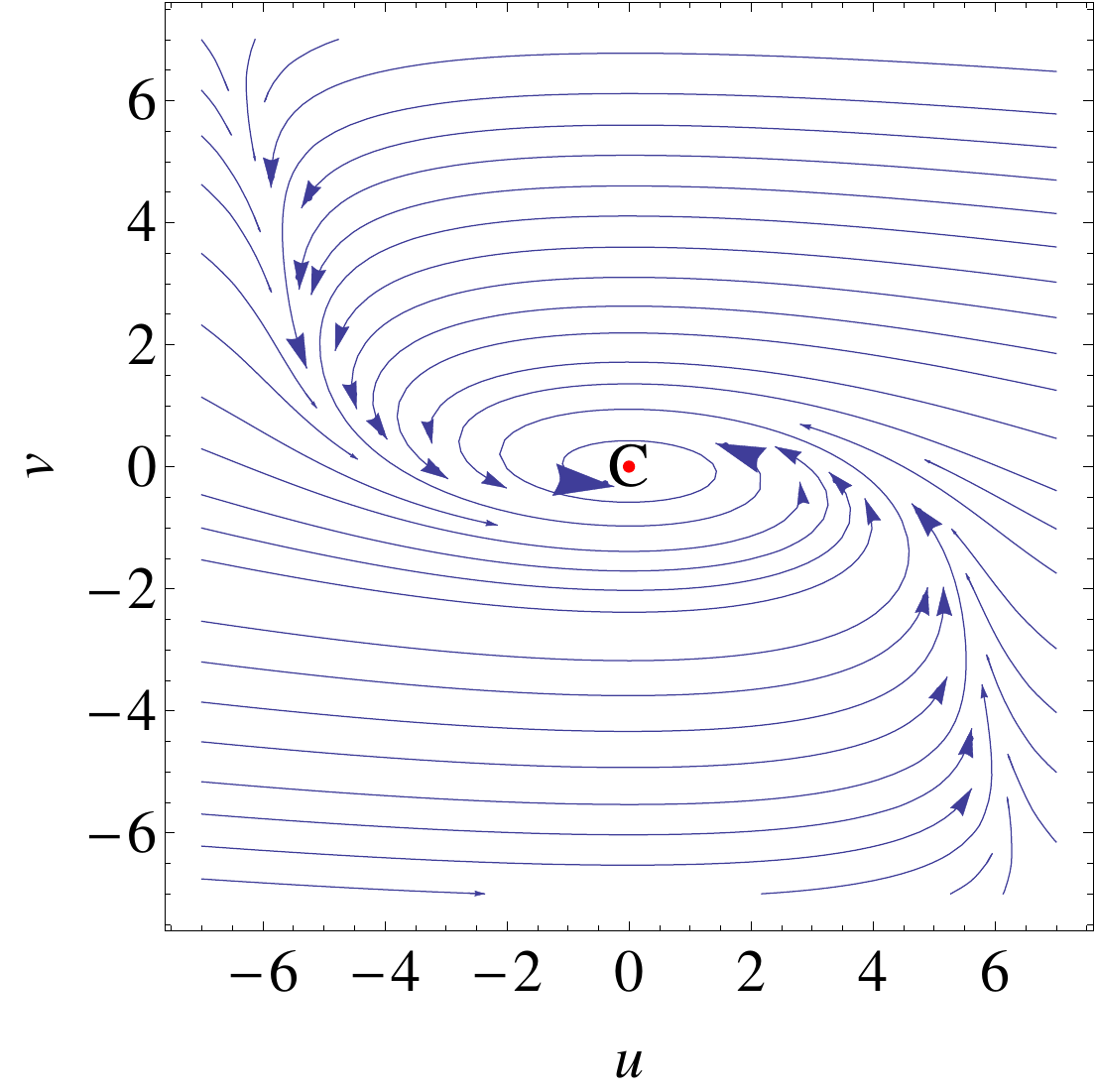}}\hspace{2cm}
\subfigure[\label{2monodromy-potential-1-a}  $(b, f, \delta, \mu)= (0.99, 0.09, 0, 0.9)$]{\includegraphics[scale=0.4]{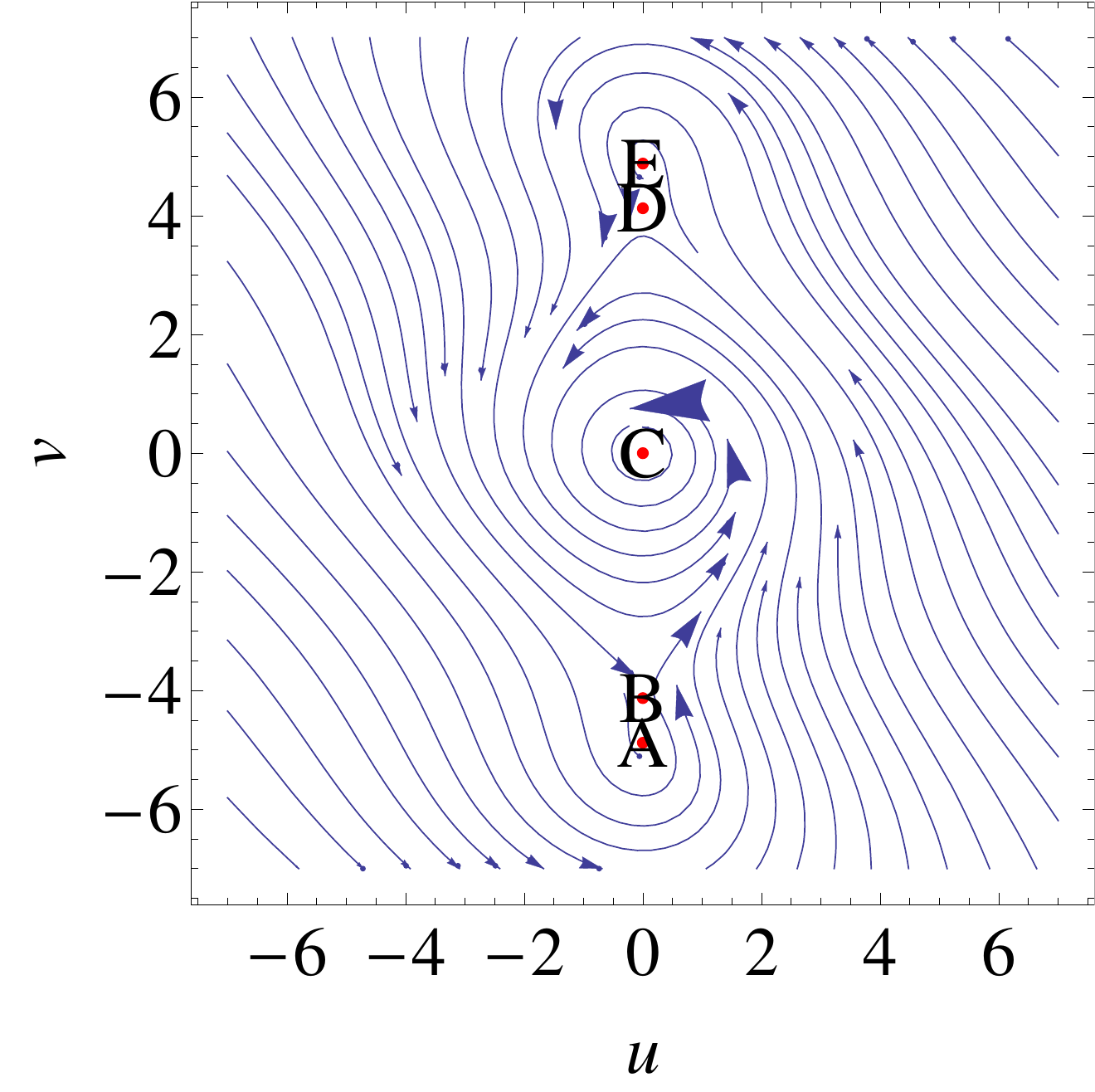}}
\caption{Phase portrait of equations    \eqref{monodronomyeqs1ab} for some choices of parameters $(b, f, \delta, \mu)$.}
\end{center}
\end{figure}

The origin $(u,v)=(0,0)$ is an equilibrium point if $\delta=0$. Then, the eigenvalues of the linearization are 
$\left\{-\sqrt{-k-1},\sqrt{-k-1}\right\}$. The  origin is a saddle for $k<-1$ and a center for $k>-1$.

Now, for $k\neq 0$ and $|k v_c|\leq 1$, we have the equilibrium points $(u,v)=(0,v_c)$ such as $-k v-\sin (\delta +v)=0$. We obtain a real valued linearization matrix which is additionally required  $3 \cos (\delta )-\frac{3}{2} v \sin (\delta +v)-3 \cos (\delta +v)\geq 0$.

If $\delta=0$ and $|k  v|>1$ there are no equilibrium points other than the origin.

In general, for $\delta \neq 0$ the system admits no equilibrium points $(u,v)=(0,v_c)$, apart from the origin, for $|k  v_c|>1$.

If $|k v_c|\leq 1$, we have the equilibrium points $(u,v)=(0,v_c)$ where $v_c$ are the roots of the transcendental equation $-k v-\sin (\delta +v)=0$.

For $\delta\neq 0$ and $|k  v_c|\leq 1$, we have
    $\delta =-\sin ^{-1}(k v)-v$, and we obtain the eigenvalues \\
$\Big\{-\frac{1}{8} \sqrt{6 b^2 k^2 \mu ^2 \left(-2 \sqrt{1-k^2 v^2}+k v^2+2 \cos \left(\sin ^{-1}(k v)+v\right)\right)-64
   \left(\sqrt{1-k^2 v^2}+k\right)}$\\
	$-\frac{1}{4} b k \mu  \sqrt{-3 \sqrt{1-k^2 v^2}+\frac{3 k v^2}{2}+3 \cos \left(\sin ^{-1}(k
   v)+v\right)}$,\\
	$\frac{1}{8} \sqrt{6 b^2 k^2 \mu ^2 \left(-2 \sqrt{1-k^2 v^2}+k v^2+2 \cos \left(\sin ^{-1}(k v)+v\right)\right)-64
   \left(\sqrt{1-k^2 v^2}+k\right)}$\\
	$-\frac{1}{4} b k \mu  \sqrt{-3 \sqrt{1-k^2 v^2}+\frac{3 k v^2}{2}+3 \cos \left(\sin ^{-1}(k
   v)+v\right)}\Big\}$.

For the choice of parameters $(b, f, \delta, \mu)= (0.1, 0.33, 0, 0.9)$ we have 
   $\rho_c=\frac{24057}{2000000}\approx 0.0120285, k=\frac{22}{3}\approx 7.33333$. 
The only equilibrium point is the origin  with eigenvalues
$\left\{\frac{5 i}{\sqrt{3}},-\frac{5 i}{\sqrt{3}}\right\}$. 
 In figure \ref{2monodromy-potential-1-b} is presented some orbits of the flow of \eqref{monodronomyeqs1ab} for the choice of parameters $(b, 
   f, \delta, \mu)=(0.1, 0.33, 0, 0.9)$. For this choice of parameters the hypotheses and the results of Theorems \ref{tm} and \ref{thm2.1} ($\lim_{t\rightarrow \infty } \dot\phi=0$, and $\lim_{t\rightarrow \infty } \phi=0$) are verified.

	Substituting the values $(b, 
   f, \delta, \mu)= (0.99, 0.09, 0, 0.9)$, we obtain
   $\rho_c=\frac{649539}{20000000}\approx 0.032477, k=\frac{20}{99}\approx 0.20202$. 
The transcendental equation is $-\frac{20 v}{99}-\sin (v)=0$. 
The equilibrium points are: 	
\begin{enumerate}
 \item $A: (u,v)=(0,-4.88035)$, eigenvalues $-0.140267-0.591197 i, -0.140267+0.591197 i$, stable spiral. 
 \item $B: (u,v)=(0,-4.12769)$, eigenvalues $-0.749132, 0.467117$, saddle. 
 \item $C: (u,v)=(0,0)$, eigenvalues $0.\, -1.09637 i , 0.\, +1.09637 i$, center.
 \item $D: (u,v)=(0,4.12769)$, eigenvalues $-0.749132, 0.467117$, saddle. 
 \item $E: (u,v)=(0,4.88035)$, eigenvalues  $-0.140267-0.591197 i , -0.140267+0.591197 i$, stable spiral. 
\end{enumerate}
In figure \ref{2monodromy-potential-1-a} are presented some orbits of the flow of \eqref{monodronomyeqs1ab} and for  $(b, f, \delta, \mu)= (0.99, 0.09, 0, 0.9)$ the hypotheses of Theorem \ref{tm} hold, and the result $\lim_{t\rightarrow +\infty} \dot \phi=0$ is attained.  The hypothesis {\it{$V^{\prime}(\phi)<0$ for $\phi<0$ and $V^{\prime}(\phi)>0$ for $\phi>0$}} of Theorem \ref{thm2.1} is violated and $\lim_{t\rightarrow +\infty}\phi$ can be zero, or finite. Recall this Theorem relies on the monotonicity of $V(\phi)$. Finally, the hypothesis {\it{$V^{\prime}(\phi)<0\quad \forall \phi\in\mathbb{R}$}} of Theorem \ref{thm2.2} is violated and $\lim_{t\rightarrow +\infty}\dot\phi=0, \lim_{t\rightarrow +\infty}\phi<\infty$.

\subsubsection{Oscillating regime.}
\label{Sect:2.5.2}

In this section, we investigate the potential 
$V(\phi)= \mu ^3 \left[b f \left(\cos (\delta )-\cos \left(\delta +\frac{\phi }{f}\right)\right)+\frac{\phi ^2}{\mu}\right]
$, $b\neq 0$ looking for oscillatory behavior as expected from the numerical investigations. We derive asymptotic expansions as well. As before, we define the pair
\begin{equation}
\left(\frac{\sqrt{2 \mu}\phi}{\sqrt{{\dot \phi}^2+2 \mu \phi^2}}, \quad \frac{\dot{\phi}}{\sqrt{{\dot{\phi}}^2+2 \mu \phi^2}}\right)
\end{equation}
defining a function of $t$ with values in the unit circle. Therefore, we define the angular function 
$\vartheta (t)$ that is unique under identification module $2\pi$, defined by 
\begin{equation}
\vartheta=  \tan ^{-1} \left(\frac{\dot \phi}{\sqrt{2 }\mu \phi}\right),
\end{equation}
together with
\begin{equation}
r=\sqrt{{\dot \phi}^2+2 \mu^2 \phi^2},
\end{equation} 
with inverse
\begin{equation}
\phi = \frac{r \cos (\vartheta )}{\sqrt{2} \mu },\quad \dot\phi= r \sin (\vartheta ). 
\end{equation}
They satisfy
\begin{equation}
b f \mu ^3 \left(\cos \left(\delta +\frac{r \cos (\vartheta )}{\sqrt{2} f \mu }\right)-\cos (\delta )\right)+3 H^2-\frac{r^2}{2}=0.
\end{equation}
For expanding universes ($H>0$) we obtain the equations 
\begin{subequations}
\label{syst104}
\begin{align}
& \dot r=-\sqrt{\frac{3}{2}} r \sin ^2(\vartheta ) \sqrt{2 b f \mu ^3 \left(\cos (\delta )-\cos \left(\delta +\frac{r \cos (\vartheta )}{\sqrt{2} f \mu }\right)\right)+r^2}\nonumber \\
& -b \mu ^3 \sin (\vartheta ) \sin \left(\delta +\frac{r \cos (\vartheta
   )}{\sqrt{2} f \mu }\right),\\
& \dot \vartheta=-\sqrt{2} \mu -\sqrt{\frac{3}{2}} \sin (\vartheta ) \cos (\vartheta ) \sqrt{2 b f \mu ^3 \left(\cos (\delta )-\cos \left(\delta +\frac{r \cos (\vartheta )}{\sqrt{2} f \mu }\right)\right)+r^2}\nonumber \\
& -\frac{b \mu ^3 \cos (\vartheta )  \sin \left(\delta +\frac{r \cos (\vartheta )}{\sqrt{2} f \mu }\right)}{r}.
\end{align}
\end{subequations}
Observing that for $b\rightarrow 0$, the solutions of the limiting equation admit the asymptotic expansions \cite{Rendall:2006cq}
\begin{equation}
\vartheta(t)= -\sqrt{2} \mu  t + O(\ln t), \quad r(t)= \frac{4}{\sqrt{6} t}+ O(t^{-2}\ln t). 
\end{equation}
Hence, when $b=0$, 
\begin{equation}
\phi(t)=\frac{4 \cos t}{\sqrt{6} t}+ O(t^{-2}\ln t), \quad \dot{\phi}(t)=\frac{4 \sin t}{\sqrt{6} t}+ O(t^{-2}\ln t).
\end{equation}
\begin{figure}
\begin{center}
\subfigure[]{\includegraphics[scale=0.5]{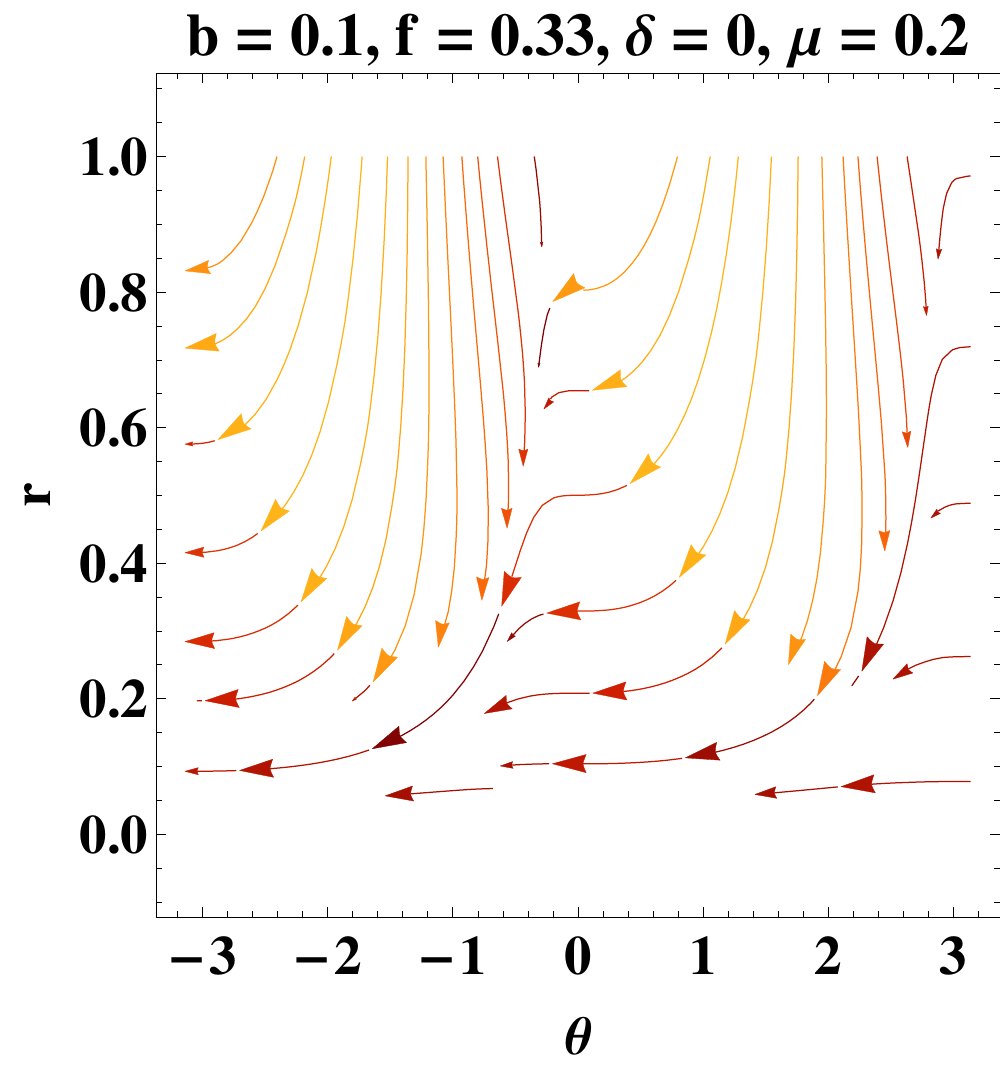} }\hspace{2cm}
\subfigure[]{\includegraphics[scale=0.5]{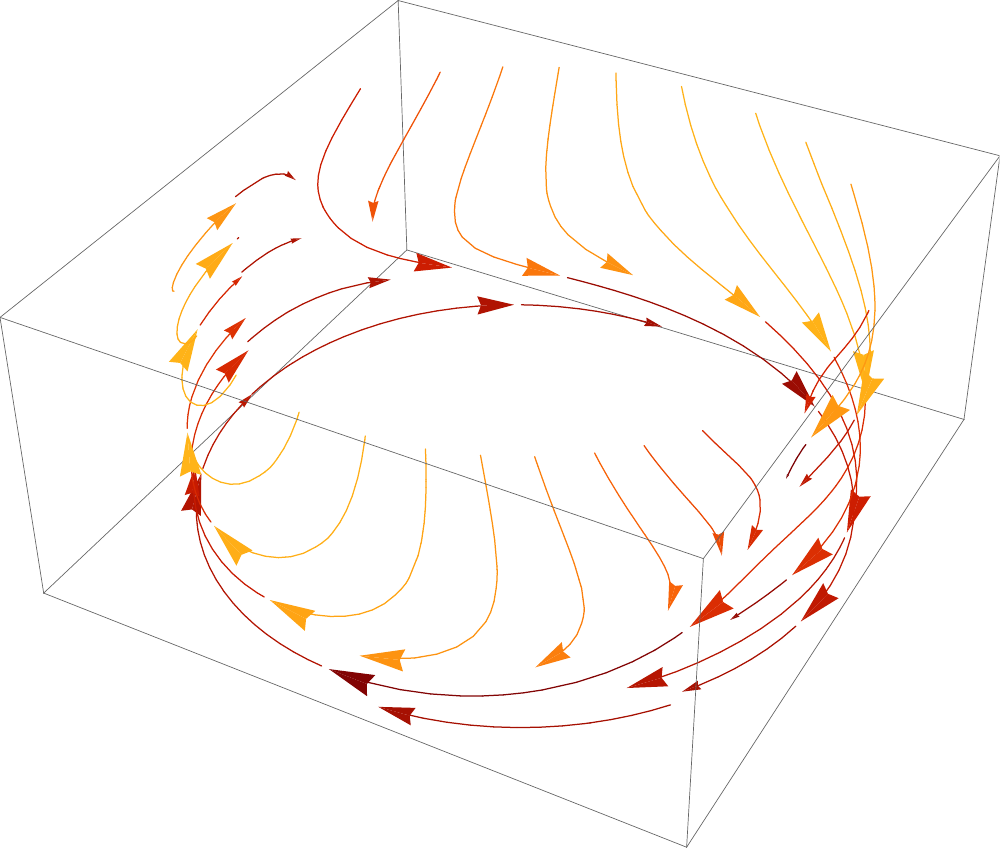}}
\subfigure[]{\includegraphics[scale=0.5]{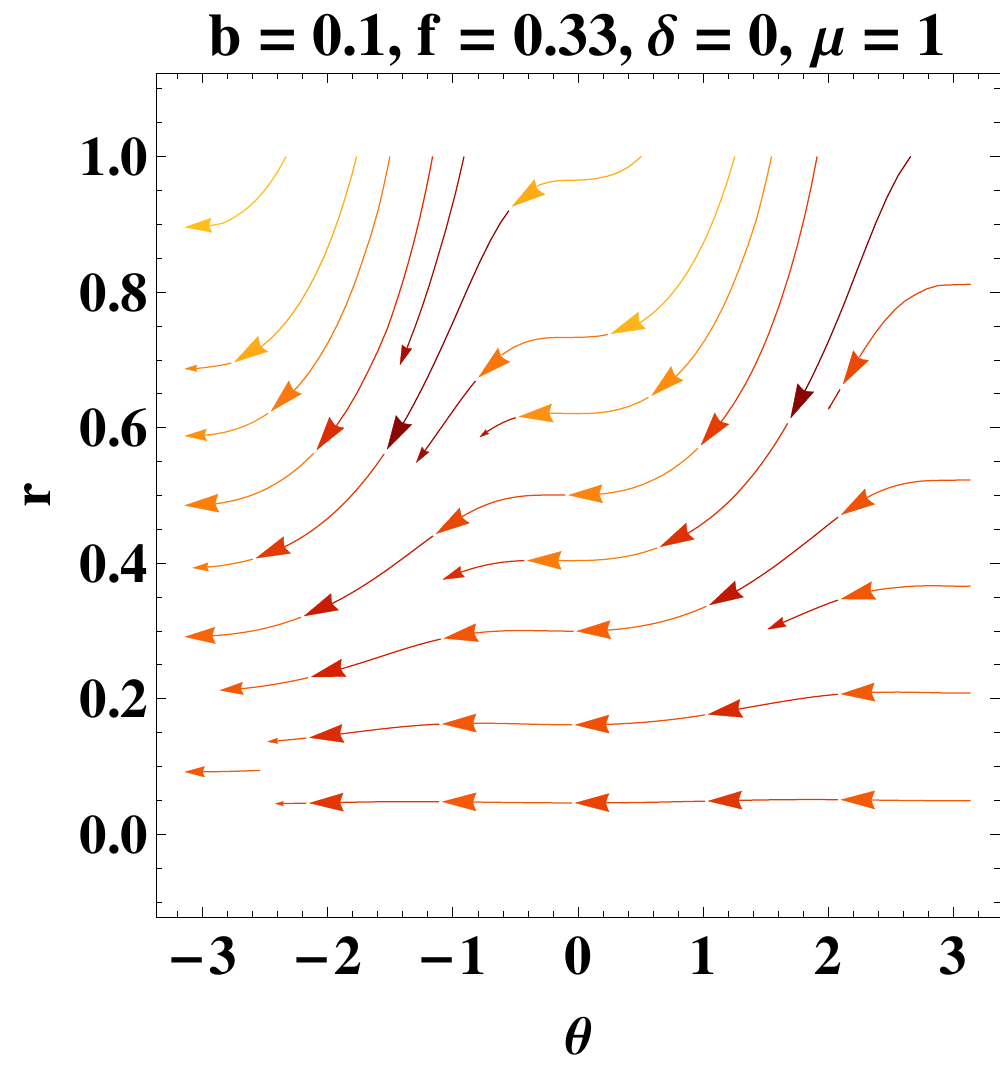} }\hspace{2cm}
\subfigure[]{\includegraphics[scale=0.5]{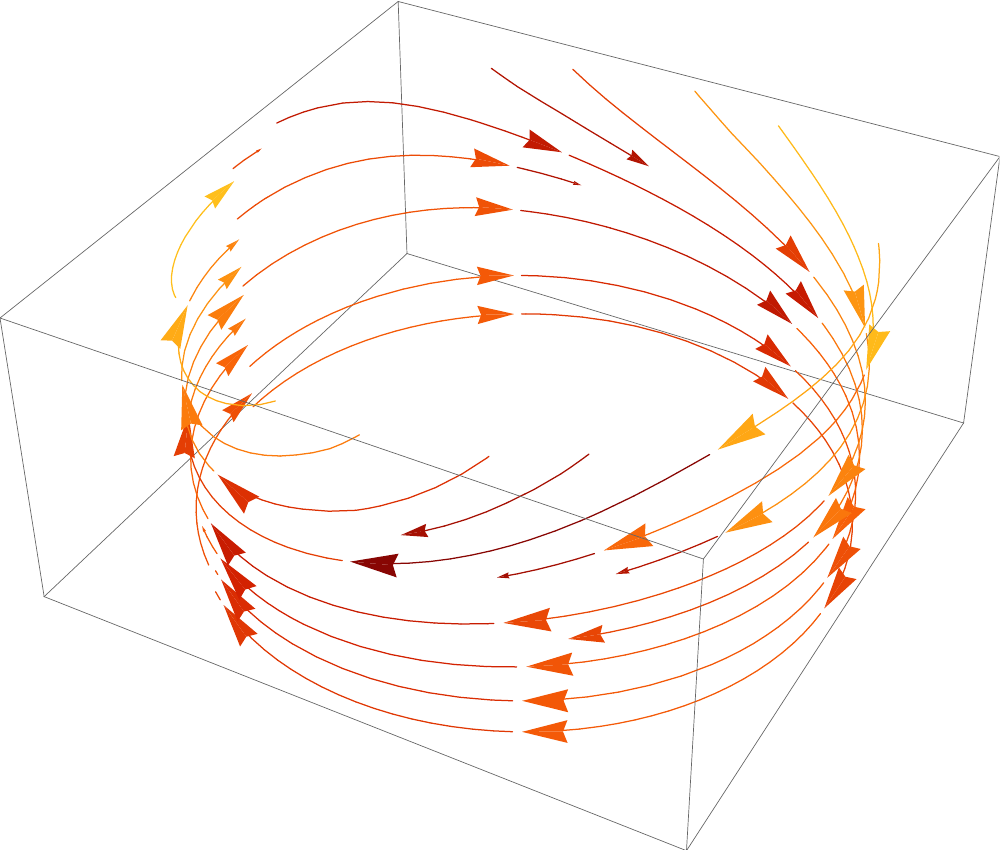}}
\end{center}\caption{\label{Fig10} Phase portrait of equations   \eqref{syst104} (left panel). Projection over the cylinder $\mathbf{S}$ (right panel) for $(b, f, \delta)= (0.1, 0.33, 0)$ and different values of $\mu$.}
\end{figure}

\begin{figure}
\begin{center}
\subfigure[]{\includegraphics[scale=0.5]{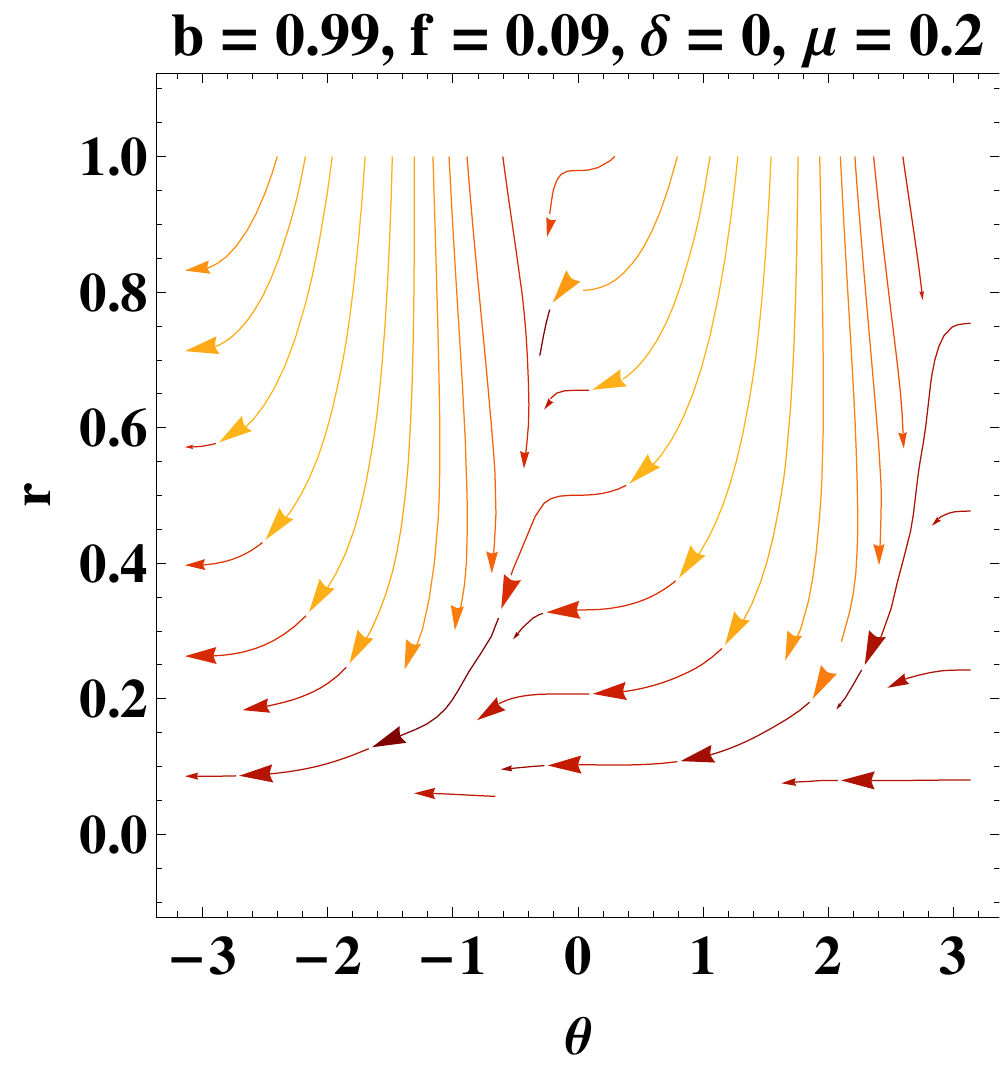} }\hspace{2cm}
\subfigure[]{\includegraphics[scale=0.5]{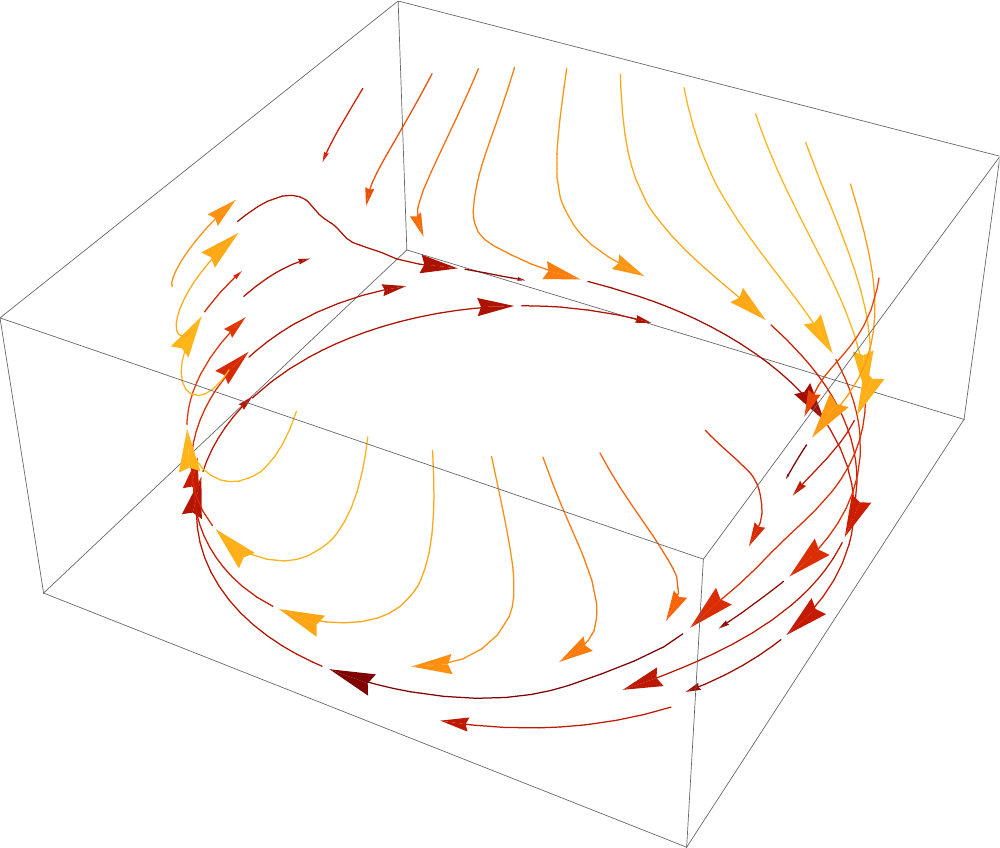}}
\subfigure[]{\includegraphics[scale=0.5]{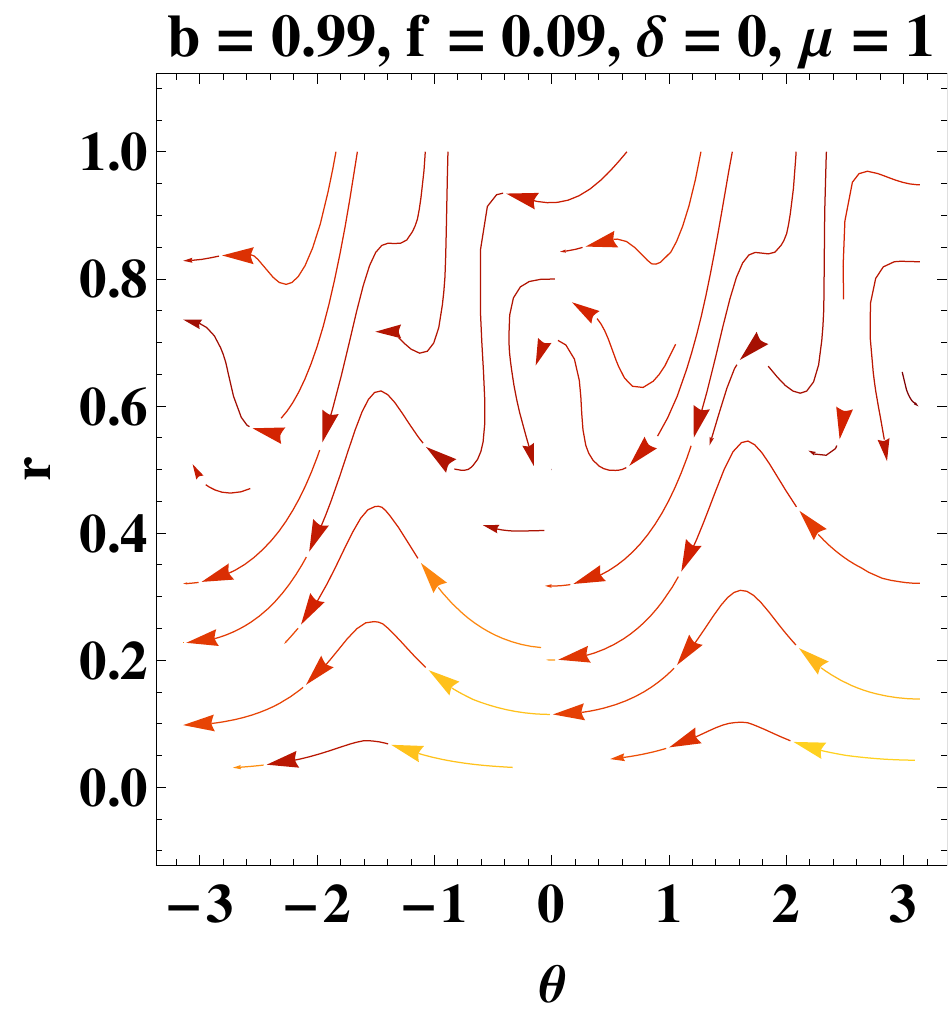} }\hspace{2cm}
\subfigure[]{\includegraphics[scale=0.5]{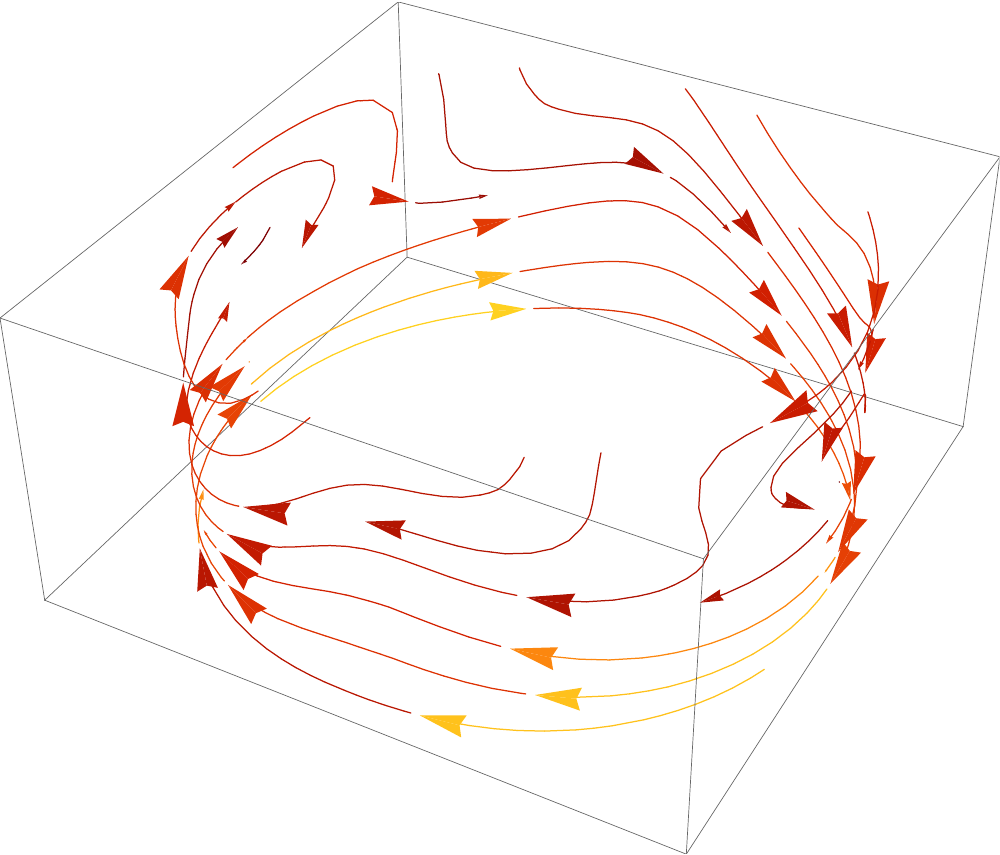}}
\end{center}\caption{\label{Fig11} Phase portrait of equations   \eqref{syst104} (left panel). Projection over the cylinder $\mathbf{S}$ (right panel) for $(b, f, \delta)= (0.99, 0.09, 0)$ and different values of $\mu$.}
\end{figure}
Now, we derive asymptotic expansions of the full problem ($b\neq 0$).
\\
Note that: 
\begin{subequations}
\begin{align}
& \dot r=  -b \mu ^3 \sin (\delta ) \sin (\vartheta )-\frac{b \mu ^2 \cos (\delta ) \cos (\theta ) \sin (\vartheta ) r}{\sqrt{2} f} \nonumber\\
& -\frac{\sqrt{3} \sqrt{b \mu ^2 \cos (\vartheta ) \sin (\delta )} \sin ^2(\vartheta )
   r^{3/2}}{\sqrt[4]{2}}+\frac{b \mu  \cos ^2(\vartheta ) \sin (\delta ) \sin (\vartheta ) r^2}{4 f^2}+O\left(r^{5/2}\right),\end{align}
	\begin{align}
& \dot \vartheta +\sqrt{2} \mu =  -\frac{\sqrt{3} \cos (\vartheta ) \sqrt{b \mu ^2 \cos (\vartheta ) \sin (\delta )} \sin (\vartheta ) \sqrt{r}}{\sqrt[4]{2}}+\frac{b \mu  \cos ^3(\vartheta ) \sin (\delta ) r}{4 f^2}
\nonumber \\
& -\frac{\left(\sqrt{3} \cos (\theta ) \left(b \mu  \cos
   (\delta ) \cos ^2(\vartheta )+2 f\right) \sin (\vartheta )\right) r^{3/2}}{4 \left(2^{3/4} f \sqrt{b \mu ^2 \cos (\theta ) \sin (\delta )}\right)}+\frac{b \cos (\delta ) \cos ^4(\vartheta ) r^2}{12 \sqrt{2}
   f^3}+O\left(r^{5/2}\right).
\end{align}
\end{subequations}
Obtaining an approximated solution near the oscillatory regime we can take the average with respect to $\vartheta$ over any orbit of period $2 \pi$, 
$\langle (\dot r, \dot \vartheta) \rangle$, leading to 
\begin{align}
& \dot r=k r^{3/2} , \quad k=\frac{2^{3/4} \sqrt{3} \sqrt{b} \mu  \left(E\left(\left.\frac{c}{2}\right|2\right)-E\left(\left.\frac{c}{2}+\pi \right|2\right)\right) \sqrt{\sin (\delta )}}{5 \pi },\\
& \dot \vartheta +\sqrt{2} \mu=\frac{b \cos (\delta ) (r-4 f \mu ) (r+4 f \mu)}{32 \sqrt{2} f^3}. 
\end{align}
where $E\left(\left.\phi\right|m\right)$ gives the elliptic integral of the second kind:
\begin{equation}
E\left(\left.\phi\right|m\right)= \int_{0}^{\phi} \left(1- m \sin^2 (\theta)\right)^{\frac{1}{2}}d\theta, \quad -\frac{\pi}{2}<\phi<\frac{\pi}{2}.
\end{equation}
The averaged equations have solutions 
\begin{equation}
r(t)=  \frac{4}{\left(c_1+k t\right){}^2}, \vartheta (t)= -\frac{b \cos (\delta ) \left(3 f^2 \mu ^2 \left(c_1+k t\right)+\frac{1}{\left(c_1+k t\right){}^3}\right)+12 f^3 \mu  \left(c_1+k t\right)}{6 \sqrt{2} f^3
   k}+\vartheta_0. 
\end{equation}
\\
Introducing along with $\vartheta$ and $r$, the new variable 
\begin{equation}
\varepsilon=\frac{H}{\mu+H}, 
\end{equation} 
with inverse
\begin{equation}
 H= \frac{\mu  \varepsilon}{1-\varepsilon},
\end{equation}
satisfying
\begin{equation}
b f \mu ^3 (\varepsilon -1)^2 \left(\cos \left(\delta +\frac{r \cos (\vartheta )}{\sqrt{2} f \mu }\right)-\cos (\delta )\right)-\frac{1}{2} r^2 (\varepsilon -1)^2+3 \mu ^2 \varepsilon ^2=0,
\end{equation}
along with the time derivative  $\hat{\tau}$ given by 
\begin{equation}
\frac{d \hat{\tau}}{d t}:= \mu+ H,
\end{equation}
we obtain the equations 
\begin{subequations}
\begin{align}
& \varepsilon'=\frac{r^2 (\varepsilon -1)^3 \sin ^2(\vartheta )}{2 \mu ^2},\\
& r'=-3 r \varepsilon  \sin ^2(\vartheta )+b \mu ^2 (\varepsilon -1) \sin (\vartheta ) \sin \left(\delta +\frac{r \cos (\vartheta )}{\sqrt{2} f \mu }\right),\\
&\vartheta'=-\sqrt{2} (1-\varepsilon)-3 \varepsilon  \sin (\vartheta ) \cos (\vartheta )+\frac{b \mu ^2 (\varepsilon -1)
   \cos (\vartheta ) \sin \left(\delta +\frac{r \cos (\vartheta )}{\sqrt{2} f \mu }\right)}{r}.
\end{align}
\end{subequations}
Obtaining an approximated solution near the oscillatory regime we take the average with respect to $\vartheta$ over any orbit of period $2 \pi$, 
$\langle (\varepsilon', r', \vartheta') \rangle$, leading to
\begin{subequations}
\begin{align}
& \varepsilon'=\frac{(\varepsilon -1)^3 r^2}{4 \mu ^2},\\
& r'=-\frac{3 \varepsilon  r}{2},\\
& \vartheta'=\frac{(\varepsilon -1) (4 f+b \mu  \cos (\delta ))}{2 \sqrt{2} f}-\frac{(b (\varepsilon -1) \cos (\delta )) r^2}{32
   \left(\sqrt{2} f^3 \mu \right)}, 
\end{align}
\end{subequations}
with the averaged constraint
\begin{equation}
3 \mu ^2 \epsilon ^2-\frac{r^2 \left((\epsilon -1)^2 (b \mu  \cos (\delta )+4 f)\right)}{8 f}=0,
\end{equation}
as $r\rightarrow 0$. 
\\
The above system is integrable yielding
\begin{subequations}
\begin{align}
& r(\varepsilon )=\frac{\sqrt{2} \sqrt{c_1 (\varepsilon -1)^2+\mu ^2 (6 \varepsilon -3)}}{1-\varepsilon },\\
& \vartheta (\varepsilon )=\frac{\mu  \left(\frac{b \cos (\delta )}{\varepsilon -1}-\frac{8 f^2 \tanh ^{-1}\left(\frac{c_1 (\varepsilon -1)+3
   \mu ^2}{\mu  \sqrt{9 \mu ^2-3 c_1}}\right) (b \mu  \cos (\delta )+4 f)}{\sqrt{9 \mu ^2-3 c_1}}\right)}{8 \sqrt{2} f^3}+c_2, 
\end{align}
and 
\begin{align}
& 3(t-t_0)= \ln \left(\frac{(1-\varepsilon )^2 \left(\frac{3 \mu  \sqrt{3 \mu ^2-c_1}+\sqrt{3} c_1 \varepsilon -\sqrt{3} c_1+3 \sqrt{3} \mu ^2}{3 \mu  \sqrt{3 \mu ^2-c_1}-\sqrt{3} c_1 \varepsilon +\sqrt{3} c_1-3 \sqrt{3} \mu ^2}\right){}^{\frac{\mu
   }{\sqrt{\mu ^2-\frac{c_1}{3}}}}}{c_1 (\varepsilon -1)^2+\mu ^2 (6 \varepsilon -3)}\right) \nonumber \\
	& \sim \ln \left(\frac{\left(\frac{2 \mu  \left(\sqrt{9 \mu ^2-3 c_1}+3 \mu \right)-c_1}{c_1}\right){}^{\frac{\mu }{\sqrt{\mu ^2-\frac{c_1}{3}}}}}{c_1-3 \mu ^2}\right)-\frac{6 \mu ^2 \varepsilon }{c_1-3 \mu ^2}+O\left(\varepsilon ^2\right),
\end{align}
\end{subequations}
as $\varepsilon\rightarrow 0$. 

Figure \ref{Fig10} shows the phase portrait of equations   \eqref{syst104} (left panel) and the projection over the cylinder $\mathbf{S}$ (right panel) for $(b, f, \delta)= (0.1, 0.33, 0)$ and different values of $\mu$. In Figure \ref{Fig11}, we present the phase portrait of equations   \eqref{syst104} (left panel) and the projection over the cylinder $\mathbf{S}$ (right panel) for $(b, f, \delta)= (0.99, 0.09, 0)$ and different values of $\mu$. The plots show the periodic nature of the solutions.

\section{Conclusions}
\label{Sect:7}

In this paper we have proved new Theorems: \ref{Proposition I}, \ref{Theorem5.2}, \ref{Proposition II}, \ref{Theorem5.3}, \ref{Proposition III} and \ref{thmIIIFINAL} valid for general situations in the context of scalar field cosmologies with arbitrary potential and/ or with arbitrary couplings to matter. Some well-known results from the literature are recovered, and they are presented as the corollaries \ref{PropositionIb}, \ref{tm}, \ref{PropositionII}, \ref{thm2.1}, \ref{Prop4Miritzis} and \ref{thm2.2}.
We presented some examples that violate one or more hypotheses of the Theorems proved.  We saw to which extent these conditions can be relaxed in order to obtain the same conclusions or provide a counterexample.  In particular, we incorporated cosine-like corrections with small phase. We have seen motivation for this kind of potential's correction in the context of inflation in loop- quantum cosmology \cite{Sharma:2018vnv}. We use both local and global dynamical system variables and smooth transformations of the scalar field to provide qualitative features of the model at hand.  We have discussed the conditions of a scalar field potential $V\in  C^{2}(\mathbb{R})$  under which $\displaystyle{\lim_{t\rightarrow  \infty} \dot \phi =0}$.  These conditions are very general: non-negativity of the potential which are zero only on the origin and the boundedness of both  $V^{\prime}(\phi)$ and  $V(\phi)$ (Theorem \ref{tm}). Additionally, we have presented some extra conditions for having $\displaystyle{\lim_{t\rightarrow  \infty} \phi(t) \in \lbrace -\infty , 0 , + \infty \rbrace }$. They are the previous conditions with the addition of $V^{\prime}(\phi)>0$ for $\phi>0$ and $V^{\prime}(\phi)<0$ for $\phi<0$ (Theorem \ref{thm2.1}). 
We have considered mild conditions under the potential (satisfied by the exponential potential with negative slope) for having  $\lim_{t \rightarrow  + \infty} \dot \phi =0$ and $\lim_{t \rightarrow  + \infty} \phi (t)= +\infty$ (Theorem \ref{thm2.2}). 

In Section \ref{SECT:6}, we provided a local dynamical systems analysis for arbitrary $V(\phi)$ and $\chi(\phi)$ using Hubble normalized equations. The analysis relies on two arbitrary functions $f(\lambda)$ and $g(\lambda)$ which encode the potential and the coupling function through the quadrature
\begin{equation*}
 \phi (\lambda )=\phi (1)-\int_1^{\lambda } \frac{1}{f(s)} \, ds, \quad
 V (\lambda )+\Lambda=W(1) e^{\int_1^{\lambda } \frac{s}{f(s)} \, ds}, \quad
 \chi (\lambda )=\chi(1) e^{\int_1^{\lambda } \frac{g(s)}{f(s)} \, ds}.
\end{equation*}
After that, we proceeded to a global dynamical systems formulation in Section \ref{SECT:3.3} using the Alho \& Uggla's approach \cite{Alho:2014fha} (first used by authors for monomial potential), which is well-suited for a global description of the phase space. We have obtained equilibrium points that represent some solutions of cosmological interest: a matter - kinetic scaling solution, a matter- scalar field scaling solution, a kinetic dominated solution representing a stiff fluid, a solution dominated by the effective energy density of the geometric term $G_0(a)$, a scaling solution where the kinetic term, and the effective energy density from $G_0(a)$, scales with the same order of magnitude, a quintessence scalar field dominated solution, the vacuum de Sitter solution associated to the minimum of the potential, and a non-interacting matter dominated solution, all of which reveal very rich cosmological behavior.
 
We have shown the results for the vacuum case and the exponential potential. The procedure was used later in Sections \ref{Sect:2.4.1} and \ref{Sect:2.5.1} to find qualitative features and also to present the asymptotic analysis as $\phi\rightarrow \infty$ for the harmonic potentials $V_1(\phi)= \mu^3 \left[\frac{\phi^2}{\mu} + b f \cos\left(\delta + \frac{\phi}{f}\right)\right]$, $b\neq 0$ and 
$V_2(\phi)= \mu ^3 \left[b f \left(\cos (\delta )-\cos \left(\delta +\frac{\phi }{f}\right)\right)+\frac{\phi ^2}{\mu}\right]
$, $b\neq 0$ in vacuum, respectively.  
In Section \ref{SECT4}, we have presented an alternative dynamical systems formulation for an scalar-field cosmology with the aforementioned generalized harmonic potentials in a vacuum. More specifically, in Section \ref{Sect2.4} we have presented a qualitative analysis for a scalar-field cosmology with generalized harmonic potential $V_1(\phi)$; whereas in Section \ref{Sect.2.5} we have presented a qualitative analysis for a scalar-field cosmology with generalized harmonic potential $V_2(\phi)$. 
In Section \ref{Sect:2.4.2}, we have investigated the oscillatory regime for the scalar field under the potential $V_1(\phi)$.  Meanwhile, in Section \ref{Sect:2.5.2} we study the oscillations of the scalar field under the potential $V_2(\phi)$.

In the first generalization of the harmonic potential $V_1(\phi)$, we find some instances which verified the hypothesis and the results of Theorems \ref{tm}, \ref{thm2.1} ($\lim_{t\rightarrow \infty } \dot\phi=0$ and $\lim_{t\rightarrow \infty } \phi=0$), as well as situations when the hypotheses {\it{$V(\phi)\geq 0$ and $V(\phi)=0$, if and only if $\phi=0$}} of Theorem \ref{tm} is violated, though the result $\lim_{t\rightarrow +\infty} \dot \phi =0$ holds. The hypotheses {\it{$V(\phi)\geq 0$ and $V(\phi)=0$ if and only if, $\phi=0$}} and {\it{$V^{\prime}(\phi)<0$ for $\phi<0$ and $V^{\prime}(\phi)>0$ for $\phi>0$}} of Theorem \ref{thm2.1} are violated, and $\lim_{t\rightarrow +\infty}\phi$ can be finite (rather than zero or infinity).  Finally, when the hypotheses {\it{$V(\phi)\geq 0$, and $V^{\prime}(\phi)<0\quad \forall \phi\in\mathbb{R}$}} of Theorem \ref{thm2.2} are violated, and $\lim_{t\rightarrow +\infty}\dot\phi=0, \lim_{t\rightarrow +\infty}\phi<\infty$. As well as, for $V_1(\phi)$ we found some instances for the potential $V_2(\phi)$ where the hypotheses and the results of Theorems are verified \ref{tm} and \ref{thm2.1} ($\lim_{t\rightarrow \infty } \dot\phi=0$, and $\lim_{t\rightarrow \infty } \phi=0$).  For other choices of the parameters, the hypothesis of Theorem \ref{tm} holds, and the result $\lim_{t\rightarrow +\infty} \dot \phi=0$ is attained.  The hypothesis {\it{$V^{\prime}(\phi)<0$ for $\phi<0$ and $V^{\prime}(\phi)>0$ for $\phi>0$}} of Theorem \ref{thm2.1} is violated and $\lim_{t\rightarrow +\infty}\phi$ can be zero, or finite. Recall this Theorem relies on the monotonicity of $V(\phi)$. The hypotheses {\it{$V^{\prime}(\phi)<0\quad \forall \phi\in\mathbb{R}$}} of Theorem \ref{thm2.2} are violated and $\lim_{t\rightarrow +\infty}\dot\phi=0, \lim_{t\rightarrow +\infty}\phi<\infty$. 
In other words,  we have discussed some simple examples that violate one or more  hypotheses of the Theorems proved, obtaining some counterexamples.

\ack

This research was funded by  Agencia Nacional de Investigaci\'on y Desarrollo - ANID for financial support through the program FONDECYT Iniciaci\'on grant no.
11180126. Ellen de los M. Fernández Flores and Joey latta are acknowledged for proofreading this manuscript and improving the English.

\appendix

\section{Existence and stability conditions of the equilibrium points of the system \eqref{systH1}}
\label{AppA}
 The equilibrium points of the system \eqref{systH1} are the following: 
\begin{enumerate}
    \item[$A_1(\hat{\lambda})$:] $(x, \Omega_m, \Omega_0, \lambda)=\left(\frac{(4-3 \gamma ) g(\hat{\lambda})}{\sqrt{6} (\gamma -2)},
   1-\frac{(4-3 \gamma )^2 g(\hat{\lambda})^2}{6 (\gamma
   -2)^2}, 0, \hat{\lambda}\right)$, where we denote by  $\hat{\lambda}$, the values of $\lambda$ for which  $f(\lambda)=0$. Exists for $1-\frac{(4-3 \gamma )^2 g(\hat{\lambda })^2}{6 (\gamma -2)^2}\geq 0$. 
   The eigenvalues are $\Bigg\{\frac{3 (\gamma -2)}{2}-\frac{(4-3 \gamma )^2 g(\hat{\lambda
   })^2}{4 (\gamma -2)}, 3 \gamma -\frac{(4-3 \gamma )^2
   g(\hat{\lambda})^2}{2 (\gamma -2)}-p, \frac{6 (\gamma -2) \gamma +(3 \gamma -4) g(\hat{\lambda}) \left((4-3 \gamma )
   g(\hat{\lambda})+2 \hat{\lambda }\right)}{2 (\gamma
   -2)},\frac{(3 \gamma -4) g(\hat{\lambda})
   f'(\hat{\lambda})}{\gamma -2}\Bigg\}$.  \\ For $1\leq \gamma< 2$, we deduce that $A_1(\hat{\lambda})$ is a sink for 
   \begin{enumerate}
    \item $3<p\leq 4, \;  1\leq \gamma <\frac{p}{3}, \;  -\sqrt{2} \sqrt{\frac{(\gamma -2) (3 \gamma -p)}{(3 \gamma -4)^2}}<g(\hat{\lambda })<0, \;  f'(\hat{\lambda })>0, \;  \hat{\lambda }>\left(\frac{3 \gamma
   }{2}-2\right) g(\hat{\lambda })-\frac{3 (\gamma -2) \gamma }{(3 \gamma -4) g(\hat{\lambda })}$, or 
   \item $3<p\leq 4, \;  1\leq \gamma <\frac{p}{3}, \;  0<g(\hat{\lambda })<\sqrt{2}
   \sqrt{\frac{(\gamma -2) (3 \gamma -p)}{(3 \gamma -4)^2}}, \;  f'(\hat{\lambda })<0, \;  \hat{\lambda }<\left(\frac{3 \gamma }{2}-2\right) g(\hat{\lambda })-\frac{3 (\gamma -2) \gamma }{(3 \gamma -4)
   g(\hat{\lambda })}$, or 
   \item $4<p\leq 6, \;  1\leq \gamma <\frac{4}{3}, \;  -\sqrt{2} \sqrt{\frac{(\gamma -2) (3 \gamma -p)}{(3 \gamma -4)^2}}<g(\hat{\lambda })<0, \;  f'\left(\hat{\lambda
   }\right)>0, \;  \hat{\lambda }>\left(\frac{3 \gamma }{2}-2\right) g(\hat{\lambda })-\frac{3 (\gamma -2) \gamma }{(3 \gamma -4) g(\hat{\lambda })}$, or 
   \item $4<p\leq 6, \;  1\leq \gamma
   <\frac{4}{3}, \;  0<g(\hat{\lambda })<\sqrt{2} \sqrt{\frac{(\gamma -2) (3 \gamma -p)}{(3 \gamma -4)^2}}, \;  f'(\hat{\lambda })<0, \;  \hat{\lambda }<\left(\frac{3 \gamma }{2}-2\right) g\left(\hat{\lambda
   }\right)-\frac{3 (\gamma -2) \gamma }{(3 \gamma -4) g(\hat{\lambda })}$, or 
   \item $4<p\leq 6, \;  \frac{4}{3}<\gamma <\frac{p}{3}, \;  -\sqrt{2} \sqrt{\frac{(\gamma -2) (3 \gamma -p)}{(3 \gamma
   -4)^2}}<g(\hat{\lambda })<0, \;  f'(\hat{\lambda })<0, \;  \hat{\lambda }<\left(\frac{3 \gamma }{2}-2\right) g(\hat{\lambda })-\frac{3 (\gamma -2) \gamma }{(3 \gamma -4) g\left(\hat{\lambda
   }\right)}$, or 
   \item $4<p\leq 6, \;  \frac{4}{3}<\gamma <\frac{p}{3}, \;  0<g(\hat{\lambda })<\sqrt{2} \sqrt{\frac{(\gamma -2) (3 \gamma -p)}{(3 \gamma -4)^2}}, \;  f'(\hat{\lambda })>0, \; 
   \hat{\lambda }>\left(\frac{3 \gamma }{2}-2\right) g(\hat{\lambda })-\frac{3 (\gamma -2) \gamma }{(3 \gamma -4) g(\hat{\lambda })}$, or 
   \item $p>6, \;  1\leq \gamma <\frac{4}{3}, \;  -\frac{\sqrt{6}
   (\gamma -2)}{3 \gamma -4}<g(\hat{\lambda })<0, \;  f'(\hat{\lambda })>0, \;  \hat{\lambda }>\left(\frac{3 \gamma }{2}-2\right) g(\hat{\lambda })-\frac{3 (\gamma -2) \gamma }{(3 \gamma -4)
   g(\hat{\lambda })}$, or 
   \item $p>6, \;  1\leq \gamma <\frac{4}{3}, \;  0<g(\hat{\lambda })<\frac{\sqrt{6} (\gamma -2)}{3 \gamma -4}, \;  f'(\hat{\lambda })<0, \;  \hat{\lambda
   }<\left(\frac{3 \gamma }{2}-2\right) g(\hat{\lambda })-\frac{3 (\gamma -2) \gamma }{(3 \gamma -4) g(\hat{\lambda })}$, or 
   \item $p>6, \;  \frac{4}{3}<\gamma <2, \;  \frac{\sqrt{6} (\gamma -2)}{3
   \gamma -4}<g(\hat{\lambda })<0, \;  f'(\hat{\lambda })<0, \;  \hat{\lambda }<\left(\frac{3 \gamma }{2}-2\right) g(\hat{\lambda })-\frac{3 (\gamma -2) \gamma }{(3 \gamma -4) g\left(\hat{\lambda
   }\right)}$, or 
   \item $p>6, \;  \frac{4}{3}<\gamma <2, \;  0<g(\hat{\lambda })<-\frac{\sqrt{6} (\gamma -2)}{3 \gamma -4}, \;  f'(\hat{\lambda })>0, \;  \hat{\lambda }>\left(\frac{3 \gamma
   }{2}-2\right) g(\hat{\lambda })-\frac{3 (\gamma -2) \gamma }{(3 \gamma -4) g(\hat{\lambda })}$.
   \end{enumerate}
If exist, it will be never a source. 
   
   \item[$A_2(\hat{\lambda})$:] $(x, \Omega_m, \Omega_0, \lambda)=\left(\frac{\sqrt{\frac{2}{3}} (p-3 \gamma )}{(3 \gamma -4) g\left(\hat{\lambda
   }\right)}, \frac{2 (6-p) (p-3 \gamma )}{3 (4-3 \gamma )^2
   g(\hat{\lambda})^2}, \frac{2 (p-3 \gamma ) (\gamma
   -2)}{(4-3 \gamma )^2 g(\hat{\lambda})^2}+1, \hat{\lambda}\right)$. Exists for $\frac{2 (6-p) (p-3 \gamma )}{3 (4-3 \gamma )^2
   g(\hat{\lambda})^2}\geq 0$. The eigenvalues are \\
   $\Bigg\{-\frac{1}{4} \left(6-p-\frac{\sqrt{p-6} \sqrt{3 (4-3 \gamma )^2
   g(\hat{\lambda})^2 (-8 \gamma +3 p-2)+16 (\gamma -2) (p-3
   \gamma )^2}}{(3 \gamma -4) g\left(\hat{\lambda
   }\right)}\right)$,\\
   $-\frac{1}{4} \left(6-p+\frac{\sqrt{p-6} \sqrt{3 (4-3 \gamma )^2
   g(\hat{\lambda})^2 (-8 \gamma +3 p-2)+16 (\gamma -2) (p-3
   \gamma )^2}}{(3 \gamma -4) g\left(\hat{\lambda
   }\right)}\right), p-\frac{2 \hat{\lambda } (p-3 \gamma )}{(3 \gamma
   -4) g(\hat{\lambda})},-\frac{2 (p-3 \gamma )
   f'(\hat{\lambda})}{(3 \gamma -4) g\left(\hat{\lambda
   }\right)}\Bigg\}$. 
\\ For $1\leq \gamma< 2$, we deduce that
$A_2(\hat{\lambda})$ is a sink for 
\begin{enumerate}
    \item $1<\gamma <\frac{4}{3}, \;  3 \gamma <p\leq \frac{2}{3} (4 \gamma +1), \;  g(\hat{\lambda })<-\sqrt{2} \sqrt{\frac{(\gamma -2) (3 \gamma -p)}{(3 \gamma -4)^2}}, \;  f'(\hat{\lambda })>0, \; 
   \hat{\lambda }>\frac{(3 \gamma -4) p g(\hat{\lambda })}{2 (p-3 \gamma )}$, or 
   \item $1<\gamma <\frac{4}{3}, \;  3 \gamma <p\leq \frac{2}{3} (4 \gamma +1), \;  g(\hat{\lambda })>\sqrt{2}
   \sqrt{\frac{(\gamma -2) (3 \gamma -p)}{(3 \gamma -4)^2}}, \;  f'(\hat{\lambda })<0, \;  \hat{\lambda }<\frac{(3 \gamma -4) p g(\hat{\lambda })}{2 (p-3 \gamma )}$, or 
   \item $1<\gamma <\frac{4}{3}, \; 
   \frac{2}{3} (4 \gamma +1)<p<6, \;  -\frac{4 \sqrt{-\frac{(\gamma -2) (p-3 \gamma )^2}{(3 \gamma -4)^2 (-8 \gamma +3 p-2)}}}{\sqrt{3}}\leq g(\hat{\lambda })<-\sqrt{2} \sqrt{\frac{(\gamma -2) (3 \gamma -p)}{(3 \gamma
   -4)^2}}, \;  f'(\hat{\lambda })>0, \;  \hat{\lambda }>\frac{(3 \gamma -4) p g(\hat{\lambda })}{2 (p-3 \gamma )}$, or 
   \item $1<\gamma <\frac{4}{3}, \;  \frac{2}{3} (4 \gamma +1)<p<6, \;  \sqrt{2}
   \sqrt{\frac{(\gamma -2) (3 \gamma -p)}{(3 \gamma -4)^2}}<g(\hat{\lambda })\leq \frac{4 \sqrt{-\frac{(\gamma -2) (p-3 \gamma )^2}{(3 \gamma -4)^2 (-8 \gamma +3 p-2)}}}{\sqrt{3}}, \;  f'\left(\hat{\lambda
   }\right)<0, \;  \hat{\lambda }<\frac{(3 \gamma -4) p g(\hat{\lambda })}{2 (p-3 \gamma )}$, or 
   \item $\frac{4}{3}<\gamma <2, \;  3 \gamma <p\leq \frac{2}{3} (4 \gamma +1), \;  g\left(\hat{\lambda
   }\right)<-\sqrt{2} \sqrt{\frac{(\gamma -2) (3 \gamma -p)}{(3 \gamma -4)^2}}, \;  f'(\hat{\lambda })<0, \;  \hat{\lambda }<\frac{(3 \gamma -4) p g(\hat{\lambda })}{2 (p-3 \gamma )}$, or
   \item $\frac{4}{3}<\gamma <2, \;  3 \gamma <p\leq \frac{2}{3} (4 \gamma +1), \;  g(\hat{\lambda })>\sqrt{2} \sqrt{\frac{(\gamma -2) (3 \gamma -p)}{(3 \gamma -4)^2}}, \;  f'(\hat{\lambda })>0, \; 
   \hat{\lambda }>\frac{(3 \gamma -4) p g(\hat{\lambda })}{2 (p-3 \gamma )}$, or 
   \item $\frac{4}{3}<\gamma <2, \;  \frac{2}{3} (4 \gamma +1)<p<6, \;  -\frac{4 \sqrt{-\frac{(\gamma -2) (p-3 \gamma )^2}{(3 \gamma
   -4)^2 (-8 \gamma +3 p-2)}}}{\sqrt{3}}\leq g(\hat{\lambda })<-\sqrt{2} \sqrt{\frac{(\gamma -2) (3 \gamma -p)}{(3 \gamma -4)^2}}, \;  f'(\hat{\lambda })<0, \;  \hat{\lambda }<\frac{(3 \gamma -4) p
   g(\hat{\lambda })}{2 (p-3 \gamma )}$, or 
   \item $\frac{4}{3}<\gamma <2, \;  \frac{2}{3} (4 \gamma +1)<p<6, \;  \sqrt{2} \sqrt{\frac{(\gamma -2) (3 \gamma -p)}{(3 \gamma -4)^2}}<g(\hat{\lambda })\leq
   \frac{4 \sqrt{-\frac{(\gamma -2) (p-3 \gamma )^2}{(3 \gamma -4)^2 (-8 \gamma +3 p-2)}}}{\sqrt{3}}, \;  f'(\hat{\lambda })>0, \;  \hat{\lambda }>\frac{(3 \gamma -4) p g(\hat{\lambda })}{2 (p-3 \gamma)}$.
  
\end{enumerate}
It is nonhyperbolic for $p=6$, saddle for $p=2$. If exists, it will be never a source. 
   
   \item[$A_3(\hat{\lambda})$:] $(x, \Omega_m, \Omega_0, \lambda)=\left(\frac{\sqrt{6} \gamma }{(4-3 \gamma ) g(\hat{\lambda})+2
   \hat{\lambda }}, \frac{2 (4-3 \gamma ) g(\hat{\lambda})
   \hat{\lambda }+4 \left(\hat{\lambda }^2-3 \gamma \right)}{\left((3
   \gamma -4) g(\hat{\lambda})-2 \hat{\lambda }\right)^2}, 0
   , \hat{\lambda}\right)$. Assuming $1\leq \gamma \leq 2$, it exists for 
   \begin{enumerate}
       \item $1\leq \gamma \leq 2, \; p>0, \; \hat{\lambda }\leq \frac{1}{4} \left(-4 g(\hat{\lambda })-\sqrt{48 \gamma +9 \gamma ^2 g(\hat{\lambda })^2-24 \gamma  g(\hat{\lambda })^2+16 g(\hat{\lambda })^2}+3 \gamma 
   g(\hat{\lambda })\right)$, or 
    \item $1\leq \gamma \leq 2, \; p>0, \; \hat{\lambda }\geq \frac{1}{4} \left(-4 g(\hat{\lambda })+\sqrt{48 \gamma +9 \gamma ^2 g(\hat{\lambda })^2-24 \gamma 
   g(\hat{\lambda })^2+16 g(\hat{\lambda })^2}+3 \gamma  g(\hat{\lambda })\right)$.
   \end{enumerate}
  The eigenvalues are
   ${\left\{\frac{p (4-3 \gamma ) g(\hat{\lambda})+2 (p-3 \gamma ) \hat{\lambda }}{(-4+3 \gamma ) g(\hat{\lambda})-2
\hat{\lambda }},\frac{1}{(-8+6 \gamma ) g(\hat{\lambda})-4 \hat{\lambda }}\left((12-9 \gamma ) g(\hat{\lambda})-\right.\right.}\\
{3 (-2+\gamma ) \hat{\lambda }-\sqrt{3} \surd \left(2 (-4+3 \gamma )^3 g(\hat{\lambda})^3 \hat{\lambda }+(4-3 \gamma )^2 g(\hat{\lambda})^2 \left(3+12 \gamma -8 \hat{\lambda }^2\right)-\right.}\\
{\left.\left.2 (-4+3 \gamma ) g(\hat{\lambda}) \hat{\lambda } \left(6-3 \gamma +6 \gamma ^2-4 \hat{\lambda }^2\right)-3 (-2+\gamma
) \left(24 \gamma ^2+(2-9 \gamma ) \hat{\lambda }^2\right)\right)\right),}\\
{\frac{1}{(-8+6 \gamma ) g(\hat{\lambda})-4 \hat{\lambda }}\left((12-9 \gamma ) g(\hat{\lambda})-3 (-2+\gamma ) \hat{\lambda
}+\right.}\\
{\sqrt{3} \surd \left(2 (-4+3 \gamma )^3 g(\hat{\lambda})^3 \hat{\lambda }+(4-3 \gamma )^2 g(\hat{\lambda})^2 \left(3+12
\gamma -8 \hat{\lambda }^2\right)-2 (-4+3 \gamma ) g(\hat{\lambda}) \hat{\lambda } \right.}\\
{\left.\left.\left.\left(6-3 \gamma +6 \gamma ^2-4 \hat{\lambda }^2\right)-3 (-2+\gamma ) \left(24 \gamma ^2+(2-9 \gamma ) \hat{\lambda }^2\right)\right)\right),\frac{6
\gamma  f'(\hat{\lambda})}{(-4+3 \gamma ) g(\hat{\lambda})-2 \hat{\lambda }}\right\}}$. \\
In this case, we can proceed semi-analytically, that is, for non-minimal coupling $g\equiv 0$, the eigenvalues reduce to \\
$\Bigg\{-\frac{-3 \sqrt{(2-\gamma ) \left(24 \gamma ^2+(2-9 \gamma ) \hat{\lambda }^2\right)}-3 (\gamma -2) \hat{\lambda }}{4 \hat{\lambda }},-\frac{3 \sqrt{(2-\gamma ) \left(24 \gamma ^2+(2-9 \gamma ) \hat{\lambda
   }^2\right)}-3 (\gamma -2) \hat{\lambda }}{4 \hat{\lambda }}$,\\
   $3 \gamma -p, -\frac{3 \gamma  f'(\hat{\lambda })}{\hat{\lambda }}\Bigg\}$. 
Hence, assuming $1\leq \gamma<2$, $A_3(\hat{\lambda})$ is a sink for 
 \begin{enumerate}
     \item $1\leq \gamma <2, p>3 \gamma, -\frac{2 \sqrt{6} \gamma }{\sqrt{9 \gamma -2}}\leq \hat{\lambda }<-\sqrt{3} \sqrt{\gamma }, f'(\hat{\lambda })<0$, or 
     \item $1\leq \gamma <2, p>3 \gamma,
   \sqrt{3} \sqrt{\gamma }<\hat{\lambda }\leq \frac{2 \sqrt{6} \gamma }{\sqrt{9 \gamma -2}}, f'(\hat{\lambda })>0$. 
 \end{enumerate}
  Otherwise, it is a saddle. 
 For non-minimal coupling the analysis is more complicate, so we rely on numerical elaboration. 
  
   \item[$A_4(\hat{\lambda})$:] $(x, \Omega_m, \Omega_0, \lambda)=\left(-1, 0, 0,\hat{\lambda}\right)$. The eigenvalues are\\
   $\left\{6-p,-3 \gamma +\sqrt{\frac{3}{2}} (3 \gamma -4)
   g(\hat{\lambda})+6,\sqrt{6} \hat{\lambda }+6,\sqrt{6}
   f'(\hat{\lambda})\right\}$. 
   
   $A_4(\hat{\lambda})$ is a source for 
   \begin{enumerate}
       \item $1\leq \gamma <\frac{4}{3},  0\leq p<6,  g(\hat{\lambda })<\frac{\sqrt{6} (\gamma -2)}{3 \gamma -4},  f'(\hat{\lambda })>0,  \hat{\lambda }>-\sqrt{6}$, or 
       \item $\gamma =\frac{4}{3},   0\leq p<6,  f'(\hat{\lambda })>0,  \hat{\lambda }>-\sqrt{6}$, or 
   \item $\frac{4}{3}<\gamma <2,  0\leq p<6,  g(\hat{\lambda })>\frac{\sqrt{6} (\gamma -2)}{3 \gamma -4}, 
   f'(\hat{\lambda })>0,  \hat{\lambda }>-\sqrt{6}$.
   \end{enumerate}

   $A_4(\hat{\lambda})$ is a sink for
   \begin{enumerate}
       \item $1\leq \gamma <\frac{4}{3}, p>6, g(\hat{\lambda })>\frac{\sqrt{6} (\gamma -2)}{3 \gamma -4}, f'(\hat{\lambda })<0, \hat{\lambda }<-\sqrt{6}$, or 
       \item $\frac{4}{3}<\gamma <2, p>6, g(\hat{\lambda })<\frac{\sqrt{6} (\gamma -2)}{3 \gamma -4}, f'(\hat{\lambda })<0, \hat{\lambda }<-\sqrt{6}$.
   \end{enumerate}
   
   \item[$A_5(\hat{\lambda})$:] $(x, \Omega_m, \Omega_0, \lambda)=\left(0, 0, 1, \hat{\lambda}\right)$. The eigenvalues are 
   $\left\{0,\frac{p-6}{2},p,p-3 \gamma \right\}$.  Nonhyperbolic, 3D unstable manifold for $p>6$.
   
   \item[$A_6(\hat{\lambda})$:] $(x, \Omega_m, \Omega_0, \lambda)=\left(1, 0, 0, \hat{\lambda}\right)$. The eigenvalues are \\
   $\left\{6-p,-3 \gamma +\sqrt{\frac{3}{2}} (4-3 \gamma )
   g(\hat{\lambda})+6,6-\sqrt{6} \hat{\lambda },-\sqrt{6}
   f'(\hat{\lambda})\right\}$.
   
   $A_6(\hat{\lambda})$ is a source for
   \begin{enumerate}
       \item $1\leq \gamma <\frac{4}{3},  p<6,  g(\hat{\lambda })>-\frac{\sqrt{6} (\gamma -2)}{3 \gamma -4},  f'(\hat{\lambda })<0,  \hat{\lambda }<\sqrt{6}$, or 
       \item $\gamma =\frac{4}{3},  p<6,    f'(\hat{\lambda })<0,  \hat{\lambda }<\sqrt{6}$, or 
   \item $\frac{4}{3}<\gamma<2,  p<6,  g(\hat{\lambda })<-\frac{\sqrt{6} (\gamma -2)}{3 \gamma -4},  f'(\hat{\lambda })<0, 
   \hat{\lambda }<\sqrt{6}$.
   \end{enumerate}
   
   $A_6(\hat{\lambda})$ is a  sink for 
    \begin{enumerate}
        \item $1\leq \gamma <\frac{4}{3},  p>6,  g(\hat{\lambda })<-\frac{\sqrt{6} (\gamma -2)}{3 \gamma -4},  f'(\hat{\lambda })>0,  \hat{\lambda }>\sqrt{6}$, or 
        \item $1\leq \gamma >\frac{4}{3},  p>6,   g(\hat{\lambda })>-\frac{\sqrt{6} (\gamma -2)}{3 \gamma -4},  f'(\hat{\lambda })>0,  \hat{\lambda }>\sqrt{6}$.
    \end{enumerate}
   
   \item[$A_7(\hat{\lambda})$:] $(x, \Omega_m, \Omega_0, \lambda)=\left(\frac{p}{\sqrt{6} \hat{\lambda }}, 0, 1-\frac{p}{\hat{\lambda }^2} , \hat{\lambda}\right)$. Exists for $\hat{\lambda }<0, 0\leq p\leq 6$, or $\hat{\lambda }>0, 0\leq p\leq 6$.  The eigenvalues are $\Bigg\{-\frac{1}{4} \left(6-p-\frac{\sqrt{\left(p-6\right)\left(\hat{\lambda }^2 (9 p-6)-8 p^2\right)}}{\hat{\lambda
   }}\right)$,\\
   $-\frac{1}{4} \left(6-p+\frac{\sqrt{\left(p-6\right)\left(\hat{\lambda }^2 (9 p-6)-8 p^2\right)}}{\hat{\lambda
   }}\right), -3 \gamma +\frac{(4-3 \gamma ) p g(\hat{\lambda})}{2
   \hat{\lambda }}+p,-\frac{p
   f'(\hat{\lambda})}{\hat{\lambda }}\Bigg\}$.

For $1\leq \gamma\leq 2$, $A_7(\hat{\lambda})$, it is a sink for:
   \begin{enumerate}
   \item $1\leq \gamma <\frac{4}{3}, \;  \hat{\lambda }\leq -\sqrt{6}, \;  0<p\leq \frac{1}{16} \left(9 \hat{\lambda }^2-\sqrt{3} \sqrt{\hat{\lambda }^2 \left(27 \hat{\lambda }^2-64\right)}\right), \;  g(\hat{\lambda})>\frac{2 \hat{\lambda } (p-3 \gamma )}{(3 \gamma -4) p}, \;  f'(\hat{\lambda })<0$, or 
   \item $1\leq \gamma <\frac{4}{3}, \;  \hat{\lambda }\geq \sqrt{6}, \;  0<p\leq \frac{1}{16} \left(9 \hat{\lambda
   }^2-\sqrt{3} \sqrt{\hat{\lambda }^2 \left(27 \hat{\lambda }^2-64\right)}\right), \;  g(\hat{\lambda })<\frac{2 \hat{\lambda } (p-3 \gamma )}{(3 \gamma -4) p}, \;  f'(\hat{\lambda })>0$, or
   \item $1\leq \gamma <\frac{4}{3}, \;  \hat{\lambda }>\frac{8}{3 \sqrt{3}}, \;  0<p\leq \frac{1}{16} \left(9 \hat{\lambda }^2-\sqrt{3} \sqrt{\hat{\lambda }^2 \left(27 \hat{\lambda }^2-64\right)}\right), \;  g(\hat{\lambda})<\frac{2 \hat{\lambda } (p-3 \gamma )}{(3 \gamma -4) p}, \;  f'(\hat{\lambda })>0$, or 
   \item $1\leq \gamma <\frac{4}{3}, \;  -\sqrt{6}<\hat{\lambda }<-\frac{8}{3 \sqrt{3}}, \;  0<p\leq \frac{1}{16}
   \left(9 \hat{\lambda }^2-\sqrt{3} \sqrt{\hat{\lambda }^2 \left(27 \hat{\lambda }^2-64\right)}\right), \;  g(\hat{\lambda })>\frac{2 \hat{\lambda } (p-3 \gamma )}{(3 \gamma -4) p}, \;  f'\left(\hat{\lambda
   }\right)<0$, or 
   \item $1\leq \gamma <\frac{4}{3}, \;  -\frac{8}{3 \sqrt{3}}\leq \hat{\lambda }<0, \;  0<p<\hat{\lambda }^2, \;  g(\hat{\lambda })>\frac{2 \hat{\lambda } (p-3 \gamma )}{(3 \gamma -4) p}, \; 
   f'(\hat{\lambda })<0$, or 
   \item $1\leq \gamma <\frac{4}{3}, \;  -\sqrt{6}<\hat{\lambda }<-\frac{8}{3 \sqrt{3}}, \;  \frac{1}{16} \left(9 \hat{\lambda }^2+\sqrt{3} \sqrt{\hat{\lambda }^2 \left(27 \hat{\lambda
   }^2-64\right)}\right)\leq p<\hat{\lambda }^2, \;  g(\hat{\lambda })>\frac{2 \hat{\lambda } (p-3 \gamma )}{(3 \gamma -4) p}, \;  f'(\hat{\lambda })<0$, or 
   \item $1\leq \gamma <\frac{4}{3}, \; 
   0<\hat{\lambda }\leq \frac{8}{3 \sqrt{3}}, \;  0<p<\hat{\lambda }^2, \;  g(\hat{\lambda })<\frac{2 \hat{\lambda } (p-3 \gamma )}{(3 \gamma -4) p}, \;  f'(\hat{\lambda })>0$, or 
   \item $1\leq \gamma
   <\frac{4}{3}, \;  \frac{8}{3 \sqrt{3}}<\hat{\lambda }<\sqrt{6}, \;  \frac{1}{16} \left(9 \hat{\lambda }^2+\sqrt{3} \sqrt{\hat{\lambda }^2 \left(27 \hat{\lambda }^2-64\right)}\right)\leq p<\hat{\lambda }^2, \; 
   g(\hat{\lambda })<\frac{2 \hat{\lambda } (p-3 \gamma )}{(3 \gamma -4) p}, \;  f'(\hat{\lambda })>0$, or 
   \item $\gamma =\frac{4}{3}, \;  \hat{\lambda }\geq \frac{8}{\sqrt{15}}, \;  0<p\leq
   \frac{1}{16} \left(9 \hat{\lambda }^2-\sqrt{3} \sqrt{\hat{\lambda }^2 \left(27 \hat{\lambda }^2-64\right)}\right), \;  f'(\hat{\lambda })>0$, or 
   \item $\gamma =\frac{4}{3}, \;  \hat{\lambda }>2, \;  0<p\leq
   \frac{1}{16} \left(9 \hat{\lambda }^2-\sqrt{3} \sqrt{\hat{\lambda }^2 \left(27 \hat{\lambda }^2-64\right)}\right), \;  f'(\hat{\lambda })>0$, or 
   \item $\gamma =\frac{4}{3}, \;  \hat{\lambda }\leq
   -\frac{8}{\sqrt{15}}, \;  0<p\leq \frac{1}{16} \left(9 \hat{\lambda }^2-\sqrt{3} \sqrt{\hat{\lambda }^2 \left(27 \hat{\lambda }^2-64\right)}\right), \;  f'(\hat{\lambda })<0$, or 
   \item $\gamma =\frac{4}{3}, \; 
   \frac{8}{3 \sqrt{3}}<\hat{\lambda }\leq 2, \;  0<p\leq \frac{1}{16} \left(9 \hat{\lambda }^2-\sqrt{3} \sqrt{\hat{\lambda }^2 \left(27 \hat{\lambda }^2-64\right)}\right), \;  f'(\hat{\lambda })>0$, or \item $\gamma =\frac{4}{3}, \;  0<\hat{\lambda }\leq \frac{8}{3 \sqrt{3}}, \;  0<p<\hat{\lambda }^2, \;  f'(\hat{\lambda })>0$, or 
   \item $\gamma =\frac{4}{3}, \;  2<\hat{\lambda }<\frac{8}{\sqrt{15}}, \; 
   \frac{1}{16} \left(9 \hat{\lambda }^2+\sqrt{3} \sqrt{\hat{\lambda }^2 \left(27 \hat{\lambda }^2-64\right)}\right)\leq p<4, \;  f'(\hat{\lambda })>0$, or 
   \item $\gamma =\frac{4}{3}, \;  \frac{8}{3
   \sqrt{3}}<\hat{\lambda }\leq 2, \;  \frac{1}{16} \left(9 \hat{\lambda }^2+\sqrt{3} \sqrt{\hat{\lambda }^2 \left(27 \hat{\lambda }^2-64\right)}\right)\leq p<\hat{\lambda }^2, \;  f'(\hat{\lambda })>0$, or
   \item $\gamma =\frac{4}{3}, \;  -\frac{8}{\sqrt{15}}<\hat{\lambda }<-2, \;  0<p\leq \frac{1}{16} \left(9 \hat{\lambda }^2-\sqrt{3} \sqrt{\hat{\lambda }^2 \left(27 \hat{\lambda }^2-64\right)}\right), \;  f'(\hat{\lambda})<0$, or 
   \item $\gamma =\frac{4}{3}, \;  -2\leq \hat{\lambda }<-\frac{8}{3 \sqrt{3}}, \;  0<p\leq \frac{1}{16} \left(9 \hat{\lambda }^2-\sqrt{3} \sqrt{\hat{\lambda }^2 \left(27 \hat{\lambda
   }^2-64\right)}\right), \;  f'(\hat{\lambda })<0$, or 
   \item $\gamma =\frac{4}{3}, \;  -\frac{8}{3 \sqrt{3}}\leq \hat{\lambda }<0, \;  0<p<\hat{\lambda }^2, \;  f'(\hat{\lambda })<0$, or
   \item $\gamma =\frac{4}{3}, \;  -\frac{8}{\sqrt{15}}<\hat{\lambda }<-2, \;  \frac{1}{16} \left(9 \hat{\lambda }^2+\sqrt{3} \sqrt{\hat{\lambda }^2 \left(27 \hat{\lambda }^2-64\right)}\right)\leq p<4, \;  f'\left(\hat{\lambda
   }\right)<0$, or 
   \item $\gamma =\frac{4}{3}, \;  -2\leq \hat{\lambda }<-\frac{8}{3 \sqrt{3}}, \;  \frac{1}{16} \left(9 \hat{\lambda }^2+\sqrt{3} \sqrt{\hat{\lambda }^2 \left(27 \hat{\lambda }^2-64\right)}\right)\leq
   p<\hat{\lambda }^2, \;  f'(\hat{\lambda })<0$, or 
   \item $\frac{4}{3}<\gamma <2, \;  \hat{\lambda }\geq \sqrt{6}, \;  0<p\leq \frac{1}{16} \left(9 \hat{\lambda }^2-\sqrt{3} \sqrt{\hat{\lambda }^2 \left(27
   \hat{\lambda }^2-64\right)}\right), \;  g(\hat{\lambda })>\frac{2 \hat{\lambda } (p-3 \gamma )}{(3 \gamma -4) p}, \;  f'(\hat{\lambda })>0$, or 
   \item $\frac{4}{3}<\gamma <2, \;  \hat{\lambda }=2
   \sqrt{6} \sqrt{\frac{\gamma ^2}{9 \gamma -2}}, \;  0<p\leq \frac{1}{16} \left(9 \hat{\lambda }^2-\sqrt{3} \sqrt{\hat{\lambda }^2 \left(27 \hat{\lambda }^2-64\right)}\right), \;  g(\hat{\lambda })>\frac{2
   \hat{\lambda } (p-3 \gamma )}{(3 \gamma -4) p}, \;  f'(\hat{\lambda })>0$, or 
   \item $\frac{4}{3}<\gamma <2, \;  \hat{\lambda }>2 \sqrt{6} \sqrt{\frac{\gamma ^2}{9 \gamma -2}}, \;  0<p\leq \frac{1}{16} \left(9
   \hat{\lambda }^2-\sqrt{3} \sqrt{\hat{\lambda }^2 \left(27 \hat{\lambda }^2-64\right)}\right), \;  g(\hat{\lambda })>\frac{2 \hat{\lambda } (p-3 \gamma )}{(3 \gamma -4) p}, \;  f'\left(\hat{\lambda
   }\right)>0$, or 
   \item $\frac{4}{3}<\gamma <2, \;  \hat{\lambda }=2 \sqrt{6} \sqrt{\frac{\gamma ^2}{9 \gamma -2}}, \;  3 \gamma \leq p<\hat{\lambda }^2, \;  g(\hat{\lambda })>\frac{2 \hat{\lambda } (p-3 \gamma
   )}{(3 \gamma -4) p}, \;  f'(\hat{\lambda })>0$, or 
   \item $\frac{4}{3}<\gamma <2, \;  \hat{\lambda }\leq -\sqrt{6}, \;  0<p\leq \frac{1}{16} \left(9 \hat{\lambda }^2-\sqrt{3} \sqrt{\hat{\lambda }^2 \left(27
   \hat{\lambda }^2-64\right)}\right), \;  g(\hat{\lambda })<\frac{2 \hat{\lambda } (p-3 \gamma )}{(3 \gamma -4) p}, \;  f'(\hat{\lambda })<0$, or 
   \item $\frac{4}{3}<\gamma <2, \;  \hat{\lambda }=-2
   \sqrt{6} \sqrt{\frac{\gamma ^2}{9 \gamma -2}}, \;  0<p\leq \frac{1}{16} \left(9 \hat{\lambda }^2-\sqrt{3} \sqrt{\hat{\lambda }^2 \left(27 \hat{\lambda }^2-64\right)}\right), \;  g(\hat{\lambda })<\frac{2
   \hat{\lambda } (p-3 \gamma )}{(3 \gamma -4) p}, \;  f'(\hat{\lambda })<0$, or 
   \item $\frac{4}{3}<\gamma <2, \;  \hat{\lambda }=-2 \sqrt{6} \sqrt{\frac{\gamma ^2}{9 \gamma -2}}, \;  3 \gamma \leq p<\hat{\lambda
   }^2, \;  g(\hat{\lambda })<\frac{2 \hat{\lambda } (p-3 \gamma )}{(3 \gamma -4) p}, \;  f'(\hat{\lambda })<0$, or 
   \item $\frac{4}{3}<\gamma <2, \;  \frac{8}{3 \sqrt{3}}<\hat{\lambda }<2 \sqrt{6}
   \sqrt{\frac{\gamma ^2}{9 \gamma -2}}, \;  0<p\leq \frac{1}{16} \left(9 \hat{\lambda }^2-\sqrt{3} \sqrt{\hat{\lambda }^2 \left(27 \hat{\lambda }^2-64\right)}\right), \;  g(\hat{\lambda })>\frac{2 \hat{\lambda } (p-3
   \gamma )}{(3 \gamma -4) p}, \;  f'(\hat{\lambda })>0$, or 
   \item $\frac{4}{3}<\gamma <2, \;  0<\hat{\lambda }\leq \frac{8}{3 \sqrt{3}}, \;  0<p<\hat{\lambda }^2, \;  g(\hat{\lambda })>\frac{2
   \hat{\lambda } (p-3 \gamma )}{(3 \gamma -4) p}, \;  f'(\hat{\lambda })>0$, or 
   \item $\frac{4}{3}<\gamma <2, \;  \frac{8}{3 \sqrt{3}}<\hat{\lambda }<2 \sqrt{6} \sqrt{\frac{\gamma ^2}{9 \gamma -2}}, \; 
   \frac{1}{16} \left(9 \hat{\lambda }^2+\sqrt{3} \sqrt{\hat{\lambda }^2 \left(27 \hat{\lambda }^2-64\right)}\right)\leq p<\hat{\lambda }^2, \;  g(\hat{\lambda })>\frac{2 \hat{\lambda } (p-3 \gamma )}{(3 \gamma -4)
   p}, \;  f'(\hat{\lambda })>0$, or 
   \item $\frac{4}{3}<\gamma <2, \;  2 \sqrt{6} \sqrt{\frac{\gamma ^2}{9 \gamma -2}}<\hat{\lambda }<\sqrt{6}, \;  \frac{1}{16} \left(9 \hat{\lambda }^2+\sqrt{3} \sqrt{\hat{\lambda
   }^2 \left(27 \hat{\lambda }^2-64\right)}\right)\leq p<\hat{\lambda }^2, \;  g(\hat{\lambda })>\frac{2 \hat{\lambda } (p-3 \gamma )}{(3 \gamma -4) p}, \;  f'(\hat{\lambda })>0$, or
   \item $\frac{4}{3}<\gamma <2, \;  -\sqrt{6}<\hat{\lambda }<-2 \sqrt{6} \sqrt{\frac{\gamma ^2}{9 \gamma -2}}, \;  0<p\leq \frac{1}{16} \left(9 \hat{\lambda }^2-\sqrt{3} \sqrt{\hat{\lambda }^2 \left(27 \hat{\lambda
   }^2-64\right)}\right), \;  g(\hat{\lambda })<\frac{2 \hat{\lambda } (p-3 \gamma )}{(3 \gamma -4) p}, \;  f'(\hat{\lambda })<0$, or 
   \item $\frac{4}{3}<\gamma <2, \;  -2 \sqrt{6} \sqrt{\frac{\gamma
   ^2}{9 \gamma -2}}<\hat{\lambda }<-\frac{8}{3 \sqrt{3}}, \;  0<p\leq \frac{1}{16} \left(9 \hat{\lambda }^2-\sqrt{3} \sqrt{\hat{\lambda }^2 \left(27 \hat{\lambda }^2-64\right)}\right), \;  g(\hat{\lambda })<\frac{2
   \hat{\lambda } (p-3 \gamma )}{(3 \gamma -4) p}, \;  f'(\hat{\lambda })<0$, or 
   \item $\frac{4}{3}<\gamma <2, \;  -\frac{8}{3 \sqrt{3}}\leq \hat{\lambda }<0, \;  0<p<\hat{\lambda }^2, \;  g\left(\hat{\lambda
   }\right)<\frac{2 \hat{\lambda } (p-3 \gamma )}{(3 \gamma -4) p}, \;  f'(\hat{\lambda })<0$, or 
   \item $\frac{4}{3}<\gamma <2, \;  -\sqrt{6}<\hat{\lambda }<-2 \sqrt{6} \sqrt{\frac{\gamma ^2}{9 \gamma -2}}, \; 
   \frac{1}{16} \left(9 \hat{\lambda }^2+\sqrt{3} \sqrt{\hat{\lambda }^2 \left(27 \hat{\lambda }^2-64\right)}\right)\leq p<\hat{\lambda }^2, \;  g(\hat{\lambda })<\frac{2 \hat{\lambda } (p-3 \gamma )}{(3 \gamma -4)
   p}, \;  f'(\hat{\lambda })<0$, or 
   \item $\frac{4}{3}<\gamma <2, \;  -2 \sqrt{6} \sqrt{\frac{\gamma ^2}{9 \gamma -2}}<\hat{\lambda }<-\frac{8}{3 \sqrt{3}}, \;  \frac{1}{16} \left(9 \hat{\lambda }^2+\sqrt{3}
   \sqrt{\hat{\lambda }^2 \left(27 \hat{\lambda }^2-64\right)}\right)\leq p<\hat{\lambda }^2, \;  g(\hat{\lambda })<\frac{2 \hat{\lambda } (p-3 \gamma )}{(3 \gamma -4) p}, \;  f'(\hat{\lambda })<0$, or
   \item $\gamma =2, \;  \hat{\lambda }\geq \sqrt{6}, \;  0<p\leq \frac{1}{16} \left(9 \hat{\lambda }^2-\sqrt{3} \sqrt{\hat{\lambda }^2 \left(27 \hat{\lambda }^2-64\right)}\right), \;  g(\hat{\lambda})>\frac{\hat{\lambda } (p-6)}{p}, \;  f'(\hat{\lambda })>0$, or 
   \item $\gamma =2, \;  \hat{\lambda }>\frac{8}{3 \sqrt{3}}, \;  0<p\leq \frac{1}{16} \left(9 \hat{\lambda }^2-\sqrt{3} \sqrt{\hat{\lambda
   }^2 \left(27 \hat{\lambda }^2-64\right)}\right), \;  g(\hat{\lambda })>\frac{\hat{\lambda } (p-6)}{p}, \;  f'(\hat{\lambda })>0$, or 
   \item $\gamma =2, \;  \hat{\lambda }\leq -\sqrt{6}, \;  0<p\leq
   \frac{1}{16} \left(9 \hat{\lambda }^2-\sqrt{3} \sqrt{\hat{\lambda }^2 \left(27 \hat{\lambda }^2-64\right)}\right), \;  g(\hat{\lambda })<\frac{\hat{\lambda } (p-6)}{p}, \;  f'(\hat{\lambda })<0$, or
   \item $\gamma =2, \;  0<\hat{\lambda }\leq \frac{8}{3 \sqrt{3}}, \;  0<p<\hat{\lambda }^2, \;  g(\hat{\lambda })>\frac{\hat{\lambda } (p-6)}{p}, \;  f'(\hat{\lambda })>0$, or 
   \item $\gamma =2, \; 
   \frac{8}{3 \sqrt{3}}<\hat{\lambda }<\sqrt{6}, \;  \frac{1}{16} \left(9 \hat{\lambda }^2+\sqrt{3} \sqrt{\hat{\lambda }^2 \left(27 \hat{\lambda }^2-64\right)}\right)\leq p<\hat{\lambda }^2, \;  g\left(\hat{\lambda
   }\right)>\frac{\hat{\lambda } (p-6)}{p}, \;  f'(\hat{\lambda })>0$, or 
   \item $\gamma =2, \;  -\sqrt{6}<\hat{\lambda }<-\frac{8}{3 \sqrt{3}}, \;  0<p\leq \frac{1}{16} \left(9 \hat{\lambda }^2-\sqrt{3}
   \sqrt{\hat{\lambda }^2 \left(27 \hat{\lambda }^2-64\right)}\right), \;  g(\hat{\lambda })<\frac{\hat{\lambda } (p-6)}{p}, \;  f'(\hat{\lambda })<0$, or 
   \item $\gamma =2, \;  -\frac{8}{3 \sqrt{3}}\leq
   \hat{\lambda }<0, \;  0<p<\hat{\lambda }^2, \;  g(\hat{\lambda })<\frac{\hat{\lambda } (p-6)}{p}, \;  f'(\hat{\lambda })<0$, or 
   \item $\gamma =2, \;  -\sqrt{6}<\hat{\lambda }<-\frac{8}{3
   \sqrt{3}}, \;  \frac{1}{16} \left(9 \hat{\lambda }^2+\sqrt{3} \sqrt{\hat{\lambda }^2 \left(27 \hat{\lambda }^2-64\right)}\right)\leq p<\hat{\lambda }^2, \;  g(\hat{\lambda })<\frac{\hat{\lambda } (p-6)}{p}, \;    f'(\hat{\lambda })<0$.
   \end{enumerate}
   It is never a source.
   
   \item[$A_8(\hat{\lambda})$:] $(x, \Omega_m, \Omega_0, \lambda)=\left(\frac{\hat{\lambda }}{\sqrt{6}}, 0, 0, \hat{\lambda}\right)$. Exists for $\hat{\lambda }^2\leq 6$. The eigenvalues are \\
   $\left\{\frac{1}{2} \left(\hat{\lambda }^2-6\right),\hat{\lambda }^2-p,-3
   \gamma +\frac{1}{2} (4-3 \gamma ) \hat{\lambda } g(\hat{\lambda
   })+\hat{\lambda }^2,-\hat{\lambda } f'(\hat{\lambda
   })\right\}$.
   
  $A_8(\hat{\lambda})$ is a sink for 
   \begin{enumerate}
       \item $1\leq \gamma <\frac{4}{3},  p>\hat{\lambda }^2,  f'(\hat{\lambda })>0,  0<\hat{\lambda }<\sqrt{6},  g(\hat{\lambda })<\frac{6 \gamma -2 \hat{\lambda }^2}{4 \hat{\lambda }-3 \gamma  \hat{\lambda}}$, or 
   \item $1\leq \gamma <\frac{4}{3},  p>\hat{\lambda }^2,  g(\hat{\lambda })>\frac{6 \gamma -2 \hat{\lambda }^2}{4 \hat{\lambda }-3 \gamma  \hat{\lambda }},  -\sqrt{6}<\hat{\lambda }<0, 
   f'(\hat{\lambda })<0$, or 
   \item $\gamma=\frac{4}{3},  p>\hat{\lambda }^2,  -2<\hat{\lambda }<0,  f'(\hat{\lambda })<0$, or 
   \item $\gamma=\frac{4}{3},  p>\hat{\lambda }^2,  f'\hat{\lambda})>0,  0<\hat{\lambda }<2$, or 
   \item $\frac{4}{3}<\gamma <2,  p>\hat{\lambda }^2,  g(\hat{\lambda })>\frac{6 \gamma -2 \hat{\lambda }^2}{4 \hat{\lambda }-3 \gamma  \hat{\lambda }}, 
   f'(\hat{\lambda })>0,  0<\hat{\lambda }<\sqrt{6}$, or 
   \item $\frac{4}{3}<\gamma <2,  p>\hat{\lambda }^2,  -\sqrt{6}<\hat{\lambda }<0,  g(\hat{\lambda })<\frac{6 \gamma -2 \hat{\lambda
   }^2}{4 \hat{\lambda }-3 \gamma  \hat{\lambda }},  f'(\hat{\lambda })<0$.
   \end{enumerate}
 It is never a source.
   
   \item[$A_9$:] $(x, \Omega_m, \Omega_0, \lambda)=\left(0, 0, 0, 0\right)$. The eigenvalues are\\
   $\left\{-p,-3 \gamma ,-\frac{1}{2} \left(3+\sqrt{9-12 f(0)}\right),-\frac{1}{2} \left(3-\sqrt{9-12 f(0)}\right)\right\}$. Sink for $p>0, \gamma>0, f(0)>0$. Otherwise, it is a saddle. 
   
   \item[$A_{10}(\tilde{\lambda})$:] $(x, \Omega_m, \Omega_0, \lambda)=\left(0, 1, 0, \tilde{\lambda}\right)$, where we denote by  $\tilde{\lambda}$, the values of $\lambda$ for which  $g(\lambda)=0$. The eigenvalues are
   $\Bigg\{\frac{1}{4} \left(3 \gamma -6-\sqrt{24 (4-3 \gamma )
   f(\tilde{\lambda }) g'(\tilde{\lambda })+9
   (\gamma -2)^2}\right)$,\\
   $\frac{1}{4} \left(3 \gamma -6+\sqrt{24 (4-3
   \gamma ) f(\tilde{\lambda }) g'(\tilde{\lambda
   })+9 (\gamma -2)^2}\right), 3 \gamma ,3 \gamma -p\Bigg\}$. It is a saddle. 
\end{enumerate}

\section{Existence and stability conditions of the equilibrium points of system \eqref{31-syst} as  $\phi\rightarrow \infty$}
\label{AppB}
The equilibrium points of system \eqref{31-syst} with $\varphi=0$  (i.e., corresponding to $\phi\rightarrow \infty$) are the following:
\begin{enumerate}
    \item[$B_1$:] $\left(0,\tan ^{-1}\left[\sqrt{1-\frac{N^2}{6}},-\frac{N}{\sqrt{6}}\right]+2 \pi  c_1,0,0,0\right), c_1\in \mathbb{Z}$, with eigenvalues \\ $\left\{0,\frac{N^2}{2},\frac{1}{2} \left(N^2-6\right),N^2-p,N (2 M+N)-\frac{3}{2} \gamma  (M N+2)\right\}$. It exists for $N^2\leq 6$. It is always a nonhyperbolic saddle for $N^2\leq 6$.
    
    \item[$B_2$:] $\left(1,\tan ^{-1}\left[\sqrt{1-\frac{N^2}{6}},-\frac{N}{\sqrt{6}}\right]+2 \pi  c_1,0,0,0\right), c_1\in \mathbb{Z}$, with eigenvalues \\ $\left\{0,-\frac{N^2}{2},\frac{1}{2} \left(N^2-6\right),N^2-p,N (2 M+N)-\frac{3}{2} \gamma  (M N+2)\right\}$. It exists for $N^2\leq 6$. The case of physical interest is when it is nonhyperbolic with a 4D stable manifold for 
    \begin{enumerate}
        \item $M\in \mathbb{R},   -2<N<0,   p>N^2,   \gamma =\frac{4}{3}$, or 
        \item $M\in \mathbb{R},   0<N<2,   p>N^2,   \gamma =\frac{4}{3}$, or
        \item $-\sqrt{6}<N\leq -2,   p>N^2,   1\leq \gamma <\frac{4}{3},   M>\frac{2 (N^2-3\gamma)}{N(3 \gamma  -4)}$, or
        \item $-2<N<0,   p>N^2,   1\leq \gamma <\frac{4}{3},   M>\frac{2 (N^2-3\gamma)}{N(3 \gamma  -4)}$, or 
        \item $0<N<2,   p>N^2,   \frac{4}{3}<\gamma \leq 2,   M>\frac{2 (N^2-3\gamma)}{N(3 \gamma  -4)}$, or 
        \item $2\leq N<\sqrt{6},   p>N^2,   \frac{4}{3}<\gamma \leq 2,   M>\frac{2 (N^2-3\gamma)}{N(3 \gamma  -4)}$, or
        \item $-\sqrt{6}<N\leq -2,   p>N^2,   \frac{4}{3}<\gamma \leq 2,   M<\frac{2 (N^2-3\gamma)}{N(3 \gamma  -4)}$, or 
        \item $0<N<2,   p>N^2,   1\leq \gamma <\frac{4}{3},   M<\frac{2 (N^2-3\gamma)}{N(3 \gamma  -4)}$, or 
        \item $2\leq N<\sqrt{6},   p>N^2,   1\leq \gamma <\frac{4}{3},   M<\frac{2 (N^2-3\gamma)}{N(3 \gamma  -4)}$, or 
        \item $-2<N<0,   p>N^2,   \frac{4}{3}<\gamma \leq 2,   M<\frac{2 (N^2-3 \gamma)}{N (3 \gamma-4)}$.
    \end{enumerate}
    
        \item[$B_3$:] $\left(0,2 \pi  c_1,0,1,0\right), c_1\in \mathbb{Z}$, with eigenvalues  $\left\{0,\frac{p-6}{2},\frac{p}{2},p,p-3 \gamma \right\}$. The physical interesting situation is when it is nonhyperbolic with a 4D unstable manifold for $p>6, 1\leq \gamma \leq 2$. It is a nonhyperbolic saddle otherwise. 
        
        \item[$B_4$:] $\left(1,2 \pi  c_1,0,1,0\right), c_1\in \mathbb{Z}$, with eigenvalues $\left\{0,\frac{p-6}{2},-\frac{p}{2},p,p-3 \gamma \right\}$.  It is a nonhyperbolic saddle. 
    
        \item[$B_5$:] $\left(0,\tan ^{-1}\left[\sqrt{1-\frac{p^2}{6 N^2}},-\frac{p}{\sqrt{6} N}\right]+2 \pi  c_1,0,1-\frac{p}{N^2},0\right), c_1\in \mathbb{Z}$, with eigenvalues \\ $\Big\{0,\frac{p}{2},\frac{p ((4-3 \gamma ) M+2 N)}{2 N}-3 \gamma, \frac{1}{4} \left(p-6+\frac{\sqrt{\left(p-6 \right)\left(N^2 (9 p-6)-8 p^2\right)}}{N}\right)$, \\
    $\frac{1}{4} \left(p-6-\frac{\sqrt{\left(p-6 \right)\left(N^2 (9 p-6)-8 p^2\right)}}{N}\right) \Big\}$. It exists for $p>0, N^2>\frac{p^2}{6}$. Furthermore, it satisfies $\Omega_0\geq 0$  for $0<p<6, N^2\geq p$, or $p\geq 6, N^2>\frac{p^2}{6}$. It is a nonhyperbolic saddle. 
    
        \item[$B_6$:] $\left(1,\tan ^{-1}\left[\sqrt{1-\frac{p^2}{6 N^2}},-\frac{p}{\sqrt{6} N}\right]+2 \pi  c_1,0,1-\frac{p}{N^2},0\right), c_1\in \mathbb{Z}$, with eigenvalues \\
   $\Big\{0,-\frac{p}{2},\frac{p ((4-3 \gamma ) M+2 N)}{2 N}-3 \gamma,\frac{1}{4} \left(p-6+\frac{\sqrt{\left(p-6\right)\left(N^2 (9 p-6)-8 p^2\right)}}{N}\right)$,\\
   $\frac{1}{4} \left(p-6-\frac{\sqrt{\left(p-6\right)\left(N^2 (9 p-6)-8 p^2\right)}}{N}\right) \Big\}$. It exists for $p>0, N^2>\frac{p^2}{6}$. Furthermore, it satisfies $\Omega_0\geq 0$  for $0<p<6, N^2\geq p$, or $p\geq 6, N^2>\frac{p^2}{6}$. The situation of physical interest is when it is nohyperbolic with a 4D stable manifold for 
   \begin{enumerate}
    \item $N>\frac{8}{\sqrt{15}}, \; 0<p\leq \frac{1}{16} \left(9 N^2-\sqrt{3} \sqrt{N^2 \left(27 N^2-64\right)}\right), \; \gamma =\frac{4}{3}$, or 
       
    \item $N<-\frac{8}{3 \sqrt{3}}, \; 0<p\leq \frac{1}{16} \left(9 N^2-\sqrt{3}
   \sqrt{N^2 \left(27 N^2-64\right)}\right), \; \gamma =\frac{4}{3}$, or 
   
   \item $N=\frac{8}{\sqrt{15}}, \; 0<p\leq \frac{4}{5}, \; \gamma =\frac{4}{3}$, or 
   
   \item $0<N\leq \frac{8}{3 \sqrt{3}}, \; 0<p<N^2, \; \gamma
   =\frac{4}{3}$, or 
   
   \item $-\frac{8}{3 \sqrt{3}}\leq N<0, \; 0<p<N^2, \; \gamma =\frac{4}{3}$, or 
   
   \item $\frac{8}{3 \sqrt{3}}<N<\frac{8}{\sqrt{15}}, \; 0<p\leq \frac{1}{16} \left(9 N^2-\sqrt{3} \sqrt{N^2 \left(27
   N^2-64\right)}\right), \; \gamma =\frac{4}{3}$, or 
   
   \item $\frac{8}{3 \sqrt{3}}<N\leq 2, \; \frac{1}{16} \left(9 N^2+\sqrt{3} \sqrt{N^2 \left(27 N^2-64\right)}\right)\leq p<N^2, \; \gamma =\frac{4}{3}$, or 
   
   \item $
   -2<N<-\frac{8}{3 \sqrt{3}}, \; \frac{1}{16} \left(9 N^2+\sqrt{3} \sqrt{N^2 \left(27 N^2-64\right)}\right)\leq p<N^2, \; \gamma =\frac{4}{3}$, or 
   
   \item $2<N<\frac{8}{\sqrt{15}}, \; \frac{1}{16} \left(9
   N^2+\sqrt{3} \sqrt{N^2 \left(27 N^2-64\right)}\right)\leq p<4, \; \gamma =\frac{4}{3}$, or 
   
   \item $-\frac{8}{\sqrt{15}}<N\leq -2, \; \frac{1}{16} \left(9 N^2+\sqrt{3} \sqrt{N^2 \left(27 N^2-64\right)}\right)\leq p<4, \;
   \gamma =\frac{4}{3}$, or 
   
   \item $\frac{8}{3 \sqrt{3}}<N<\frac{8}{\sqrt{15}}, \; 0<p\leq \frac{1}{16} \left(9 N^2-\sqrt{3} \sqrt{N^2 \left(27 N^2-64\right)}\right), \; 1\leq \gamma <\frac{4}{3}, \; M<\frac{2 N (p-3
   \gamma )}{(3 \gamma -4) p}$, or 
   
   \item $\frac{8}{3 \sqrt{3}}<N<\frac{8}{\sqrt{15}}, \; 0<p\leq \frac{1}{16} \left(9 N^2-\sqrt{3} \sqrt{N^2 \left(27 N^2-64\right)}\right), \; \frac{4}{3}<\gamma \leq 2, \; M>\frac{2 N
   (p-3 \gamma )}{(3 \gamma -4) p}$, or 
   
   \item $N<-\frac{8}{3 \sqrt{3}}, \; 0<p\leq \frac{1}{16} \left(9 N^2-\sqrt{3} \sqrt{N^2 \left(27 N^2-64\right)}\right), \; 1\leq \gamma <\frac{4}{3}, \; M>\frac{2 N (p-3 \gamma )}{(3
   \gamma -4) p}$, or 
   
   \item $N<-\frac{8}{3 \sqrt{3}}, \; 0<p\leq \frac{1}{16} \left(9 N^2-\sqrt{3} \sqrt{N^2 \left(27 N^2-64\right)}\right), \; \frac{4}{3}<\gamma \leq 2, \; M<\frac{2 N (p-3 \gamma )}{(3 \gamma -4)
   p}$, or 
   
   \item $N>\frac{8}{\sqrt{15}}, \; 0<p\leq \frac{1}{16} \left(9 N^2-\sqrt{3} \sqrt{N^2 \left(27 N^2-64\right)}\right), \; 1\leq \gamma <\frac{4}{3}, \; M<\frac{2 N (p-3 \gamma )}{(3 \gamma -4) p}$, or 
   
   \item $N>\frac{8}{\sqrt{15}}, \; 0<p\leq \frac{1}{16} \left(9 N^2-\sqrt{3} \sqrt{N^2 \left(27 N^2-64\right)}\right), \; \frac{4}{3}<\gamma \leq 2, \; M>\frac{2 N (p-3 \gamma )}{(3 \gamma -4) p}$, or 
   
   \item $N=\frac{8}{\sqrt{15}}, \; 0<p\leq \frac{4}{5}, \; 1\leq \gamma <\frac{4}{3}, \; M<\frac{16 (p-3 \gamma )}{\sqrt{15} (3 \gamma -4) p}$, or 
   
   \item $N=\frac{8}{\sqrt{15}}, \; 0<p\leq \frac{4}{5}, \;
   \frac{4}{3}<\gamma \leq 2, \; M>\frac{16 (p-3 \gamma )}{\sqrt{15} (3 \gamma -4) p}$, or 
   
   \item $N=\frac{8}{\sqrt{15}}, \; 4\leq p<\frac{64}{15}, \; 1\leq \gamma <\frac{4}{3}, \; M<\frac{16 (p-3 \gamma )}{\sqrt{15} (3
   \gamma -4) p}$, or 
   
   \item $N=\frac{8}{\sqrt{15}}, \; 4\leq p<\frac{64}{15}, \; \frac{4}{3}<\gamma \leq 2, \; M>\frac{16 (p-3 \gamma )}{\sqrt{15} (3 \gamma -4) p}$, or 
   
   \item $-2<N<-\frac{8}{3 \sqrt{3}}, \;
   \frac{1}{16} \left(9 N^2+\sqrt{3} \sqrt{N^2 \left(27 N^2-64\right)}\right)\leq p<N^2, \; 1\leq \gamma <\frac{4}{3}, \; M>\frac{2 N (p-3 \gamma )}{(3 \gamma -4) p}$, or 
   
   \item $-\sqrt{6}<N\leq -\frac{8}{\sqrt{15}}, \;
   \frac{1}{16} \left(9 N^2+\sqrt{3} \sqrt{N^2 \left(27 N^2-64\right)}\right)\leq p<N^2, \; 1\leq \gamma <\frac{4}{3}, \; M>\frac{2 N (p-3 \gamma )}{(3 \gamma -4) p}$, or 
   
   \item $\frac{8}{3 \sqrt{3}}<N\leq 2, \;
   \frac{1}{16} \left(9 N^2+\sqrt{3} \sqrt{N^2 \left(27 N^2-64\right)}\right)\leq p<N^2, \; \frac{4}{3}<\gamma \leq 2, \; M>\frac{2 N (p-3 \gamma )}{(3 \gamma -4) p}$, or 
   
   \item $\frac{8}{\sqrt{15}}<N<\sqrt{6}, \;
   \frac{1}{16} \left(9 N^2+\sqrt{3} \sqrt{N^2 \left(27 N^2-64\right)}\right)\leq p<N^2, \; \frac{4}{3}<\gamma \leq 2, \; M>\frac{2 N (p-3 \gamma )}{(3 \gamma -4) p}$, or 
   
   \item $\frac{8}{3 \sqrt{3}}<N\leq 2, \;
   \frac{1}{16} \left(9 N^2+\sqrt{3} \sqrt{N^2 \left(27 N^2-64\right)}\right)\leq p<N^2, \; 1\leq \gamma <\frac{4}{3}, \; M<\frac{2 N (p-3 \gamma )}{(3 \gamma -4) p}$, or 
   
   \item $\frac{8}{\sqrt{15}}<N<\sqrt{6}, \;
   \frac{1}{16} \left(9 N^2+\sqrt{3} \sqrt{N^2 \left(27 N^2-64\right)}\right)\leq p<N^2, \; 1\leq \gamma <\frac{4}{3}, \; M<\frac{2 N (p-3 \gamma )}{(3 \gamma -4) p}$, or 
   
   \item $-2<N<-\frac{8}{3 \sqrt{3}}, \; \frac{1}{16}
   \left(9 N^2+\sqrt{3} \sqrt{N^2 \left(27 N^2-64\right)}\right)\leq p<N^2, \; \frac{4}{3}<\gamma \leq 2, \; M<\frac{2 N (p-3 \gamma )}{(3 \gamma -4) p}$, or 
   
   \item $-\sqrt{6}<N\leq -\frac{8}{\sqrt{15}}, \; \frac{1}{16}
   \left(9 N^2+\sqrt{3} \sqrt{N^2 \left(27 N^2-64\right)}\right)\leq p<N^2, \; \frac{4}{3}<\gamma \leq 2, \; M<\frac{2 N (p-3 \gamma )}{(3 \gamma -4) p}$, or 
   
   \item $2<N<\frac{8}{\sqrt{15}}, \; \frac{1}{16} \left(9
   N^2+\sqrt{3} \sqrt{N^2 \left(27 N^2-64\right)}\right)\leq p<4, \; \frac{4}{3}<\gamma \leq 2, \; M>\frac{2 N (p-3 \gamma )}{(3 \gamma -4) p}$, or 
   
   \item $-\frac{8}{\sqrt{15}}<N\leq -2, \; \frac{1}{16} \left(9
   N^2+\sqrt{3} \sqrt{N^2 \left(27 N^2-64\right)}\right)\leq p<4, \; 1\leq \gamma <\frac{4}{3}, \; M>\frac{2 N (p-3 \gamma )}{(3 \gamma -4) p}$, or 
   
   \item $2<N<\frac{8}{\sqrt{15}}, \; \frac{1}{16} \left(9 N^2+\sqrt{3}
   \sqrt{N^2 \left(27 N^2-64\right)}\right)\leq p<4, \; 1\leq \gamma <\frac{4}{3}, \; M<\frac{2 N (p-3 \gamma )}{(3 \gamma -4) p}$, or 
   
   \item $-\frac{8}{\sqrt{15}}<N\leq -2, \; \frac{1}{16} \left(9 N^2+\sqrt{3} \sqrt{N^2
   \left(27 N^2-64\right)}\right)\leq p<4, \; \frac{4}{3}<\gamma \leq 2, \; M<\frac{2 N (p-3 \gamma )}{(3 \gamma -4) p}$, or 
   
   \item $-\frac{8}{3 \sqrt{3}}\leq N<0, \; 0<p<N^2, \; 1\leq \gamma <\frac{4}{3}, \; M>\frac{2 N
   (p-3 \gamma )}{(3 \gamma -4) p}$, or 
   
   \item $0<N\leq \frac{8}{3 \sqrt{3}}, \; 0<p<N^2, \; \frac{4}{3}<\gamma \leq 2, \; M>\frac{2 N (p-3 \gamma )}{(3 \gamma -4) p}$, or 
   
   \item $2<N<\frac{8}{\sqrt{15}}, \; 4\leq
   p<N^2, \; \frac{4}{3}<\gamma \leq 2, \; M>\frac{2 N (p-3 \gamma )}{(3 \gamma -4) p}$, or 
   
   \item $-\frac{8}{\sqrt{15}}<N<-2, \; 4\leq p<N^2, \; 1\leq \gamma <\frac{4}{3}, \; M>\frac{2 N (p-3 \gamma )}{(3 \gamma -4)
   p}$, or 
   
   \item $-\frac{8}{3 \sqrt{3}}\leq N<0, \; 0<p<N^2, \; \frac{4}{3}<\gamma \leq 2, \; M<\frac{2 N (p-3 \gamma )}{(3 \gamma -4) p}$, or 
   
   \item $0<N\leq \frac{8}{3 \sqrt{3}}, \; 0<p<N^2, \; 1\leq \gamma
   <\frac{4}{3}, \; M<\frac{2 N (p-3 \gamma )}{(3 \gamma -4) p}$, or \item $2<N<\frac{8}{\sqrt{15}}, \; 4\leq p<N^2, \; 1\leq \gamma <\frac{4}{3}, \; M<\frac{2 N (p-3 \gamma )}{(3 \gamma -4) p}$, or 
   
   \item $-\frac{8}{\sqrt{15}}<N<-2, \; 4\leq p<N^2, \; \frac{4}{3}<\gamma \leq 2, \; M<\frac{2 N (p-3 \gamma )}{(3 \gamma -4) p}$.
   
   \end{enumerate}
    
    \item[$B_7$:] $\left(0,\tan ^{-1}\left[\sqrt{1-\frac{2 (p-3 \gamma )^2}{3 M^2(4-3 \gamma )^2}},\frac{2(p-3 \gamma)}{\sqrt{6}M(4 -3 \gamma)}\right]+2 \pi  c_1,\frac{2 (6-p) (p-3 \gamma )}{3 (4-3 \gamma )^2 M^2},\frac{(4-3 \gamma )^2
   M^2+2 (\gamma -2) (p-3 \gamma )}{(4-3 \gamma )^2 M^2},0\right), c_1\in \mathbb{Z}$, with eigenvalues \\ $\Big\{0, \frac{p}{2}, \frac{-3 \gamma  M p+4 M p-6 \gamma  N+2 N p}{M(4-3\gamma)},  \frac{1}{4} \left(p-6+\frac{\sqrt{\left(p-6\right) \left(3 (4-3 \gamma )^2 M^2 (-8 \gamma +3 p-2)+16 (\gamma -2) (p-3 \gamma )^2\right)}}{(3 \gamma -4) M}\right)$, \\ $\frac{1}{4} \left(p-6-\frac{\sqrt{\left(p-6\right) \left(3 (4-3 \gamma )^2 M^2 (-8 \gamma +3 p-2)+16 (\gamma -2) (p-3 \gamma )^2\right)}}{(3 \gamma -4) M}\right)\Big\}$. Exists for
   \begin{enumerate}
    \item $1\leq \gamma <\frac{4}{3}, \; p=3 \gamma , \; M>0$, or 
    \item $\frac{4}{3}<\gamma \leq 2, \; p=3 \gamma , \; M>0$, or 
    \item $1\leq \gamma <\frac{4}{3}, \; p=3 \gamma , \; M>0$, or
    \item $\frac{4}{3}<\gamma \leq 2, \; p=3 \gamma , \; M>0$, or 
    \item $1\leq \gamma <\frac{4}{3}, \; p=3 \gamma , \; M<0$, or 
    \item $\frac{4}{3}<\gamma \leq 2, \; p=3 \gamma , \; M<0$, or 
    \item $1\leq \gamma
   <\frac{4}{3}, \; p=3 \gamma , \; M<0$, or 
   \item $\frac{4}{3}<\gamma \leq 2, \; p=3 \gamma , \; M<0$, or 
   \item $1\leq \gamma <\frac{4}{3}, \; 3 \gamma <p\leq 6, \; M\geq \frac{\sqrt{\frac{2}{3}} (3 \gamma -p)}{3
   \gamma -4}$, or 
   \item $\frac{4}{3}<\gamma <2, \; 3 \gamma <p\leq 6, \; M\geq -\frac{\sqrt{\frac{2}{3}} (3 \gamma -p)}{3 \gamma -4}$, or 
   \item $1\leq \gamma <\frac{4}{3}, \; 3 \gamma <p\leq 6, \; M\geq
   \frac{\sqrt{\frac{2}{3}} (3 \gamma -p)}{3 \gamma -4}$, or 
   \item $\frac{4}{3}<\gamma <2, \; 3 \gamma <p\leq 6, \; M\geq -\frac{\sqrt{\frac{2}{3}} (3 \gamma -p)}{3 \gamma -4}$, or 
   \item $1\leq \gamma
   <\frac{4}{3}, \; 3 \gamma <p\leq 6, \; M\leq -\frac{\sqrt{\frac{2}{3}} (3 \gamma -p)}{3 \gamma -4}$, or 
   \item $\frac{4}{3}<\gamma <2, \; 3 \gamma <p\leq 6, \; M\leq \frac{\sqrt{\frac{2}{3}} (3 \gamma -p)}{3 \gamma
   -4}$, or 
   \item $1\leq \gamma <\frac{4}{3}, \; 3 \gamma <p\leq 6, \; M\leq -\frac{\sqrt{\frac{2}{3}} (3 \gamma -p)}{3 \gamma -4}$, or 
   \item $\frac{4}{3}<\gamma <2, \; 3 \gamma <p\leq 6, \; M\leq
   \frac{\sqrt{\frac{2}{3}} (3 \gamma -p)}{3 \gamma -4}$.
   \end{enumerate}
 It is a nonhyperbolic saddle.

    \item[$B_8$:] $\left(1,\tan ^{-1}\left[\sqrt{1-\frac{2 (p-3 \gamma )^2}{3 M^2(4-3 \gamma )^2}},\frac{2(p-3 \gamma)}{\sqrt{6}M(4 -3 \gamma)}\right]+2 \pi  c_1,\frac{2 (6-p) (p-3 \gamma )}{3 (4-3 \gamma )^2 M^2},\frac{(4-3 \gamma )^2
   M^2+2 (\gamma -2) (p-3 \gamma )}{(4-3 \gamma )^2 M^2},0\right), c_1\in \mathbb{Z}$, with eigenvalues \\
   $\Big\{0,-\frac{p}{2},\frac{-3 \gamma  M p+4 M p-6 \gamma  N+2 N p}{M(4-3\gamma)},\frac{1}{4} \left(p-6+\frac{\sqrt{\left(p-6\right) \left(3 (4-3 \gamma )^2 M^2 (-8 \gamma +3 p-2)+16 (\gamma -2) (p-3 \gamma )^2\right)}}{(3 \gamma -4)
   M}\right)$,\\
   $\frac{1}{4} \left(p-6-\frac{\sqrt{\left(p-6\right) \left(3 (4-3 \gamma )^2 M^2 (-8 \gamma +3 p-2)+16 (\gamma -2) (p-3 \gamma )^2\right)}}{(3 \gamma -4) M}\right)\Big\}$.
   Exists for
   \begin{enumerate}
    \item $1\leq \gamma <\frac{4}{3}, \; p=3 \gamma , \; M>0$, or 
    \item $\frac{4}{3}<\gamma \leq 2, \; p=3 \gamma , \; M>0$, or 
    \item $1\leq \gamma <\frac{4}{3}, \; p=3 \gamma , \; M>0$, or
    \item $\frac{4}{3}<\gamma \leq 2, \; p=3 \gamma , \; M>0$, or 
    \item $1\leq \gamma <\frac{4}{3}, \; p=3 \gamma , \; M<0$, or 
    \item $\frac{4}{3}<\gamma \leq 2, \; p=3 \gamma , \; M<0$, or 
    \item $1\leq \gamma
   <\frac{4}{3}, \; p=3 \gamma , \; M<0$, or 
   \item $\frac{4}{3}<\gamma \leq 2, \; p=3 \gamma , \; M<0$, or 
   \item $1\leq \gamma <\frac{4}{3}, \; 3 \gamma <p\leq 6, \; M\geq \frac{\sqrt{\frac{2}{3}} (3 \gamma -p)}{3
   \gamma -4}$, or 
   \item $\frac{4}{3}<\gamma <2, \; 3 \gamma <p\leq 6, \; M\geq -\frac{\sqrt{\frac{2}{3}} (3 \gamma -p)}{3 \gamma -4}$, or 
   \item $1\leq \gamma <\frac{4}{3}, \; 3 \gamma <p\leq 6, \; M\geq
   \frac{\sqrt{\frac{2}{3}} (3 \gamma -p)}{3 \gamma -4}$, or 
   \item $\frac{4}{3}<\gamma <2, \; 3 \gamma <p\leq 6, \; M\geq -\frac{\sqrt{\frac{2}{3}} (3 \gamma -p)}{3 \gamma -4}$, or 
   \item $1\leq \gamma
   <\frac{4}{3}, \; 3 \gamma <p\leq 6, \; M\leq -\frac{\sqrt{\frac{2}{3}} (3 \gamma -p)}{3 \gamma -4}$, or 
   \item $\frac{4}{3}<\gamma <2, \; 3 \gamma <p\leq 6, \; M\leq \frac{\sqrt{\frac{2}{3}} (3 \gamma -p)}{3 \gamma
   -4}$, or 
   \item $1\leq \gamma <\frac{4}{3}, \; 3 \gamma <p\leq 6, \; M\leq -\frac{\sqrt{\frac{2}{3}} (3 \gamma -p)}{3 \gamma -4}$, or 
   \item $\frac{4}{3}<\gamma <2, \; 3 \gamma <p\leq 6, \; M\leq
   \frac{\sqrt{\frac{2}{3}} (3 \gamma -p)}{3 \gamma -4}$.
   \end{enumerate}
 The situation of physical interest is when it is nohyperbolic with a 4D stable manifold for 
   \begin{enumerate}
    \item $N<-3 \sqrt{6}, \; p=\frac{1}{16} \left(27 N^2-\sqrt{3} \sqrt{N^2 \left(243 N^2-1216\right)}\right), \; \frac{4}{3}<\gamma <\frac{2 p \left(N^2-p\right)}{6 N^2-p^2}, \; \frac{2 N (p-3 \gamma )}{(3 \gamma -4)
   p}<M<-\sqrt{2} \sqrt{\frac{(\gamma -2) (3 \gamma -p)}{(3 \gamma -4)^2}}$, or 
   \item $N<-3 \sqrt{6}, \; N^2-\sqrt{N^2 \left(N^2-6\right)}<p\leq \frac{1}{8} \left(9 N^2-\sqrt{3} \sqrt{N^2 \left(27
   N^2-160\right)}\right), \; \gamma <\frac{2 p \left(N^2-p\right)}{6 N^2-p^2}, \; -\sqrt{2} \sqrt{\frac{(\gamma -2) (3 \gamma -p)}{(3 \gamma -4)^2}}<M<\frac{2 N (p-3 \gamma )}{(3 \gamma -4) p}$, or 
   \item $N<-3
   \sqrt{6}, \; \frac{1}{2} \left(3 N^2-\sqrt{3} \sqrt{N^2 \left(3 N^2-16\right)}\right)<p<\frac{1}{16} \left(27 N^2-\sqrt{3} \sqrt{N^2 \left(243 N^2-1216\right)}\right), \; \frac{4}{3}<\gamma <\frac{2 p \left(N^2-p\right)}{6
   N^2-p^2}, \; \frac{2 N (p-3 \gamma )}{(3 \gamma -4) p}<M<-\sqrt{2} \sqrt{\frac{(\gamma -2) (3 \gamma -p)}{(3 \gamma -4)^2}}$, or 
   \item $N<-3 \sqrt{6}, \; \frac{1}{8} \left(9 N^2-\sqrt{3} \sqrt{N^2 \left(27
   N^2-160\right)}\right)<p\leq \frac{1}{2} \left(3 N^2-\sqrt{3} \sqrt{N^2 \left(3 N^2-16\right)}\right), \; \frac{9 N^2 p-6 N^2-8 p^2}{4 \left(6 N^2-p^2\right)}\leq \gamma <\frac{2 p \left(N^2-p\right)}{6 N^2-p^2}, \;
   -\sqrt{2} \sqrt{\frac{(\gamma -2) (3 \gamma -p)}{(3 \gamma -4)^2}}<M<\frac{2 N (p-3 \gamma )}{(3 \gamma -4) p}$, or 
   \item $-3 \sqrt{6}\leq N<-\sqrt{6}, \; p=\frac{1}{16} \left(27 N^2-\sqrt{3} \sqrt{N^2 \left(243
   N^2-1216\right)}\right), \; \frac{4}{3}<\gamma <\frac{2 p \left(N^2-p\right)}{6 N^2-p^2}, \; \frac{2 N (p-3 \gamma )}{(3 \gamma -4) p}<M<-\sqrt{2} \sqrt{\frac{(\gamma -2) (3 \gamma -p)}{(3 \gamma -4)^2}}$, or 
   \item $-3
   \sqrt{6}\leq N<-\sqrt{6}, \; N^2-\sqrt{N^2 \left(N^2-6\right)}<p\leq \frac{1}{8} \left(9 N^2-\sqrt{3} \sqrt{N^2 \left(27 N^2-160\right)}\right), \; \gamma <\frac{2 p \left(N^2-p\right)}{6 N^2-p^2}, \; -\sqrt{2}
   \sqrt{\frac{(\gamma -2) (3 \gamma -p)}{(3 \gamma -4)^2}}<M<\frac{2 N (p-3 \gamma )}{(3 \gamma -4) p}$, or 
   \item $-3 \sqrt{6}\leq N<-\sqrt{6}, \; \frac{1}{2} \left(3 N^2-\sqrt{3} \sqrt{N^2 \left(3
   N^2-16\right)}\right)<p<\frac{1}{16} \left(27 N^2-\sqrt{3} \sqrt{N^2 \left(243 N^2-1216\right)}\right), \; \frac{4}{3}<\gamma <\frac{2 p \left(N^2-p\right)}{6 N^2-p^2}, \; \frac{2 N (p-3 \gamma )}{(3 \gamma -4)
   p}<M<-\sqrt{2} \sqrt{\frac{(\gamma -2) (3 \gamma -p)}{(3 \gamma -4)^2}}$, or 
   \item $-3 \sqrt{6}\leq N<-\sqrt{6}, \; \frac{1}{8} \left(9 N^2-\sqrt{3} \sqrt{N^2 \left(27 N^2-160\right)}\right)<p<\frac{1}{2} \left(3
   N^2-\sqrt{3} \sqrt{N^2 \left(3 N^2-16\right)}\right), \; \frac{9 N^2 p-6 N^2-8 p^2}{4 \left(6 N^2-p^2\right)}\leq \gamma <\frac{2 p \left(N^2-p\right)}{6 N^2-p^2}, \; -\sqrt{2} \sqrt{\frac{(\gamma -2) (3 \gamma -p)}{(3
   \gamma -4)^2}}<M<\frac{2 N (p-3 \gamma )}{(3 \gamma -4) p}$, or 
   \item $N>\sqrt{6}, \; p=\frac{1}{16} \left(27 N^2-\sqrt{3} \sqrt{N^2 \left(243 N^2-1216\right)}\right), \; \frac{4}{3}<\gamma <\frac{2 p
   \left(N^2-p\right)}{6 N^2-p^2}, \; \sqrt{2} \sqrt{\frac{(\gamma -2) (3 \gamma -p)}{(3 \gamma -4)^2}}<M<\frac{2 N (p-3 \gamma )}{(3 \gamma -4) p}$, or 
   \item $N>\sqrt{6}, \; N^2-\sqrt{N^2 \left(N^2-6\right)}<p\leq
   \frac{1}{8} \left(9 N^2-\sqrt{3} \sqrt{N^2 \left(27 N^2-160\right)}\right), \; \gamma <\frac{2 p \left(N^2-p\right)}{6 N^2-p^2}, \; \frac{2 N (p-3 \gamma )}{(3 \gamma -4) p}<M<\sqrt{2} \sqrt{\frac{(\gamma -2) (3 \gamma
   -p)}{(3 \gamma -4)^2}}$, or 
   \item $N>\sqrt{6}, \; \frac{1}{2} \left(3 N^2-\sqrt{3} \sqrt{N^2 \left(3 N^2-16\right)}\right)<p<\frac{1}{16} \left(27 N^2-\sqrt{3} \sqrt{N^2 \left(243 N^2-1216\right)}\right), \;
   \frac{4}{3}<\gamma <\frac{2 p \left(N^2-p\right)}{6 N^2-p^2}, \; \sqrt{2} \sqrt{\frac{(\gamma -2) (3 \gamma -p)}{(3 \gamma -4)^2}}<M<\frac{2 N (p-3 \gamma )}{(3 \gamma -4) p}$.
   \end{enumerate}
   
      \item[$B_9$:] $\left(0,\tan ^{-1}\left[ \frac{\sqrt{6 (2-\gamma )^2-(4-3 \gamma )^2 M^2}}{\sqrt{6}(2-\gamma )},\frac{(4-3 \gamma ) M}{\sqrt{6}(2-\gamma)}\right]+2 \pi  c_1,\frac{6 (2-\gamma )^2-(4-3 \gamma )^2 M^2}{6 (2-\gamma )^2},0,0\right), c_1\in \mathbb{Z}$, with eigenvalues \\
   $\Big\{0,\frac{6 (\gamma -2)^2-(4-3 \gamma )^2 M^2}{4 (\gamma -2)},-\frac{(4-3 \gamma )^2 M^2-6 (\gamma -2) \gamma }{4 (\gamma -2)},\gamma  \left(3-\frac{9 M^2}{2}\right)+\frac{2 M (N-M)}{\gamma -2}+3 M (M+N),-\frac{(4-3 \gamma
   )^2 M^2+2 (\gamma -2) (p-3 \gamma )}{2 (\gamma -2)}\Big\}$. Exists for 
   \begin{enumerate}
       \item $\gamma =\frac{4}{3}, \;  p>0$, or 
       \item $1\leq \gamma <\frac{4}{3}, \;  \frac{2 \sqrt{6}-\sqrt{6} \gamma }{3 \gamma -4}\leq M\leq \frac{\sqrt{6} \gamma -2 \sqrt{6}}{3 \gamma -4}, \;  p>0$, or
       \item $\frac{4}{3}<\gamma <2, \;  \frac{\sqrt{6} \gamma -2 \sqrt{6}}{3 \gamma -4}\leq M\leq \frac{2 \sqrt{6}-\sqrt{6} \gamma }{3 \gamma -4}, \;  p>0$.
   \end{enumerate}
 It is a nonhyperbolic saddle. 
   
      \item[$B_{10}$:] $\left(1,\tan ^{-1}\left[ \frac{\sqrt{6 (2-\gamma )^2-(4-3 \gamma )^2 M^2}}{\sqrt{6}(2-\gamma )},\frac{(4-3 \gamma ) M}{\sqrt{6}(2-\gamma)}\right]+2 \pi  c_1,\frac{6 (2-\gamma )^2-(4-3 \gamma )^2 M^2}{6 (2-\gamma )^2},0,0\right), c_1\in \mathbb{Z}$, with eigenvalues 
     $\Big\{0,\frac{6 (\gamma -2)^2-(4-3 \gamma )^2 M^2}{4 (\gamma -2)},\frac{(4-3
   \gamma )^2 M^2-6 (\gamma -2) \gamma }{4 (\gamma -2)},\gamma  \left(3-\frac{9 M^2}{2}\right)+\frac{2 M (N-M)}{\gamma -2}+3 M (M+N), -\frac{(4-3 \gamma )^2 M^2+2 (\gamma -2) (p-3 \gamma )}{2 (\gamma -2)}\Big\}$.
    Exists for 
   \begin{enumerate}
       \item $\gamma =\frac{4}{3}, \;  p>0$, or 
       \item $1\leq \gamma <\frac{4}{3}, \;  \frac{2 \sqrt{6}-\sqrt{6} \gamma }{3 \gamma -4}\leq M\leq \frac{\sqrt{6} \gamma -2 \sqrt{6}}{3 \gamma -4}, \;  p>0$, or
       \item $\frac{4}{3}<\gamma <2, \;  \frac{\sqrt{6} \gamma -2 \sqrt{6}}{3 \gamma -4}\leq M\leq \frac{2 \sqrt{6}-\sqrt{6} \gamma }{3 \gamma -4}, \;  p>0$.
   \end{enumerate}
  The situation of physical interest is when it is nohyperbolic with a 4D stable manifold for 
   \begin{enumerate}
    \item $1\leq \gamma <\frac{4}{3}, \;  N<-\sqrt{6}, \;  \frac{N}{3 \gamma -4}-\sqrt{\frac{6 \gamma ^2-12 \gamma +N^2}{(3 \gamma -4)^2}}<M<\frac{\sqrt{6} \gamma -2 \sqrt{6}}{3 \gamma -4}, \;  p>\frac{6 \gamma ^2-12 \gamma -9 \gamma
   ^2 M^2+24 \gamma  M^2-16 M^2}{2 \gamma -4}$, or 
   \item $\frac{4}{3}<\gamma <2, \;  N<-\sqrt{6}, \;  \frac{\sqrt{6} \gamma -2 \sqrt{6}}{3 \gamma -4}<M<\sqrt{\frac{6 \gamma ^2-12 \gamma +N^2}{(3 \gamma -4)^2}}+\frac{N}{3
   \gamma -4}, \;  p>\frac{6 \gamma ^2-12 \gamma -9 \gamma ^2 M^2+24 \gamma  M^2-16 M^2}{2 \gamma -4}$, or 
   \item $1\leq \gamma <\frac{4}{3}, \;  N>\sqrt{6}, \;  \frac{2 \sqrt{6}-\sqrt{6} \gamma }{3 \gamma -4}<M<\sqrt{\frac{6
   \gamma ^2-12 \gamma +N^2}{(3 \gamma -4)^2}}+\frac{N}{3 \gamma -4}, \;  p>\frac{6 \gamma ^2-12 \gamma -9 \gamma ^2 M^2+24 \gamma  M^2-16 M^2}{2 \gamma -4}$, or 
   \item $\frac{4}{3}<\gamma <2, \;  N>\sqrt{6}, \;  \frac{N}{3
   \gamma -4}-\sqrt{\frac{6 \gamma ^2-12 \gamma +N^2}{(3 \gamma -4)^2}}<M<\frac{2 \sqrt{6}-\sqrt{6} \gamma }{3 \gamma -4}, \;  p>\frac{6 \gamma ^2-12 \gamma -9 \gamma ^2 M^2+24 \gamma  M^2-16 M^2}{2 \gamma -4}$.
   \end{enumerate}
     
      \item[$B_{11}$:] $\left(0,\tan^{-1}\left[\sqrt{1-\frac{6 \gamma ^2}{\left(2 N+M (4-3 \gamma )\right)^2}},-\frac{\sqrt{6}
\gamma }{2 N+M (4-3 \gamma )}\right]+2 \pi  c_1, \frac{4 N (2 M+N)-6 (2+M N) \gamma }{(2 N+M (4-3 \gamma ))^2},0,0\right), c_1\in \mathbb{Z}$, with eigenvalues 
   ${\left\{0,\frac{3 N \gamma }{4 M+2 N-3 M \gamma },\frac{M p (4-3 \gamma )+2 N (p-3 \gamma )}{-2 N+M (-4+3 \gamma )},\right.}\\
{\frac{1}{2 (2 N+M (4-3 \gamma ))^2}(3 (2 N+M (4-3 \gamma )) (N (-2+\gamma )+M (-4+3 \gamma ))+}\\
{\sqrt{3} \surd \left((2 N+M (4-3 \gamma ))^2 \left(-M^2 \left(-3+8 N^2-12 \gamma \right) (4-3 \gamma )^2+2 M^3 N (-4+3 \gamma )^3+\right.\right.}\\
{\left.\left.\left.2 M N (-4+3 \gamma ) \left(-6+4 N^2+3 \gamma -6 \gamma ^2\right)-3 (-2+\gamma ) \left(N^2 (2-9 \gamma )+24 \gamma ^2\right)\right)\right)\right),}\\
{\frac{1}{2 (2 N+M (4-3 \gamma ))^2}(3 (2 N+M (4-3 \gamma )) (N (-2+\gamma )+M (-4+3 \gamma ))-}\\
{\sqrt{3} \surd \left((2 N+M (4-3 \gamma ))^2 \left(-M^2 \left(-3+8 N^2-12 \gamma \right) (4-3 \gamma )^2+2 M^3 N (-4+3 \gamma )^3+\right.\right.}\\
{\left.\left.\left.\left.2 M N (-4+3 \gamma ) \left(-6+4 N^2+3 \gamma -6 \gamma ^2\right)-3 (-2+\gamma ) \left(N^2 (2-9 \gamma )+24 \gamma ^2\right)\right)\right)\right)\right\}}$. Exists for 
\begin{enumerate}
   \item $M\in \mathbb{R}, \; \gamma =\frac{4}{3}, \; N<-2$, or 
   \item $M\in \mathbb{R}, \; \gamma =\frac{4}{3}, \; N>2$, or 
   \item $\frac{4}{3}<\gamma \leq 2, \; N=-\sqrt{6}, \; M>\frac{\sqrt{6} (\gamma -2)}{3 \gamma
   -4}$, or 
   \item $\frac{4}{3}<\gamma \leq 2, \; 0<N\leq \sqrt{6}, \; M<\frac{2 \left(N^2-3 \gamma \right)}{(3 \gamma -4) N}$, or 
   \item $\frac{4}{3}<\gamma \leq 2, \; N<-\sqrt{6}, \; M\geq \frac{\sqrt{6} \gamma +2
   N}{3 \gamma -4}$, or 
   \item $\frac{4}{3}<\gamma \leq 2, \; -\sqrt{6}<N<0, \; M>\frac{2 \left(N^2-3 \gamma \right)}{(3 \gamma -4) N}$, or 
   \item $\frac{4}{3}<\gamma \leq 2, \; N>\sqrt{6}, \; M\leq -\frac{\sqrt{6}
   \gamma -2 N}{3 \gamma -4}$, or 
   \item $1\leq \gamma <\frac{4}{3}, \; N=-\sqrt{6}, \; M<\frac{\sqrt{6} (\gamma -2)}{3 \gamma -4}$, or 
   \item $1\leq \gamma <\frac{4}{3}, \; N>\sqrt{6}, \; M\geq -\frac{\sqrt{6} \gamma
   -2 N}{3 \gamma -4}$, or 
   \item $1\leq \gamma <\frac{4}{3}, \; N<-\sqrt{6}, \; M\leq \frac{\sqrt{6} \gamma +2 N}{3 \gamma -4}$, or 
   \item $1\leq \gamma <\frac{4}{3}, \; 0<N\leq \sqrt{6}, \; M>\frac{2 \left(N^2-3
   \gamma \right)}{(3 \gamma -4) N}$, or 
   \item $1\leq \gamma <\frac{4}{3}, \; -\sqrt{6}<N<0, \; M<\frac{2 \left(N^2-3 \gamma \right)}{(3 \gamma -4) N}$
\end{enumerate}
It is an hyperbolic saddle. 
     \item[$B_{12}$:] $\left(1,\tan^{-1}\left[\sqrt{1-\frac{6 \gamma ^2}{\left(2 N+M (4-3 \gamma )\right)^2}},-\frac{\sqrt{6}
\gamma }{2 N+M (4-3 \gamma )}\right]+2 \pi  c_1,
\frac{4 N (2 M+N)-6 (2+M N) \gamma }{(2 N+M (4-3 \gamma ))^2},0,0\right), c_1\in \mathbb{Z}$, with eigenvalues 
${\left\{0,-\frac{3 N \gamma }{4 M+2 N-3 M \gamma },\frac{M p (4-3 \gamma )+2 N (p-3 \gamma )}{-2 N+M (-4+3 \gamma )},\right.}\\
{\frac{1}{4 N+M (8-6 \gamma )}(3 N (-2+\gamma )+3 M (-4+3 \gamma )+}\\
{\surd \left(-3 M^2 \left(-3+8 N^2-12 \gamma \right) (4-3 \gamma )^2+6 M^3 N (-4+3 \gamma )^3+\right.}\\
{\left.\left.\left.6 M N (-4+3 \gamma ) \left(-6+4 N^2+3 \gamma -6 \gamma ^2\right)-9 (-2+\gamma ) \left(N^2 (2-9 \gamma )+24 \gamma ^2\right)\right)\right)\right\},}\\
{\frac{1}{4 N+M (8-6 \gamma )}(3 N (-2+\gamma )+3 M (-4+3 \gamma )-}\\
{\surd \left(-3 M^2 \left(-3+8 N^2-12 \gamma \right) (4-3 \gamma )^2+6 M^3 N (-4+3 \gamma )^3+\right.}\\
{\left.\left.\left.6 M N (-4+3 \gamma ) \left(-6+4 N^2+3 \gamma -6 \gamma ^2\right)-9 (-2+\gamma ) \left(N^2 (2-9 \gamma )+24 \gamma ^2\right)\right)\right)\right\}}$.
 Exists for 
\begin{enumerate}
   \item $M\in \mathbb{R}, \; \gamma =\frac{4}{3}, \; N<-2$, or 
   \item $M\in \mathbb{R}, \; \gamma =\frac{4}{3}, \; N>2$, or 
   \item $\frac{4}{3}<\gamma \leq 2, \; N=-\sqrt{6}, \; M>\frac{\sqrt{6} (\gamma -2)}{3 \gamma
   -4}$, or 
   \item $\frac{4}{3}<\gamma \leq 2, \; 0<N\leq \sqrt{6}, \; M<\frac{2 \left(N^2-3 \gamma \right)}{(3 \gamma -4) N}$, or 
   \item $\frac{4}{3}<\gamma \leq 2, \; N<-\sqrt{6}, \; M\geq \frac{\sqrt{6} \gamma +2
   N}{3 \gamma -4}$, or 
   \item $\frac{4}{3}<\gamma \leq 2, \; -\sqrt{6}<N<0, \; M>\frac{2 \left(N^2-3 \gamma \right)}{(3 \gamma -4) N}$, or 
   \item $\frac{4}{3}<\gamma \leq 2, \; N>\sqrt{6}, \; M\leq -\frac{\sqrt{6}
   \gamma -2 N}{3 \gamma -4}$, or 
   \item $1\leq \gamma <\frac{4}{3}, \; N=-\sqrt{6}, \; M<\frac{\sqrt{6} (\gamma -2)}{3 \gamma -4}$, or 
   \item $1\leq \gamma <\frac{4}{3}, \; N>\sqrt{6}, \; M\geq -\frac{\sqrt{6} \gamma
   -2 N}{3 \gamma -4}$, or 
   \item $1\leq \gamma <\frac{4}{3}, \; N<-\sqrt{6}, \; M\leq \frac{\sqrt{6} \gamma +2 N}{3 \gamma -4}$, or 
   \item $1\leq \gamma <\frac{4}{3}, \; 0<N\leq \sqrt{6}, \; M>\frac{2 \left(N^2-3
   \gamma \right)}{(3 \gamma -4) N}$, or 
   \item $1\leq \gamma <\frac{4}{3}, \; -\sqrt{6}<N<0, \; M<\frac{2 \left(N^2-3 \gamma \right)}{(3 \gamma -4) N}$.
\end{enumerate}
The situation of physical interest is when it is nohyperbolic with a 4D stable manifold for 
\begin{enumerate}
    \item $\gamma =\frac{4}{3}, \;  2<N\leq 2.06559, \;  p>4$, or 
    \item $\gamma =\frac{4}{3}, \;  -2.06559\leq N<-2, \;  p>4$, or 
    \item $\gamma=1, \;  N=-2.42061, \;  M=M_{1,1}, \;  p>\frac{6 N}{M+2 N}$, or
   \item $\gamma=1, \;  N=2.42061, \;  M=-2.13014, \;  p>\frac{6 N}{M+2 N}$, or 
   \item $\gamma=1, \;  N=2.44949, \;  \frac{6\, -2 N^2}{N}<M\leq M_{1,3}, \;  p>\frac{6 N}{M+2
   N}$, or 
   \item $\gamma=1, \;  N\leq -2.44949, \;  M_{1,1}\leq M<-\sqrt{N^2-6}-N, \;  p>\frac{6 N}{M+2 N}$, or 
   \item $\gamma=1, \;  N>2.44949, \;  \sqrt{N^2-6}-N<M\leq
   M_{1,1}, \;  p>\frac{6 N}{M+2 N}$, or 
   \item $\gamma=1, \;  2.42061<N<2.44949, \;  M_{1,2}\leq M\leq M_{1,3}, \; 
   p>\frac{6 N}{M+2 N}$, or 
   \item $\gamma=1, \;  -2.44949<N<-2.42061, \;  M_{1,1}\leq M\leq M_{1,2}, \;  p>\frac{6 N}{M+2 N}$, or 
   \item $\gamma  =1, \;  -2.42061<N<0, \;  M_{1,1}\leq M<\frac{6\, -2 N^2}{N}, \;  p>\frac{6 N}{M+2 N}$, or 
   \item $\gamma=1, \;  -2.44949<N\leq -2.42061, \; 
   M_{1,3}\leq M<\frac{6\, -2 N^2}{N}, \;  p>\frac{6 N}{M+2 N}$, or 
   \item $\gamma=1, \;  0<N<2.44949, \;  \frac{6\, -2 N^2}{N}<M\leq M_{1,1}, \; 
   p>\frac{6 N}{M+2 N}$, or 
   \item $\gamma =1.62562, \;  N>2.44949, \;  M_{2,1}\leq M<\frac{N}{3\gamma -4 }- \sqrt{\frac{6 \gamma ^2-12\gamma +N^2}{(3\gamma -4 )^2}}, \; 
   p>\frac{6 \gamma  N}{M(4-3\gamma)+2 N}$, or 
   \item $1<\gamma <\frac{9}{8}, \;  N=N_{3,2}, \;  M=M_{2,1}, \;  p>\frac{6 \gamma  N}{M(4-3\gamma) +2
   N}$, or 
   \item $1<\gamma <\frac{9}{8}, \;  N=N_{3,3}, \;  M=M_{2,2}, \;  p>\frac{6 \gamma  N}{M(4-3\gamma)+2 N}$, or 
   \item $\gamma =1.62562, \; 
   N<N_{3,1}, \;  \sqrt{\frac{6 \gamma ^2-12\gamma +N^2}{(3\gamma -4 )^2}}+\frac{N}{3\gamma -4 }<M\leq M_{2,1}, \;  p>\frac{6 \gamma  N}{-3\gamma  M+4
   M+2 N}$, or 
   \item $\gamma =1.62562, \;  -2<N<0, \;  \frac{2 N^2-6 \gamma }{3\gamma  N-4 N}<M\leq M_{2,1}, \;  p>\frac{6 \gamma  N}{M(4-3\gamma)+2 N}$, or
   \item $\gamma =1.62562, \;  N_{3,1}\leq N<-2.44949, \;  \sqrt{\frac{6 \gamma ^2-12\gamma +N^2}{(3\gamma -4 )^2}}+\frac{N}{3\gamma -4 }<M\leq M_{2,3}, \; 
   p>\frac{6 \gamma  N}{M(4-3\gamma)+2 N}$, or 
   \item $\gamma =1.62562, \;  0<N<2.44949, \;  M_{2,1}\leq M<\frac{2 N^2-6 \gamma }{N(3 \gamma -4)}, \;  p>\frac{6 \gamma  N}{-3
   \gamma  M+4 M+2 N}$, or 
   \item $1<\gamma <\frac{9}{8}, \;  N=-2.44949, \;  M_{2,1}\leq M<\frac{2 N^2-6 \gamma }{N(3 \gamma -4)}, \;  p>\frac{6 \gamma  N}{M(4-3\gamma) +2
   N}$, or 
   \item $\frac{4}{3}<\gamma <1.62562, \;  N>2.44949, \;  M_{2,1}\leq M<\frac{N}{3\gamma -4 }-\sqrt{\frac{6 \gamma ^2-12\gamma +N^2}{(3\gamma -4 )^2}}, \;  p>\frac{6 \gamma 
   N}{M(4-3\gamma)+2 N}$, or 
   \item $1.62562<\gamma <2,   \;  N>2.44949, \;  M_{2,1}\leq M<\frac{N}{3\gamma -4 }-\sqrt{\frac{6 \gamma ^2-12\gamma +N^2}{(3\gamma -4 )^2}}, \; 
   p>\frac{6 \gamma  N}{M(4-3\gamma)+2 N}$, or 
   \item $1<\gamma <\frac{9}{8}, \;  N<-2.44949, \;  M_{2,1}\leq M<\frac{N}{3\gamma -4 }-\sqrt{\frac{6 \gamma ^2-12\gamma +N^2}{(3
   \gamma -4)^2}}, \;  p>\frac{6 \gamma  N}{M(4-3\gamma)+2 N}$, or 
   \item $\frac{9}{8}<\gamma <\frac{4}{3}, \;  N<-2.44949, \;  M_{2,1}\leq M<\frac{N}{3\gamma -4 }-\sqrt{\frac{6 \gamma
   ^2-12\gamma +N^2}{(3\gamma -4 )^2}}, \;  p>\frac{6 \gamma  N}{M(4-3\gamma)+2 N}$, or
   \item $1<\gamma <\frac{9}{8}, \;  N=N_{3,2}, \;  M_{2,3}\leq
   M<\frac{2 N^2-6 \gamma }{N(3 \gamma -4)}, \;  p>\frac{6 \gamma  N}{M(4-3\gamma)+2 N}$, or 
   \item $\gamma =1.62562, \;  -2.44949<N\leq -2, \;  \frac{2 N^2-6 \gamma }{N(3 \gamma -4)}<M\leq
   M_{2,3}, \;  p>\frac{6 \gamma  N}{M(4-3\gamma)+2 N}$, or 
   \item $1<\gamma <\frac{9}{8}, \;  N>N_{3,4}, \;  \sqrt{\frac{6 \gamma ^2-12\gamma +N^2}{(3
   \gamma -4)^2}}+\frac{N}{3\gamma -4 }<M\leq M_{2,1}, \;  p>\frac{6 \gamma  N}{M(4-3\gamma)+2 N}$, or 
   \item $1<\gamma <\frac{9}{8}, \;  N=N_{3,3}, \; 
   \frac{2 N^2-6 \gamma }{N(3 \gamma -4)}<M\leq M_{2,1}, \;  p>\frac{6 \gamma  N}{M(4-3\gamma)+2 N}$, or 
   \item $1<\gamma <\frac{9}{8}, \;  N=2.44949, \;  \frac{2 N^2-6 \gamma }{3
   \gamma  N-4 N}<M\leq M_{2,3}, \;  p>\frac{6 \gamma  N}{M(4-3\gamma)+2 N}$, or 
   \item $\frac{9}{8}<\gamma <\frac{4}{3}, \;  N>N_{3,4}, \;  \sqrt{\frac{6
   \gamma ^2-12\gamma +N^2}{(3\gamma -4 )^2}}+\frac{N}{3\gamma -4 }<M\leq M_{2,1}, \;  p>\frac{6 \gamma  N}{M(4-3\gamma)+2 N}$, or 
   \item $\frac{4}{3}<\gamma <1.62562, \; 
   N<N_{3,1}, \;  \sqrt{\frac{6 \gamma ^2-12\gamma +N^2}{(3\gamma -4 )^2}}+\frac{N}{3\gamma -4 }<M\leq M_{2,1}, \;  p>\frac{6 \gamma  N}{-3 \gamma  M+4
   M+2 N}$, or 
   \item $1.62562<\gamma <2,   \;  N<N_{3,1}, \;  \sqrt{\frac{6 \gamma ^2-12\gamma +N^2}{(3\gamma -4 )^2}}+\frac{N}{3\gamma -4 }<M\leq
   M_{2,1}, \;  p>\frac{6 \gamma  N}{M(4-3\gamma)+2 N}$, or 
   \item $\frac{4}{3}<\gamma <1.62562, \;  N_{3,1}\leq N<-2.44949, \;  \sqrt{\frac{6 \gamma
   ^2-12\gamma +N^2}{(3\gamma -4 )^2}}+\frac{N}{3\gamma -4 }<M\leq M_{2,3}, \;  p>\frac{6 \gamma  N}{M(4-3\gamma)+2 N}$, or 
   \item $1.62562<\gamma <2,   \; 
   N_{3,1}\leq N<-2.44949, \;  \sqrt{\frac{6 \gamma ^2-12\gamma +N^2}{(3\gamma -4 )^2}}+\frac{N}{3\gamma -4 }<M\leq M_{2,3}, \;  p>\frac{6 \gamma  N}{-3
   \gamma  M+4 M+2 N}$, or 
   \item $\frac{4}{3}<\gamma <1.62562, \;  0<N<2.44949, \;  M_{2,1}\leq M<\frac{2 N^2-6 \gamma }{N(3 \gamma -4)}, \;  p>\frac{6 \gamma  N}{M(4-3\gamma) +2
   N}$, or 
   \item $1.62562<\gamma <2,   \;  0<N<2.44949, \;  M_{2,1}\leq M<\frac{2 N^2-6 \gamma }{N(3 \gamma -4)}, \;  p>\frac{6 \gamma  N}{M(4-3\gamma)+2 N}$, or
   \item $\frac{9}{8}<\gamma <\frac{4}{3}, \;  -2.44949<N<0, \;  M_{2,1}\leq M<\frac{2 N^2-6 \gamma }{N(3 \gamma -4)}, \;  p>\frac{6 \gamma  N}{M(4-3\gamma)+2 N}$, or 
   \item $1<\gamma
   <\frac{9}{8}, \;  N_{3,2}<N<0, \;  M_{2,1}\leq M<\frac{2 N^2-6 \gamma }{N(3 \gamma -4)}, \;  p>\frac{6 \gamma  N}{M(4-3\gamma)+2 N}$, or
   \item $1<\gamma <\frac{9}{8}, \;  -2.44949<N<N_{3,2}, \;  M_{2,1}\leq M\leq M_{2,2}, \;  p>\frac{6 \gamma  N}{M(4-3\gamma) +2
   N}$, or 
   \item $1<\gamma <\frac{9}{8}, \;  0<N<N_{3,3}, \;  \frac{2 N^2-6 \gamma }{N(3 \gamma -4)}<M\leq M_{2,1}, \;  p>\frac{6 \gamma  N}{-3 \gamma 
   M+4 M+2 N}$, or 
   \item $\frac{9}{8}<\gamma <\frac{4}{3}, \;  0<N<N_{3,3}, \;  \frac{2 N^2-6 \gamma }{N(3 \gamma -4)}<M\leq M_{2,1}, \;  p>\frac{6 \gamma 
   N}{M(4-3\gamma)+2 N}$, or 
   \item $1<\gamma <\frac{9}{8}, \;  -2.44949<N<N_{3,2}, \;  M_{2,3}\leq M<\frac{2 N^2-6 \gamma }{N(3 \gamma -4)}, \; 
   p>\frac{6 \gamma  N}{M(4-3\gamma)+2 N}$, or 
   \item $1<\gamma <\frac{9}{8}, \;  2.44949<N\leq N_{3,4}, \;  \sqrt{\frac{6 \gamma ^2-12\gamma +N^2}{(3\gamma -4 )^2}}+\frac{N}{3
   \gamma -4}<M\leq M_{2,3}, \;  p>\frac{6 \gamma  N}{M(4-3\gamma)+2 N}$, or 
   \item $\frac{9}{8}<\gamma <\frac{4}{3}, \;  2.44949<N\leq N_{3,4}, \; 
   \sqrt{\frac{6 \gamma ^2-12\gamma +N^2}{(3\gamma -4 )^2}}+\frac{N}{3\gamma -4 }<M\leq M_{2,3}, \;  p>\frac{6 \gamma  N}{M(4-3\gamma)+2 N}$, or 
   \item $\frac{4}{3}<\gamma
   <1.62562, \;  -2.44949<N\leq N_{3,2}, \;  \frac{2 N^2-6 \gamma }{N(3 \gamma -4)}<M\leq M_{2,3}, \;  p>\frac{6 \gamma  N}{M(4-3\gamma) +2
   N}$, or 
   \item $1.62562<\gamma <2,   \;  -2.44949<N\leq N_{3,2}, \;  \frac{2 N^2-6 \gamma }{N(3 \gamma -4)}<M\leq M_{2,3}, \;  p>\frac{6 \gamma 
   N}{M(4-3\gamma)+2 N}$, or 
   \item $1<\gamma <\frac{9}{8}, \;  N_{3,3}<N<2.44949, \;  M_{2,2}\leq M\leq M_{2,3}, \; 
   p>\frac{6 \gamma  N}{M(4-3\gamma)+2 N}$, or 
   \item $1<\gamma <\frac{9}{8}, \;  N_{3,3}<N<2.44949, \;  \frac{2 N^2-6 \gamma }{N(3 \gamma -4)}<M\leq
   M_{2,1}, \;  p>\frac{6 \gamma  N}{M(4-3\gamma)+2 N}$, or 
   \item $\frac{4}{3}<\gamma <1.62562, \;  N_{3,2}<N<0, \;  \frac{2 N^2-6 \gamma }{3 \gamma 
   N-4 N}<M\leq M_{2,1}, \;  p>\frac{6 \gamma  N}{M(4-3\gamma)+2 N}$, or 
   \item $1.62562<\gamma <2,   \;  N_{3,2}<N<0, \;  \frac{2 N^2-6 \gamma
   }{N(3 \gamma -4)}<M\leq M_{2,1}, \;  p>\frac{6 \gamma  N}{M(4-3\gamma)+2 N}$, or 
   \item $\frac{9}{8}<\gamma <\frac{4}{3}, \;  N_{3,3}\leq N<2.44949, \; 
   \frac{2 N^2-6 \gamma }{N(3 \gamma -4)}<M\leq M_{2,3}, \;  p>\frac{6 \gamma  N}{M(4-3\gamma)+2 N}$, or 
   \item $\gamma =\frac{9}{8}, \;  N>2.61227, \;  \frac{1}{5} \sqrt{64 N^2-378}-\frac{8}{5}
   N<M\leq M_{4,1}, \;  p>\frac{54 N}{5 M+16 N}$, or 
   \item $\gamma =\frac{9}{8}, \;  N<-2.44949, \;  M_{4,1}\leq M<-\frac{1}{5} \sqrt{64 N^2-378}-\frac{8}{5} N, \; 
   p>\frac{54 N}{5M+16 N}$, or 
   \item $\gamma =\frac{9}{8}, \;  2.44949<N\leq 2.61227, \;  \frac{1}{5} \sqrt{64 N^2-378}-\frac{8}{5} N<M\leq M_{4,3}, \;  p>\frac{54 N}{5. M+16 N}$, or 
   \item $\gamma
   =\frac{9}{8}, \;  -2.44949<N<0, \;  M_{4,1}\leq M<\frac{\left(54\, -16 N^2\right)}{5N}, \;  p>\frac{54 N}{5. M+16 N}$, or 
   \item $\gamma =\frac{9}{8}, \;  0<N<2.44949, \;  \frac{
   \left(54-16 N^2\right)}{5N}<M\leq M_{4,1}, \;  p>\frac{54 N}{5M+16 N}$, or 
   \item $\gamma=2, \;  0<N<2.44949, \;  \frac{\left(8N^2-27\right)}{8N}-\frac{1}{8}
   \sqrt{\frac{729 -48 N^2}{N^2}}\leq M<\frac{N^2-6}{N}, \;  p>\frac{6 N}{N-M}$, or 
   \item $\gamma=2, \;  -2.44949<N<0, \;  \frac{N^2-6}{N}<M\leq \frac{1}{8} \sqrt{\frac{729-48N^2}{N^2}}+\frac{\left(8 N^2-27\right)}{8N}, \;  p>\frac{6 N}{N-M}$. 
\end{enumerate}
Where $M_{1,1}$, $M_{1,2}$, $M_{1,3}$ are the first, the second and the third root of the polynomial
$P_1(M)=2 M^3 N+M^2 \left(8 N^2-15\right)+M \left(8 N^3-18 N\right)+21 N^2-72$, respectively. 

$M_{2,1}$, $M_{2,2}$ and $M_{2,3}$ are the first, the second and the third root of the polynomial 
$P_2(M)=-72 \gamma ^3+144 \gamma ^2+M^3 \left(54 \gamma ^3 N-216 \gamma ^2 N+288 \gamma  N-128 N\right)+M^2 \left(108 \gamma ^3-261
   \gamma ^2+120 \gamma -72 \gamma ^2 N^2+192 \gamma  N^2-128 N^2+48\right)$\\
   $+M \left(24 \gamma  N^3-32 N^3-36 \gamma ^3 N+66 \gamma ^2 N-60 \gamma  N+48 N\right)+27 \gamma ^2 N^2-60 \gamma  N^2+12 N^2$, respectively.
$N_{3,1}$, $N_{3,2}$, $N_{3,3}$ and $N_{3,4}$ are the first, the second, the third and the fourth root of the polynomial:   
   $P_3(N)=-1152 \gamma
   ^4+1440 \gamma ^3+1512 \gamma ^2+414 \gamma +64 N^4+\left(321 \gamma ^2-1176 \gamma +96\right) N^2+36$, respectively.
  Finally, $M_{4,1}$  is first root and  $M_{4,3}$ is the third root of the polynomial:  
   $P_4(M)=125 M^3 N+M^2 \left(800 N^2-1650\right)+M \left(1280 N^3-3270 N\right)+5460 N^2-20412$.
\end{enumerate}

\end{document}